\documentclass{article}
\pdfoutput=1

\usepackage[nonatbib]{arxiv}

\usepackage[utf8]{inputenc} 
\usepackage[T1]{fontenc}    
\usepackage[hidelinks]{hyperref}       
\usepackage{url}            
\usepackage{booktabs}       
\usepackage{amsfonts}       
\usepackage{nicefrac}       
\usepackage{microtype}      
\usepackage{lipsum}
\usepackage{amsmath}
\usepackage{numprint}
\usepackage{graphicx}
\usepackage{placeins}
\usepackage{supertabular}
\usepackage[draft, xcolor=dvipsnames]{changes}
\usepackage[ruled]{algorithm2e}
\SetKwInput{KwInput}{Input}
\SetKwInput{KwOutput}{Output}
\newcommand{\JF}{Bittersweet}

\newcommand{\AJ}{OliveGreen}
\definechangesauthor[name={Johan Falkenjack}, color=\JF]{JF}
\definechangesauthor[name={Mattias Villani}, color=\JF]{MV}
\definechangesauthor[name={Arne Jönsson}, color=\AJ]{AJ}
\usepackage[colorinlistoftodos]{todonotes}
\usepackage{caption}

\usepackage[
backend=bibtex,
citestyle=authoryear-icomp
]{biblatex}

\title{Modeling Text Complexity using a Multi-Scale Probit}

\author{
  Johan Falkenjack\\
  Department of Computer Science\\
  Linköping University\\
  Linköping, Sweden \\ 
  \texttt{johan.falkenjack@liu.se} \\
   \And
 Mattias Villani\\
  Department of Computer Science\\
  Linköping University\\
  and\\
  Department of Statistics\\
  Stockholm University\\
  \texttt{mattias.villani@gmail.com} \\
   \And 
 Arne Jönsson\\
  Department of Computer Science\\
  Linköping University\\
  Linköping, Sweden \\
  \texttt{arne.jonsson@liu.se} \\
   \And
}

\begin{document}
\maketitle

\begin{abstract}
We present a novel model for text complexity analysis which can be fitted to ordered categorical data measured on multiple scales, e.g. a corpus with binary responses mixed with a corpus with more than two ordered outcomes. The multiple scales are assumed to be driven by the same underlying latent variable describing the complexity of the text. We propose an easily implemented Gibbs sampler to sample from the posterior distribution by a direct extension of  established data augmentation schemes. By being able to combine multiple corpora with different annotation schemes we can get around the common problem of having more text features than annotated documents, i.e. an example of the $p>n$ problem. The predictive performance of the model is evaluated using both simulated and real world readability data with very promising results. 
\end{abstract}


\section{Introduction}
\label{sec:intro}



Text complexity is a concept inherently tied to the concept of readability. Readability meanwhile is commonly defined as "the sum total (including all the interactions) of all those elements within a given piece of printed material that affect the success a group of readers have with it"~\parencite{Dale1949TheReadability}. That is, while readability is a function of both the text and a specific group of readers, text complexity is a function only of text, or a function of text and a generalised group of readers. There are certainly differences in what makes text difficult to read for different readers, for instance the difficulties a second language learner has might be very different from the difficulties of a reader with dyslexia or aphasia. Through focusing text complexity, we attempt to create a baseline model of complexity which can later on be adapted to the needs of different reader groups.
    
Modern models of readability analysis for classification often use classification algorithms such as SVM~\parencite{Petersen2007NaturalEducation,Feng2010AAssessment,Falkenjack2013FeaturesText} which give us an assessment whether a text is easy-to-read or not. Such models have a very high accuracy, for instance, a model using 117 parameters from shallow, lexical, morphological and syntactic analyses achieves 98,9 \% accuracy~\parencite{Falkenjack2013FeaturesText}. 
However, these models do not tell us much about whether a given text is easier to read than any other text, other than the binary classification. In order to perform a more fine grained prediction we normally need to train the models using a corpus of graded texts, for an overview of such methods see~\textcite{Collins-Thompson2014ComputationalResearch}. 


There are also attempts to grade texts without an extensive corpus of graded texts \parencite{Pitler2008RevisitingQuality,Tanaka-Ishii2010SortingReadability}. \textcite{Tanaka-Ishii2010SortingReadability} present an approach which predicts relative difficulty by modelling pair-wise comparisons between texts. Another strength of this approach is that multiple corpora with text complexity annotated using different scales can be included in the same model. However, a downside of this approach is the necessity of squaring the number of training examples to model the full data set.

Texts released by different publishers of materials for readers with varying reading proficiency are often measured on different scales. Some publishers simply use an "easy-to-read" label, other label material with an intended age group, and others use a qualitative scale of difficulties where materials are placed in ordered categories. All these scales, though containing varying number of categories, in some sense measure the same thing. Text complexity can thus be viewed as a shared latent variable underlying different measures of readability. 

In this paper we propose a new statistical model we refer to as Multi-Scale Probit, based on the well established Ordered Probit model. The Ordered Probit, as well as the traditional Binary assume that the response variable is measured on a single ordinal scale, or is classified with a binary label. The Multi-Scale Probit is able to take sets of data where the response variable is measured on different scales and find the shared latent feature, even without a vignette or "Rosetta stone" translating between the scales.

This allows us to use multiple non-overlapping corpora with text complexity annotated on different scales and find the shared phenomenon of text complexity each annotation scale is based on. In other words, we can take a corpus with texts organised by target reader groups of different ages, a corpus with texts organised by degree of readability, and a corpus of easy-to-read texts, and put them all in the same model and estimate a shared model of text complexity.

In this study we apply this new model to both simulated and real data with promising results.

\FloatBarrier

\section{The Multi-Scale Probit model}
\label{sec:background}

In this section we will present our proposed model. For background we will start by introducing the original dichotomous Probit and the Ordered Probit before showing how this later model can be generalised to multiple scales.

\subsection{The Probit Model}
\label{sec:probit}

The Probit model is a well established statistical model for supervised statistical classification with some properties which makes it especially suitable to Bayesian modelling~\parencite{McCulloch:00}. The Probit model takes the following form for the $i$th observation in the sample

\begin{equation}
\label{eq:probit}
	\Pr(y_i=1 \vert \mathbf{x}_i) = \Phi(\alpha + \mathbf{x}_i^T \boldsymbol{\beta}),\hspace{0.5cm}i=1,\ldots,n,
\end{equation}

where $\Phi$ is the Cumulative Distribution Function (CDF) for the Standard Normal distribution, $y$ is the dependent variable, or label, $\alpha$ is the intercept, $\mathbf{x}$ is the vector of covariates, or features, on which $y$ depend, and $\beta$ a vector of coefficients corresponding to $\mathbf{x}$. In the text complexity case, $y$ could be an indicator value indicating whether the text is easy-to-read or not, while $\mathbf{x}$ is a vector of text feature values. The CDF $\Phi$ can be replaced by any other CDF, for example the logistic to get the Logit model, but the Probit model implies a characterisation where the underlying latent readability variable is normally distributed, which we will now explain in detail.

The Probit model can be interpreted as a \textit{latent variable model}. A latent variable model assumes one or more unobserved, or latent, variables to be the drivers of the observed data. The following model is easily seen to be equivalent to the Probit model in \eqref{eq:probit}

\begin{equation}
\label{eq:latentprobit}
	y_i =
	\begin{cases}
	2 & \text{if }y_i^\ast > 0 \\
	1 &\text{otherwise.}
	\end{cases}
    \quad \text{where} \quad
    y_i^\ast = \alpha + \mathbf{x}_i^T \boldsymbol{\beta} +  \varepsilon_i,
\end{equation}
where $y^\ast$ is the latent variable and $\varepsilon_i$ is independent $N(0,1)$ noise.

A slight reformulation of the model that is more suitable for our purposes is to not fix the threshold at 0 and model the intercept $\alpha$, but rather to model the threshold as $\gamma=-\alpha$, given the equivalent model

\begin{equation}
\label{eq:latentprobitreparam}
	y_i =
	\begin{cases}
	2 & \text{if }y_i^\ast > \gamma \\
	1 &\text{otherwise.}
	\end{cases}
    \quad \text{where} \quad
    y_i^\ast = \mathbf{x}_i^T \boldsymbol{\beta} + \varepsilon_i,
\end{equation}

A latent variable formulation allows us to view the Probit model as a linear regression over an unobserved, or latent, real valued variable which underlies the assigned labels in the classification problem. The observed variable $y$ is simply an indicator of whether $y^\ast$ is larger than the threshold, $\gamma$, see the left part of Figure \ref{fig:probitoprobit} for an illustration. This is particularly useful when different classes are defined by the degree of some linear property as in the case with easy-to-read classification where the underlying property is text complexity, which now is being indirectly modelled on an interval scale. Note also that if the relation between the features and the latent variables is expected to be nonlinear, we can always add polynomial or spline terms in the feature set \parencite{friedman2001elements}.

The latent variable formulation gives an elegant interpretation of the Probit model, but has also very attractive computational properties. The main goal in Bayesian inference is the posterior distribution of $\boldsymbol{\beta}$

\begin{equation}
    p(\boldsymbol{\beta} \vert \mathbf{y},\mathbf{X}) \propto p(\mathbf{y} \vert \boldsymbol{\beta}, \mathbf{X})p(\boldsymbol{\beta}),
\end{equation}

where $\mathbf{y}=(y_1,\ldots,y_n)^T$, $\mathbf{X}=(\mathbf{x}_1,\ldots,\mathbf{x}_n)^T$, $p(\mathbf{y} \vert \boldsymbol{\beta}, \mathbf{X})$ is the likelihood function and $p(\boldsymbol{\beta})$ is the prior distribution. This posterior distribution is mathematically intractable and the usual practice is to explore it by simulation. The latent variable formulation can here be used to obtain a very simple and effective so called Gibbs sampling algorithm. 

The Gibbs sampling algorithm is a type of Markov Chain Monte Carlo (MCMC) simulation to sample from a multivariate probability distribution. The algorithm works iteratively by drawing a new value for each variable conditional on the most recent draws of all other variables. The trick here is to \emph{augment} the observed data $\mathbf{y}$ with the unobserved latent variable $\mathbf{y}^\ast$, and to sample from the joint posterior distribution of both the latent variable $\mathbf{y}^\ast$ and $\boldsymbol{\beta}$. Pseudo code for the algorithm is given in Algorithm \ref{alg:gibbsProbit}; details are provided in the original article by \textcite{Albert1993BayesianData}.

\begin{algorithm}
\SetAlgoLined
\KwInput{response labels $\mathbf{y}$, feature data $\mathbf{X}$, initial value $\boldsymbol{\beta}^{(0)}$,  initial value $\alpha^{(0)}$, number of Gibbs iterations $M$.} 
\BlankLine
\For{$m = 1$ \KwTo $M$}{
   Draw $\mathbf{y}^\ast \vert \alpha^{(m-1)},  \boldsymbol{\beta}^{(m-1)}, \mathbf{y}, \mathbf{X}$ for each observation from truncated normal (TN) distributions.\\
   Draw $\alpha^{(m)}, \boldsymbol{\beta}^{(m)} \vert \mathbf{y}^\ast, \mathbf{X}$ using standard formulas for Bayesian Gaussian linear regression.
} 
\BlankLine
\KwOutput{autocorrelated posterior draws $\alpha^{(1)},\ldots,\alpha^{(M)}$ and $\boldsymbol{\beta}^{(1)},\ldots,\boldsymbol{\beta}^{(M)}$.}
\BlankLine
\caption{The data augmented Gibbs sampler for the Probit model in \eqref{eq:latentprobit} with an intercept.\label{alg:gibbsProbit}}
\end{algorithm}

\subsection{The Ordered Probit model}
\label{sec:oprobit}

The Ordered, or Ordinal, Probit model is an extension of the Probit model from a binary response to a response on the ordinal scale. The assumption of an ordinal response variable allows us to model more than two classes using a single latent variable by estimating different thresholds, $\boldsymbol{\gamma}$, for each class. In essence, we model the probability of belonging to class $c\in \{1,\ldots,C\}$ as

\begin{equation}
\label{eq:oprobit}
	\Pr(y_i=c \vert \mathbf{x}_i) = \Phi(\gamma_c-\mathbf{x}_i^T\boldsymbol{\beta}) - \Phi(\gamma_{c-1}-\mathbf{x}_i^T\boldsymbol{\beta})
\end{equation}

where $\gamma_c$ is the threshold for class $c$ and $\gamma_0=0$.

\begin{figure}
  \begin{minipage}[b]{0.49\linewidth}\captionsetup{width=\textwidth}
    \centering
    \includegraphics[width=\linewidth]{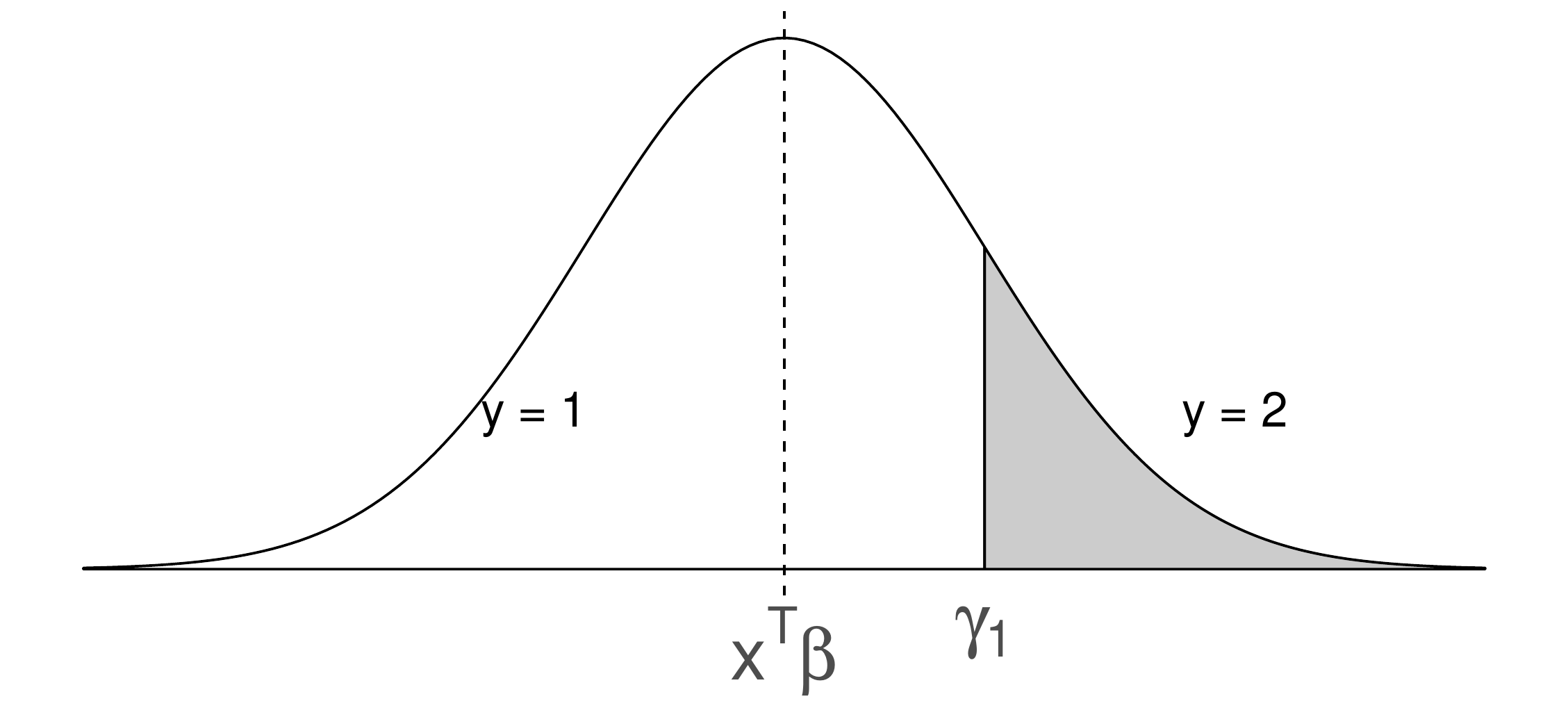} 
    \caption*{Probit} 
  \end{minipage} 
  \hfill
  \begin{minipage}[b]{0.49\linewidth}\captionsetup{width=\textwidth}
    \centering
    \includegraphics[width=\linewidth]{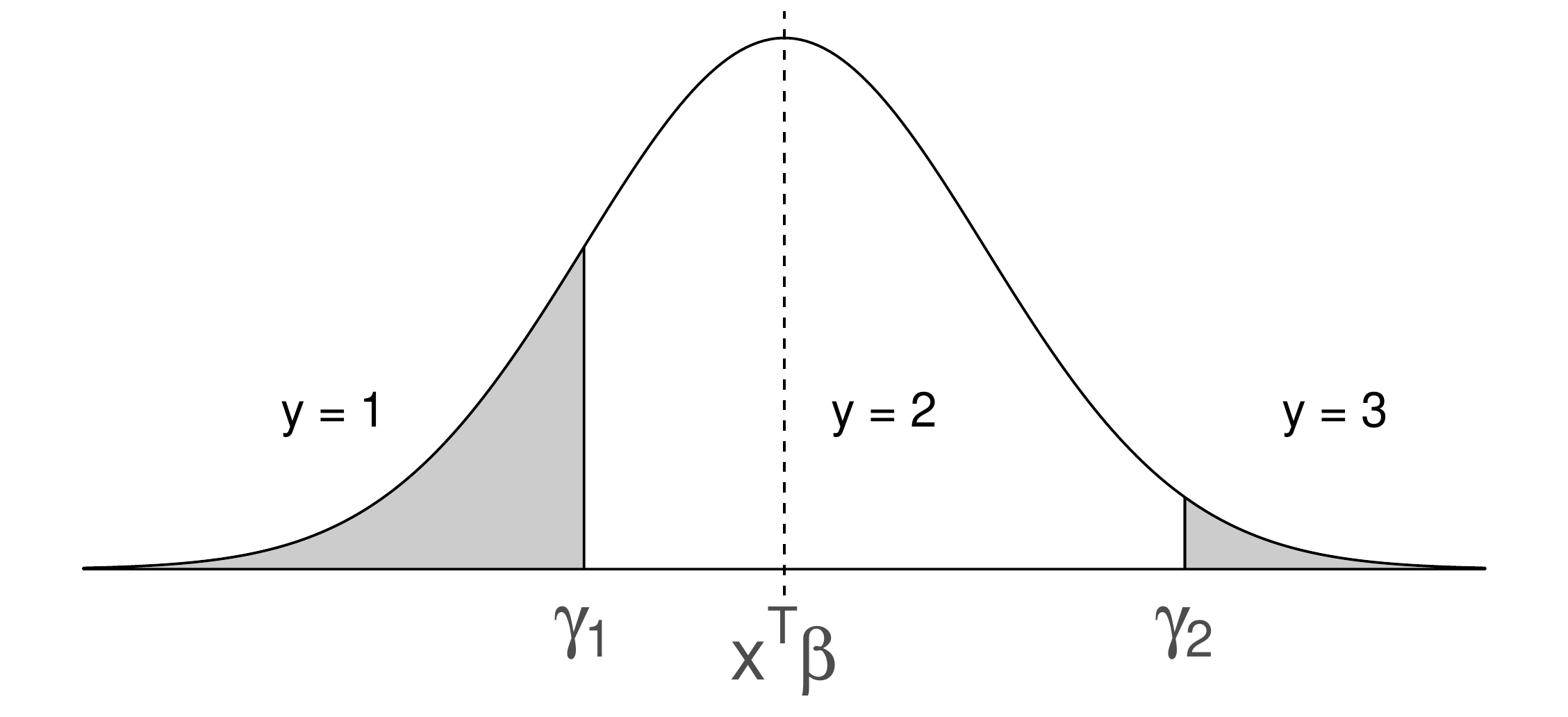} 
    \caption*{Ordered Probit} 
  \end{minipage}
    \caption{Latent variable representation of the Probit (left) and the Ordered Probit (right).
    }
  \label{fig:probitoprobit}
\end{figure}

The latent variable model in Equation \eqref{eq:latentprobitreparam} can be extended to $C>2$ classes: 

\begin{equation}
\label{eq:latentoprobit}
    y_i= \begin{cases}
    1~~ \text{if}~~y_i^\ast \le \gamma_1, \\
    2~~ \text{if}~~\gamma_1<y_i^\ast \le \gamma_2, \\
    3~~ \text{if}~~\gamma_2 <y_i^\ast \le \gamma_3 \\
    \vdots \\
    C~~ \text{if}~~ y_i^\ast > \gamma_{C-1}.
    \end{cases}
    \quad \text{where} \quad
    y_i^\ast = \mathbf{x}_i^T\boldsymbol{\beta} + \varepsilon
\end{equation}

In this case, the observed variable $y_i$ is an indicator for which interval the latent variable $y_i^\ast$ falls within or, in other words, which ordinal class $y_i$ belongs to. In the case with $C=2$, the ordered Probit reduces to a regular binary Probit which is illustrated in Figure \ref{fig:probitoprobit}.

The joint posterior of $\boldsymbol{\beta}$ and $\boldsymbol{\gamma}$ in the ordered Probit is given by the following equation

\begin{equation}
    \label{eq:oprobit:posterior}
    \pi(\boldsymbol{\beta}, \boldsymbol{\gamma} \vert \mathbf{y}, \mathbf{X}) \propto \pi(\boldsymbol{\beta}, \boldsymbol{\gamma})\prod^{n}_{i=1}\prod^{C}_{c=1}[\Phi(\gamma_c-\mathbf{x}_i^T\boldsymbol{\beta}) - \Phi(\gamma_{c-1}-\mathbf{x}_i^T\boldsymbol{\beta})]^{1(y_i = c)}
\end{equation}

where $\pi(\boldsymbol{\beta}, \boldsymbol{\gamma})$ is the prior. Similar to the regular Probit, this posterior is largely intractable, though some point estimates can be approximated \parencite{Cowles1996AcceleratingModels}. Similar to the Gibbs sampling algorithm for the binary Probit in Algorithm \ref{alg:gibbsProbit}, we can augment the data with the latent variable $\mathbf{y}^\ast$ and explore the joint posterior of $\boldsymbol{\beta}$, $\boldsymbol{\gamma}$ and $\mathbf{y}^\ast$ by a three block Gibbs sampler. However, the full conditional posterior of $\boldsymbol{\gamma}$ is intractable but can be sampled by adding a Metropolis-Hastings (MH) step to the Gibbs sampler. This is first presented in \textcite{Albert1993BayesianData} and later improved in \textcite{Cowles1996AcceleratingModels} by adding blocking and a Metropolis-Hastings-within-Gibbs step to handle problems with slow mixing of the original approach. A similar approach, again by \textcite{Albert2001SequentialData}, sidestepped the Metropolis-Hastings-within-Gibbs step of \citeauthor{Cowles1996AcceleratingModels} while retaining the same form of blocking. Here, we will present an approach using Metropolis-Hastings-within-Gibbs based on the \citeauthor{Cowles1996AcceleratingModels} sampler. The Gibbs sampler is presented in Algorithm \ref{alg:gibbsOrderedProbit}, with the necessary conditional posterior distributions obtained as follows.

For $\boldsymbol{\gamma}$ the full conditional posterior is proportional to

\begin{equation}
    \label{eq:oprobit:posterior:gamma}
    \gamma_j \mid \mathbf{y}^\ast, \mathbf{y}, \boldsymbol{\beta} \sim
    \prod^n_{i=1}\left[
        1(y_i=c)1(\gamma_{c-1} \le y^\ast_i \le \gamma_c) +
        1(y_i=c+1)1(\gamma_{c} \le y^\ast_i \le \gamma_{c+1})
    \right]
\end{equation}

which is intractable due to the unknown proportionality constant. \cite{Cowles1996AcceleratingModels} proposed a Metropolis-Hastings step using a truncated Normal (TN) proposal distribution with the $\boldsymbol{\gamma}$ values from the previous state in the chain as upper cut-off points.

For the latent response variable $y_i^\ast$, the conditional posterior is

\begin{equation}
    \label{eq:oprobit:posterior:latent}
    y^\ast_i \mid \boldsymbol{\beta}, \boldsymbol{\gamma}, y_i = j \sim \mathrm{TN}(\mathbf{x}^T_i \boldsymbol{\beta}, 1, \gamma_{j-1}, \gamma_j) 
\end{equation}

where $\mathrm{TN}(\mathbf{x}^T_i \boldsymbol{\beta}, 1, \gamma_{j-1}, \gamma_j)$ refers to the Normal distribution with mean $\mathbf{x}^T\boldsymbol{\beta}$ and variance 1, truncated to the interval $(\gamma_{j-i}, \gamma_j)$.

Lastly, given the proper conjugate prior $\text{Normal}(\mu, \boldsymbol{\Sigma})$, the conditional posterior of $\boldsymbol{\beta}$ is

\begin{equation}
    \label{eq:oprobit:posterior:beta}
    \boldsymbol{\beta} \mid \mathbf{y}^\ast, \mathbf{X} \sim 
    \text{Normal}((\boldsymbol{\Sigma}^{-1}+\mathbf{X}^T\mathbf{X})^{-1}(\boldsymbol{\Sigma}^{-1}\mu+\mathbf{X}^T\mathbf{y}^\ast),
    (\boldsymbol{\Sigma}^{-1}+\mathbf{X}^T\mathbf{X})^{-1}).
\end{equation}

That is, for each step in the chain, we draw from the posterior of a linear regression model with the current values of the latent variable as response.

\begin{algorithm}
\SetAlgoLined
\KwInput{response labels $\mathbf{y}$, feature data $\mathbf{X}$, initial values for $\boldsymbol{\beta}$ and $\boldsymbol{\gamma}$, the prior mean and precision of $\boldsymbol{\beta}$, $\boldsymbol{\mu}_0$ and $\boldsymbol{\Lambda}_0$, a tuning parameter for the MH proposal distribution $\sigma^2_\gamma$, number of Gibbs iterations $M$.} 
\BlankLine
\For{$m = 1$ \KwTo $M$}{
   \For{$c = 1$ \KwTo $C$}{
      Simulate $\gamma_c^{(m)} \vert \mathbf{y}^\ast, \boldsymbol{\beta}$ from the truncated normal distribution $\mathrm{TN}(\gamma_c^{(m-1)}, \sigma^2_\gamma)$ on the interval $[\gamma_{c-1}^{(m)},\gamma_{c+1}^{(m-1)}]$. 
   }
   Perform a Metropolis-Hastings accept/reject for $\boldsymbol{\gamma}^{(m)}$.
   
   \For{$i = 1$ \KwTo $n$}{
      Simulate $y^\ast_i | \boldsymbol{\gamma}, \boldsymbol{\beta}, \mathbf{y}, \mathbf{X}$ from the truncated normal distribution $\mathrm{TN}(\mathbf{x}_i\boldsymbol{\beta}^{(m)}, 1)$ on the interval $[\gamma_{y_i},\gamma_{y_i+1}]$.
   }
   
   Simulate $\boldsymbol{\beta}^{(m)} \vert \mathbf{y}^\ast, \mathbf{X}$ from
 the multivariate normal distribution in \eqref{eq:oprobit:posterior:beta}.
} 
\BlankLine
\KwOutput{autocorrelated posterior draws $\boldsymbol{\beta}^{(1)},\ldots,\boldsymbol{\beta}^{(M)}$ and $\boldsymbol{\gamma}^{(1)},\ldots,\boldsymbol{\gamma}^{(M)}$.}
\BlankLine
\caption{The Gibbs sampler in \textcite{Cowles1996AcceleratingModels}, adapted to the Ordered Probit model without intercept in \eqref{eq:latentoprobit}.\label{alg:gibbsOrderedProbit}}
\end{algorithm}

\subsection{The Multi-Scale Probit}
\label{sec:newprobit}

Our methodological contribution in this article stems from the observation that the latent variable formulation of the binary and ordered Probit models opens up the possibility to learn about readability from multiple corpora that each use a different ordinal scale. The same underlying latent text complexity variable is assumed to drive all of the observed readability scores in the different corpora. To fix ideas, we can imagine a data set where 20$\%$ of the examples come from the binary easy/hard labelling in the left hand side of Figure~\ref{fig:probitoprobit} and the remaining examples comes from the scale with three redability classes, easy/medium/hard, in the right part of Figure~\ref{fig:probitoprobit}. Note that, for example, 'easy' may have a different meaning in the two scales, and is something that we will learn from the data.

We propose an extension of the existing Probit framework, here referred to as Multi-Scale Probit. Assume that a total of $n$ examples are labelled on $S$ different scales. Define a variable $s_i$, for $i=1,...,n$, such that $s_i=s$ means that response label, $y_i$ is measured on scale $s$. Also, let $C^{(s)}$ denote the number of classes for scale $s$. Finally, define $\boldsymbol{\gamma}^{(s)}$ as the collection of thresholds for scale $s$. The Multi-Scale Probit is then defined as

\begin{equation}
\label{eq:latentnewprobit}
    y_i = \begin{cases}
    1~~ \text{if}~~y^\ast_i \le \gamma^{(s_i)}_{1}, \\
    2~~ \text{if}~~\gamma^{(s_i)}_{1} \le y^\ast_i \le \gamma^{(s_i)}_{2}, \\
    3~~ \text{if}~~\gamma^{(s_i)}_{2} \le y^\ast_i \le \gamma^{(s_i)}_{3} \\
    \vdots \\
    C^{(s_i)}~~ \text{if}~~ \gamma^{(s_i)}_{C^{(s_i)}-1} < y^\ast_i.
    \end{cases}
    \quad \text{where} \quad
    y^\ast_i = \mathbf{x}_i^T\boldsymbol{\beta} + \varepsilon,
\end{equation}
for $i=1,...,n$.

The above formulation gives us the ability to fit a single latent variable $y^\ast$ with coefficients $\boldsymbol{\beta}$ to the observed data. It should now be clear why no intercept is included in $\boldsymbol{\beta}$. As $\boldsymbol{\beta}$ is shared among all response variables regardless of scale, an intercept coefficient in $\boldsymbol{\beta}$ would mean that some $\gamma^{(s)}_1$ would have to be locked down to 0. This would then shift the $\boldsymbol{\gamma}^{(s)}$ vectors for all other response variables with regards to that intercept which seems counter-intuitive, it also means one response variable is treated differently from the others making the model more complex.

The joint posterior for this Probit is very similar to the posterior for the Ordered Probit. As the different $\boldsymbol{\gamma}^{(s)}$ are independent except through $\boldsymbol{\beta}$, the only difference is that we need to add a mapping from each $y_i$ to the set of thresholds $\gamma^{(s)}$ corresponding to its scale, which we, as mentioned above, denote $\gamma^{(s)}$.

\begin{equation}
    \label{eq:hprobit:posterior}
    \pi(\boldsymbol{\beta}, \boldsymbol{\gamma} \vert \mathbf{y}, \mathbf{X}) \propto \pi(\boldsymbol{\beta}, \boldsymbol{\gamma})
    \prod^{n}_{i=1}
        \prod^{S}_{s=1}
            [\Phi({\gamma^{(s)}_{y_i}}-\mathbf{x}^T_i\boldsymbol{\beta}) -
            \Phi(({\gamma^{(s)}_{y_i-1}}-\mathbf{x}^T_i\boldsymbol{\beta})]^{1(s_i=s)},
\end{equation}
where $S$ is the total number of scales. This posterior is as intractable as the posterior of the regular Ordered Probit and again we apply the Gibbs sampling approach to simulate from the joint posterior. The full conditional posteriors for the three blocks are given below.

\subsubsection*{The full conditional posterior for $\gamma^{(s)}_{j}$}

\begin{equation}
    \label{eq:hprobit:posterior:gamma}
    \pi\left({\gamma^{(s)}_{j}} \mid \boldsymbol{\gamma}^{(s)}_{-j}, \mathbf{y}^\ast, \mathbf{y}, \boldsymbol{\beta}\right) \propto
    \prod_{i:s_i=s}\left[
        1(y_i=j)1({\gamma^{(s)}_{j-1}} \le y^\ast_i \le {\gamma^{(s)}_{j}}) +
        1(y_i=j+1)1({\gamma^{(s)}_{j}} \le y^\ast_i \le {\gamma^{(s)}_{j+1}})
    \right],
\end{equation}
where $\boldsymbol{\gamma}^{(s)}_{-j}$ contains all thresholds except $\gamma^{(s)}_{j}$  for scale $s$, and the product runs over all observations from scale $s$. This distribution is not of known form, and we sample it with a Metropolis-Hastings-within-Gibbs step.

\subsubsection*{The conditional posterior for $\boldsymbol{\beta}$} only depends on $\mathbf{y}^\ast$ and thus is the same as \eqref{eq:oprobit:posterior:beta}.

\subsubsection*{The full conditional for the latent variable, $y^\ast_i$}
\begin{equation}
    \label{eq:hprobit:posterior:latent}
    y^\ast_i \mid \boldsymbol{\beta}, \boldsymbol{\gamma}, y_i = j, s_i =s \sim \mathrm{TN}(\mathbf{x}^T_i \boldsymbol{\beta}, 1, {\gamma^{(s)}_{j-1}}, {\gamma^{(s)}_{j}}) 
\end{equation}

Turning these conditionals into a Gibbs sampler is also very similar to Ordered Probit and is illustrated in Algorithm \ref{alg:gibbsMultiScaleProbit}.

\begin{algorithm}
\SetAlgoLined
\KwInput{response labels $\mathbf{y}$, feature data $\mathbf{X}$, initial values for $\boldsymbol{\beta}$ and $\boldsymbol{\gamma}^{(1)},\ldots,\boldsymbol{\gamma}^{(S)}$, the prior mean and precision of $\boldsymbol{\beta}$, $\boldsymbol{\mu}_0$ and $\boldsymbol{\Lambda}_0$, a vector of tuning parameters for the MH proposal distribution $\sigma^2_{\gamma^{(1)}},\ldots,\sigma^2_{\gamma^{(S)}}$, number of Gibbs iterations $M$.} 
\BlankLine
\For{$m = 1$ \KwTo $M$}{
   \For{$s = 1$ \KwTo $S$}{
       \For{$c = 1$ \KwTo $C^{(s)}$}{
          Simulate $\gamma_c^{(s)(m)} \vert \mathbf{y}^{\ast(s)}, \boldsymbol{\beta}$ from $\mathrm{TN}(\gamma_c^{(s)(m-1)}, \sigma^2_{\gamma^{(s)}},\gamma_{c-1}^{(s)(m)},\gamma_{c+1}^{(s)(m-1)})$. 
       }
   Perform a Metropolis-Hastings accept/reject for $\boldsymbol{\gamma}^{(s)}$.
   }
   \For{$i = 1$ \KwTo $n$}{
      Simulate $y^{\ast}_i | s_i, \boldsymbol{\gamma}, \boldsymbol{\beta}, \mathbf{y}, \mathbf{X}$ from the $\mathrm{TN}(\mathbf{x}_i^T\boldsymbol{\beta}^{(m)}, 1,\gamma_{y_i}^{(s_i)},\gamma_{y_i+1}^{(s_i)})$.
   }
   Simulate $\boldsymbol{\beta}^{(m)} \vert \mathbf{y}^\ast, \mathbf{X}$ from
 the multivariate normal distribution in \eqref{eq:oprobit:posterior:beta}.
} 
\BlankLine
\KwOutput{autocorrelated posterior draws $(\boldsymbol{\beta},\boldsymbol{\gamma}^{(1)},\ldots,\boldsymbol{\gamma}^{(S)})^{(1)},\ldots,(\boldsymbol{\beta},\boldsymbol{\gamma}^{(1)},\ldots,\boldsymbol{\gamma}^{(S)})^{(M)}$.}
\BlankLine
\caption{The Gibbs sampler for the Multi-Scale Probit model in \eqref{eq:latentnewprobit}.\label{alg:gibbsMultiScaleProbit}}
\end{algorithm}

\subsubsection*{Implementation}

Our implementation of this Gibbs sampler was built using Armadillo \parencite{Sanderson2016Armadillo:Algebra} and wrapped in R \parencite{RCoreTeam2018R:Computing} using the Rcpp \parencite{Eddelbuettel2011Rcpp:Integration} and RcppArmadillo \parencite{Eddelbuettel2014RcppArmadillo:Algebra} libraries. 

\FloatBarrier

\section{Evaluation procedure}


\subsection{Simulation setup}

To evaluate the performance of the Multi-Scale Probit model in comparison to the Binary and Ordered Probit model we first inspect the model fit and prediction performance using simulated data. For this purpose we implemented Algorithm \ref{alg:simdata} which randomly generates data using the assumptions of the Probit model.

\begin{algorithm}
\SetAlgoLined
\KwInput{the number of different scales $S$, number of observations per scale $n$, number of covariates $p$, a vector of class labels for each scale $\mathbf{c}^{(1)}, \ldots, \mathbf{c}^{(s)}$, smallest acceptable number of observations per class $k$.} 
\BlankLine
Draw $\boldsymbol{\beta} \sim N_p(0, I)$.\\
\For{$s=1$ \KwTo $S$}{
    Draw a $n \times p$ training matrix $\mathbf{X}^{(s)}$ by drawing $\mathbf{x} \sim N_p(0, I)$ $n$ times.\\
    \Repeat{$\mathbf{y}^{(s)}$ has at least $k$ instances of each class in $\mathbf{c}^{(s)}$}{
        Draw $\boldsymbol{\gamma}^{(s)} \sim N_{|\mathbf{c}^{(s)}|-1}(0, 5)$\\
        \For{$i=1$ \KwTo $n_t$}{
            Draw $y^{\ast(s)}_i \sim N( \mathbf{x}^{(s)}_i \boldsymbol{\beta}, 1)$\\
            Compute $y^{(s)}_i$ by finding the interval corresponding to $y^{\ast(s)}_i$ in $\{ (-\infty, \gamma^{(s)}_1), \ldots, [\gamma^{(s)}_c, \infty)\}$
        }
    }
}
\BlankLine
\KwOutput{covariate matrices $\mathbf{X}^{(1)},\ldots,\mathbf{X}^{(s)}$, corresponding response vectors $\mathbf{y}^{(1)},\ldots,\mathbf{y}^{(s)}$, latent variable vectors $\mathbf{y}^{\ast(1)},\ldots,\mathbf{y}^{\ast(s)}$, $\boldsymbol{\beta}$, $\boldsymbol{\gamma}^{(1)}, \ldots, \boldsymbol{\gamma}^{(s)}$.}
\BlankLine
\caption{Algorithm to simulate data to test the models.\label{alg:simdata}}
\end{algorithm}

Using such simulated data we can examine how well the different versions of the model estimate the known $\boldsymbol{\beta}$ and $\boldsymbol{\gamma}$ for each simulated data set. By computing the root-mean-square error (RMSE) for each draw of $\boldsymbol{\beta}$ and $\boldsymbol{\gamma}^{(s)}$ we can inspect the posterior distribution of these RMSEs and plot them to compare the performance of the established Probit models and our proposed Multi-Scale Probit. On real world text complexity data we evaluate the fit of a model by testing its predictive performance on a validation set.

One problem with the text complexity data we have available is that with regards to the established models the data suffers from the $p > n$ problem, i.e. that the number of covariates is larger than the number of observations. The linear regression for the latent variable $y^\ast$ requires the solution of a system of equations which in the $p > n$ case is under-determined and thus have infinitely many solutions. The Multi-Scale Probit, being able to use all three corpora in a single model, does not suffer from this problem, which is one reason for why we propose it.

We confront the $p > n$ problem by using a regularising $N(0,10)$ prior on $\boldsymbol{\beta}$. This so called $L2$ regularisation makes the regression problem tractable even in an under-determined situation \parencite{friedman2001elements}. The prior variance, i.e. the degree of regularisation, can also be estimated in a hierarchical Bayesian approach, see Section \ref{sec:p>n}.

Since the models are fit by simulating from the posterior distribution using MCMC, we can obtain the distribution of the evaluation metrics by computing them for each draw. This allows us to plot kernel density estimates of the posterior predictive distributions of some common summary statistics for classification and ranking. The kernel smoothing is only used to make the plots easier to interpret and does not impact the performance or the conclusions drawn.

We will show a few examples of in-sample performance but as there is little difference between different models on in-sample performance we will focus mainly on out-of sample performance evaluated using a validation set.

Below, we present the summary statistics for which we compute the posterior predictive distributions.

\subsection{Classification, the F-measure}

Precision for a class $C$ is the ratio of instances correctly classified as $C$ to all instances classified as $C$:

\begin{equation}
    \label{eq:precision}
    \text{Precision} = \frac{|\text{True positive}|}{|\text{True positive}|+|\text{False positive}|}
\end{equation}

Recall for the class $C$ is the ratio of instances correctly classified as $C$ to all instances of $C$ in the data:

\begin{equation}
    \label{eq:recall}
    \text{Recall} = \frac{|\text{True positive}|}{|\text{True positive}|+|\text{False negative}|}
\end{equation}

The F-measure \parencite{VanRijsbergen1974FoundationEvaluation} is a well established evaluation metric for classification algorithms consisting of the harmonic mean of \textit{precision} and \textit{recall}. In our multi-class context $F_1$ scores are computed for each class.

\begin{equation}
    \label{eq:f1}
    F = 2 \cdot \frac{\text{Precision} \cdot \text{Recall}}{\text{Precision} + \text{Recall}}
\end{equation}

\subsection{Ranking, the Kendall Rank Correlation}


Since the latent variable gives a near total ordering of all data points, the Multi-Scale Probit can be viewed as not only a model for classification but for ranking. The quality of this ranking can be assessed using the Kendall Rank Correlation Coefficient, or Kendall's $\tau$ \parencite{Kendall1955RankEd.}. As the reference will not be a total order, we use a version called $\tau_B$ which makes adjustments for ties. Kendall's $\tau_B$ takes values on the interval $-1 \leq \tau_B \leq 1$ where $1$ indicates perfect correlation, $0$ indicates no correlation and $-1$ indicates an inverse correlation.





\subsection{A combined evaluation metric}

We also compute the harmonic mean of the $F_1$ and $\tau_B$. The harmonic mean tends to put more weight on small outliers and less weight on large. The purpose of this measure is to mitigate any impact from trade-offs between single performance metrics.

\FloatBarrier

\section{Simulation experiments}
\label{sec:experiments}

The Multi-Class Probit model is explored on real data in the next section in an application to text complexity analysis.  In this section, we investigate the performance of the model and its associated Bayesian inference machinery on simulated data. The first experiment simulates data sets from the same underlying latent variable distribution but uses three different sets of thresholds $\boldsymbol{\gamma}^{(s)}$, hence producing three data sets on different scales. We compare the performance of traditional Probit and Ordered Probit on each data set with the performance of the Multi-Scale Probit applied on all data simultaneously. The second experiment repeats Experiment 1, but for the $p > n$ situation. In the following, we will refer to the Binary and Ordered Probit models as single-scale Probits to differentiate them from our new Multi-Scale Probit.

\subsection{Experiment 1: Simulated data, $p < n$}
\label{sec:experiment1}

A data set consisting of three different subsets are randomly generated using the definitions of the Probit and Ordered Probit models with the same $\boldsymbol{\beta}$ vector for each data set but using three different $\boldsymbol{\gamma^{(s)}}$ using Algorithm \ref{alg:simdata}. The parameters of these data are displayed in Table \ref{tab:simparams600}. Note that the parameters for simulating Set 2 and Set 3 are exactly the same and we thus expect similar results for single-scale models applied to these.

\begin{table}
\begin{center}
\caption{\label{tab:simparams600}Experimental parameters for testing our model under $p<n$ conditions.}
\begin{tabular}[]{l p{10cm}}
\bf{Parameter} & \bf{Value(s)} \\
\hline
No. of repetitions$^1$ & 500 \\
Total number of draws & 250 000 (product of $^1$ and $^2$) \\
\multicolumn{2}{l}{\textbf{Data}}\\
\hline
No. of covariates ($p$) & 48\\
No. of data points & 400 per $\boldsymbol{\gamma}$ vector, with at least 1 instance per class label. \\
Number of class labels per scale & $(1, 3, 3)$ \\
\multicolumn{2}{l}{\textbf{MCMC hyper-parameters}} \\
\hline
Burn in phase & 50000 steps \\
Thinning & 1 step in 100 is stored \\
No. of stored draws$^2$ & 500 \\
$\sigma^2_{\gamma^{(1)}}$ & 1.0,\\
$\sigma^2_{\gamma^{(2)}}$ & 0.3\\
$\sigma^2_{\gamma^{(3)}}$ & 0.3\\
& All empirically chosen to get a mean acceptance rate close to 0.234 as suggested by \textcite{Roberts1997WeakAlgorithms} \\
$\mu_0$ & (0, ..., 0) \\
$\Lambda_0$ & $1 \times I$ \\
\hline
\end{tabular}
\end{center}
\end{table}

These data are then fed into our Gibbs sampler, using the parameters shown in Table \ref{tab:simparams600}. Four different simulations are performed, one for each data set, i.e. one Binary Probit simulation for set 1, two Ordered Probit simulations for set 2 and 3, and one Multi-Scale Probit simulation using all three data sets at once. This procedure is repeated 500 times using different values for $\boldsymbol{\beta}$ and $\boldsymbol{\gamma}$.

\begin{figure}[h!]
    \begin{minipage}[b]{0.32\linewidth}\captionsetup{width=\textwidth}
    \centering
    \includegraphics[width=\linewidth]{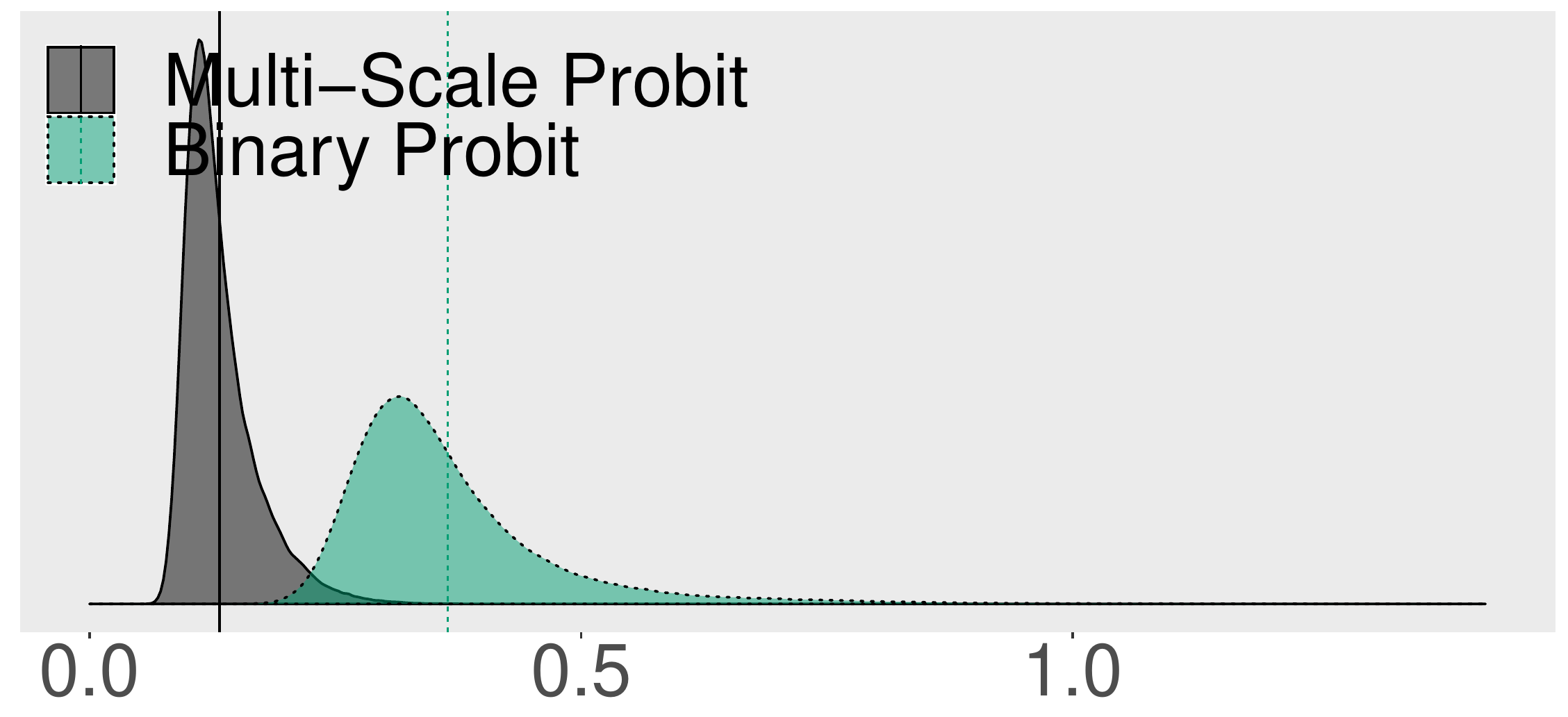} 
    \caption*{1)} 
  \end{minipage} 
  \hfill
  \begin{minipage}[b]{0.32\linewidth}\captionsetup{width=\textwidth}
    \centering
    \includegraphics[width=\linewidth]{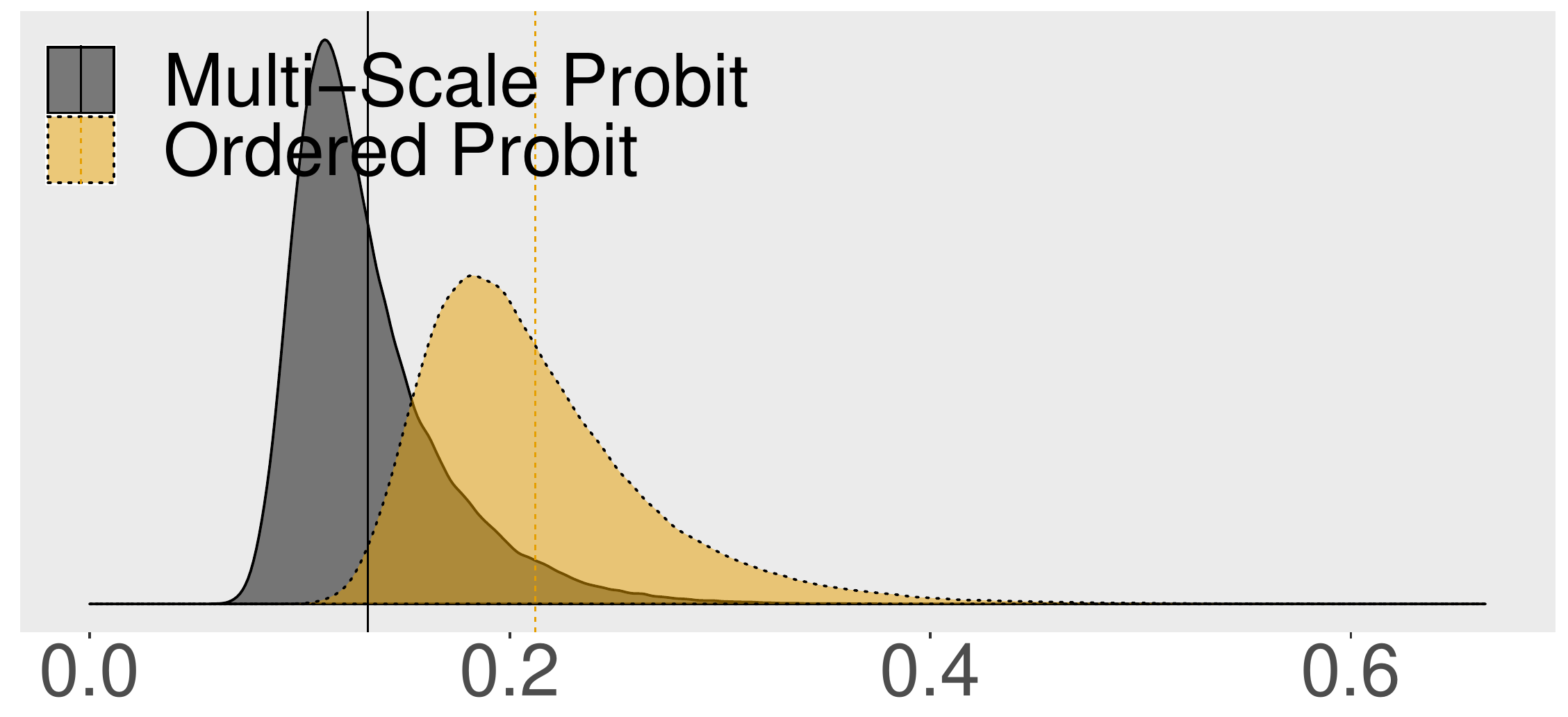} 
    \caption*{2)} 
  \end{minipage}
  \hfill
    \begin{minipage}[b]{0.32\linewidth}\captionsetup{width=\textwidth}
    \centering
    \includegraphics[width=\linewidth]{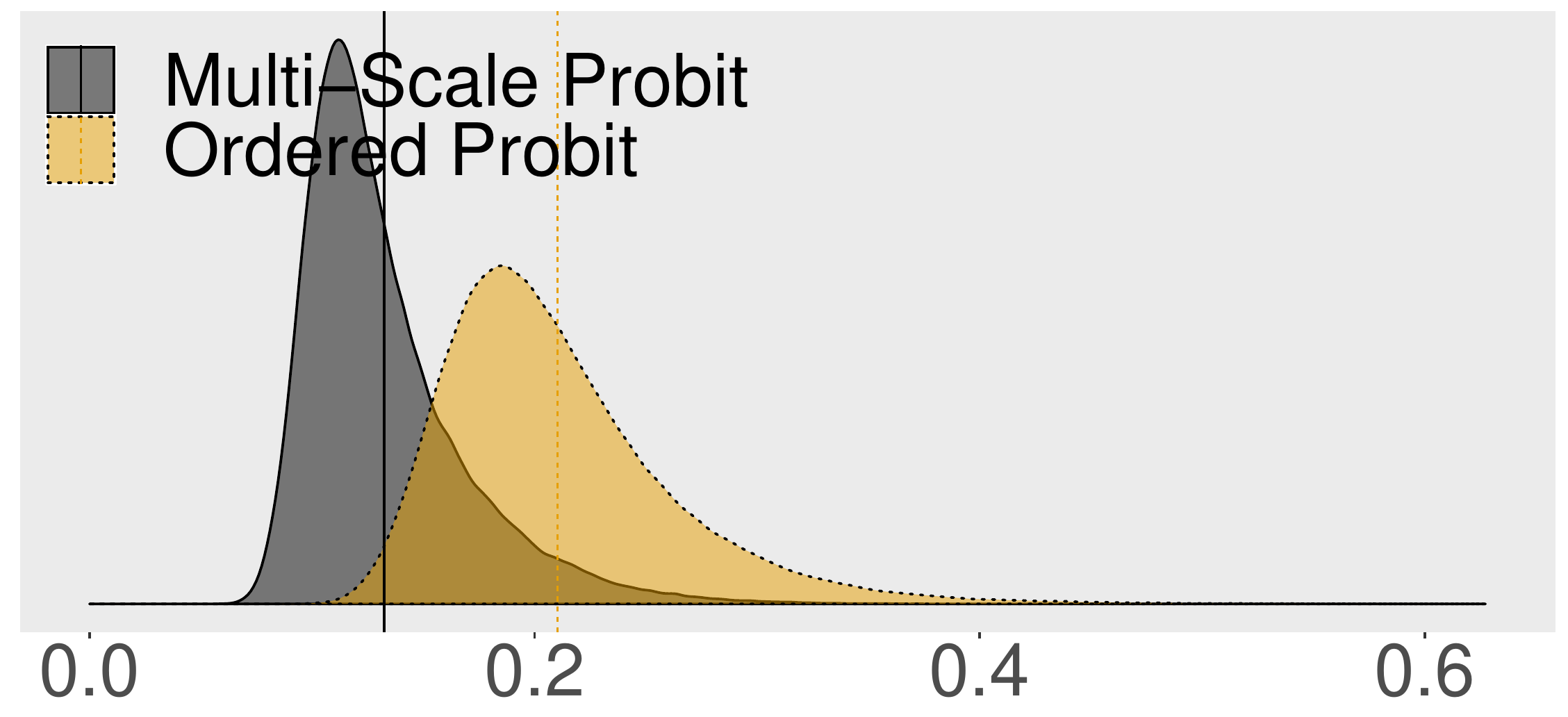} 
    \caption*{3)} 
  \end{minipage} 
  \caption{The posterior of $\boldsymbol{\beta}_{RMSE}$ on the three scales for all 500 simulated $p < n$ data sets.
  \label{fig:beta_rmse600}}
\end{figure}

After running the Gibbs sampler on our 500 different data sets, we start by inspecting the distribution of posterior root-mean-square error for $\boldsymbol{\beta}$ ($\boldsymbol{\beta}_{RMSE}$) for all 250 000 draws made during the 500 repetitions. Plots for each data set is provided in Figure \ref{fig:beta_rmse600}. In each sub-figure the posterior distribution of $\boldsymbol{\beta}$ given a single-scale model, Probit or Ordinal Probit, estimated using a single data set is compared to the posterior distribution of $\boldsymbol{\beta}$ given all three data sets. The plots indicate that the error is smaller using the Multi-Scale Probit model. This is not surprising as the value of $\boldsymbol{\beta}$ is estimated using three times as much data in the Multi-Scale compared to the single-scale models. Note that Multi-Scale model distribution is exactly the same in all three sub-plots, but the scales of the graphs differ.

\begin{figure}[h!]
    \begin{minipage}[b]{0.32\linewidth}\captionsetup{width=\textwidth}
    \centering
    \includegraphics[width=\linewidth]{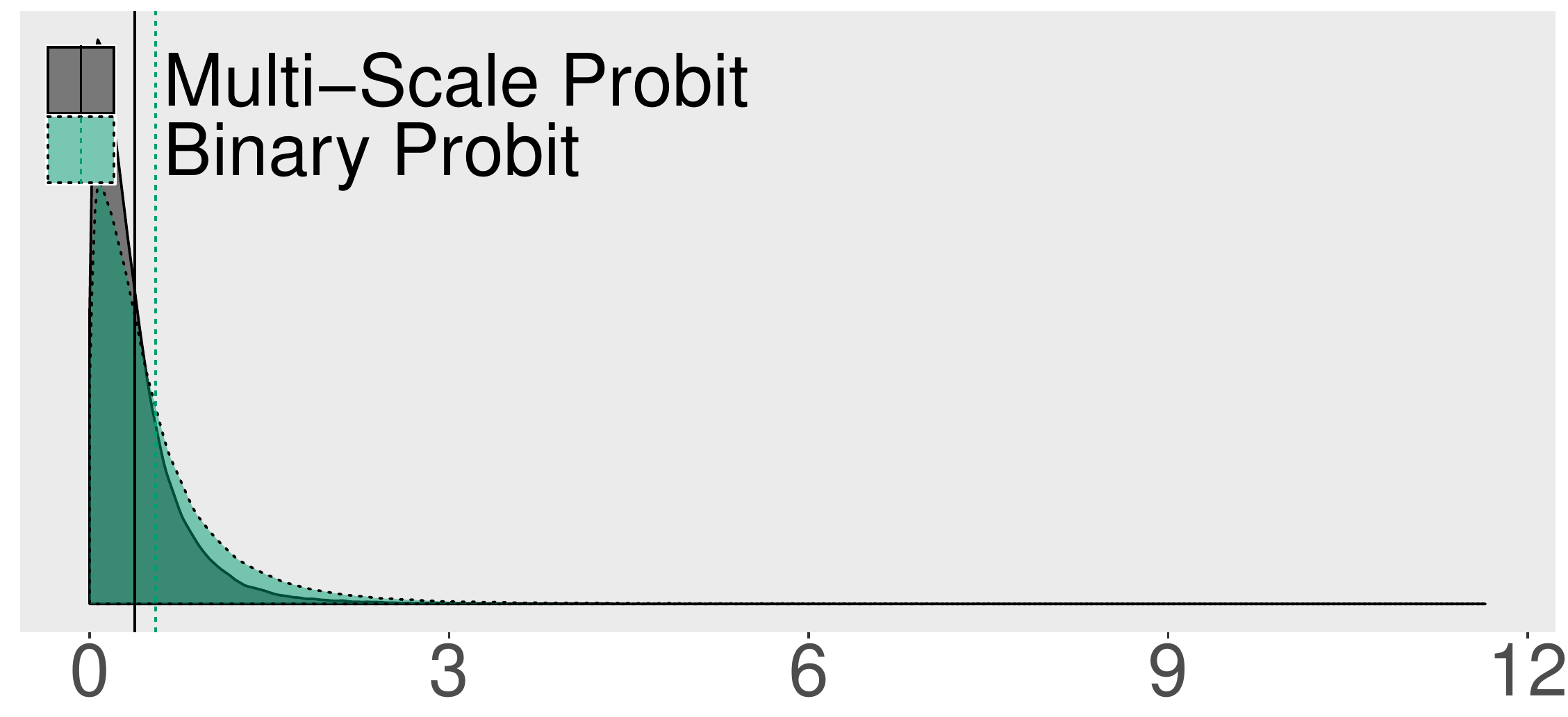} 
    \caption*{$\boldsymbol{\gamma}^{(1)}$} 
  \end{minipage} 
  \hfill
  \begin{minipage}[b]{0.32\linewidth}\captionsetup{width=\textwidth}
    \centering
    \includegraphics[width=\linewidth]{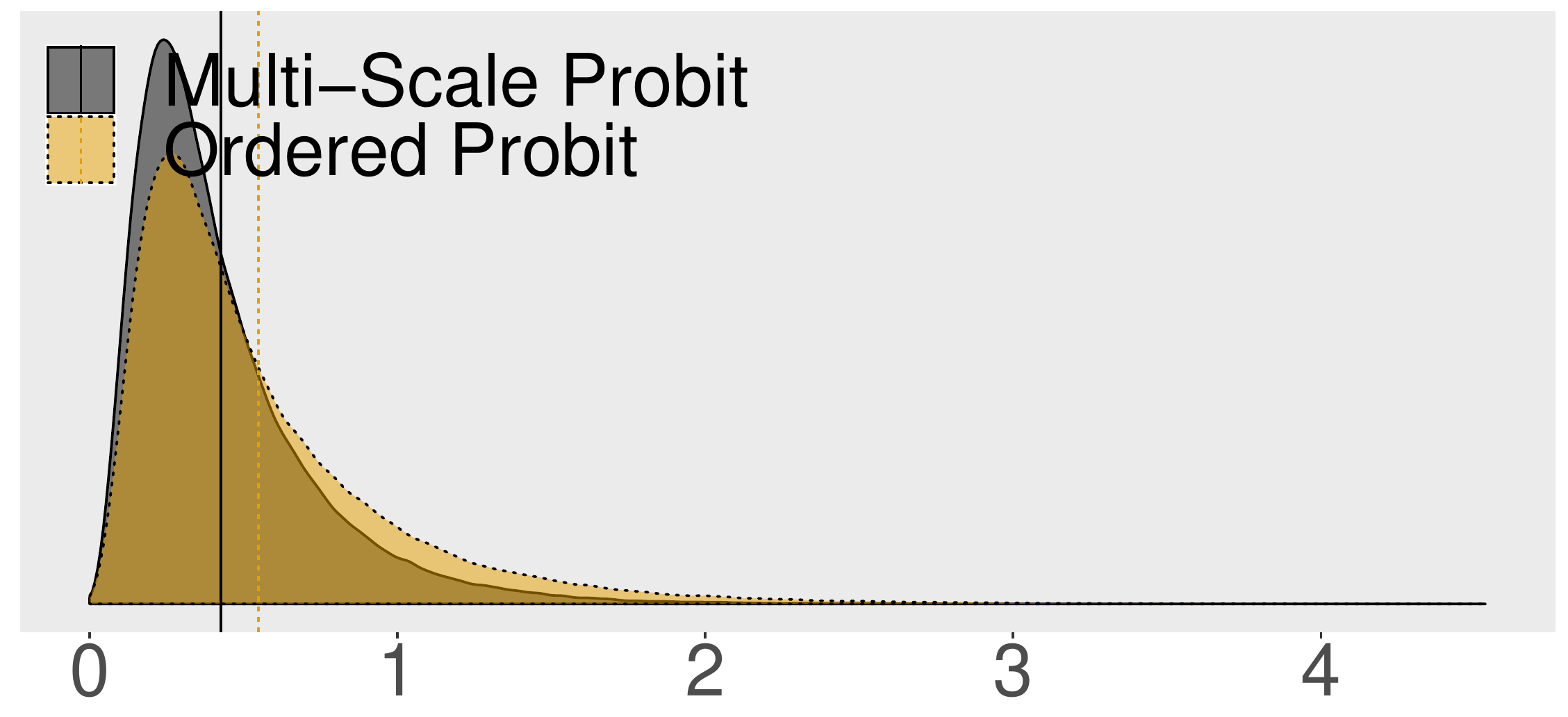} 
    \caption*{$\boldsymbol{\gamma}^{(2)}$} 
  \end{minipage}
  \hfill
    \begin{minipage}[b]{0.32\linewidth}\captionsetup{width=\textwidth}
    \centering
    \includegraphics[width=\linewidth]{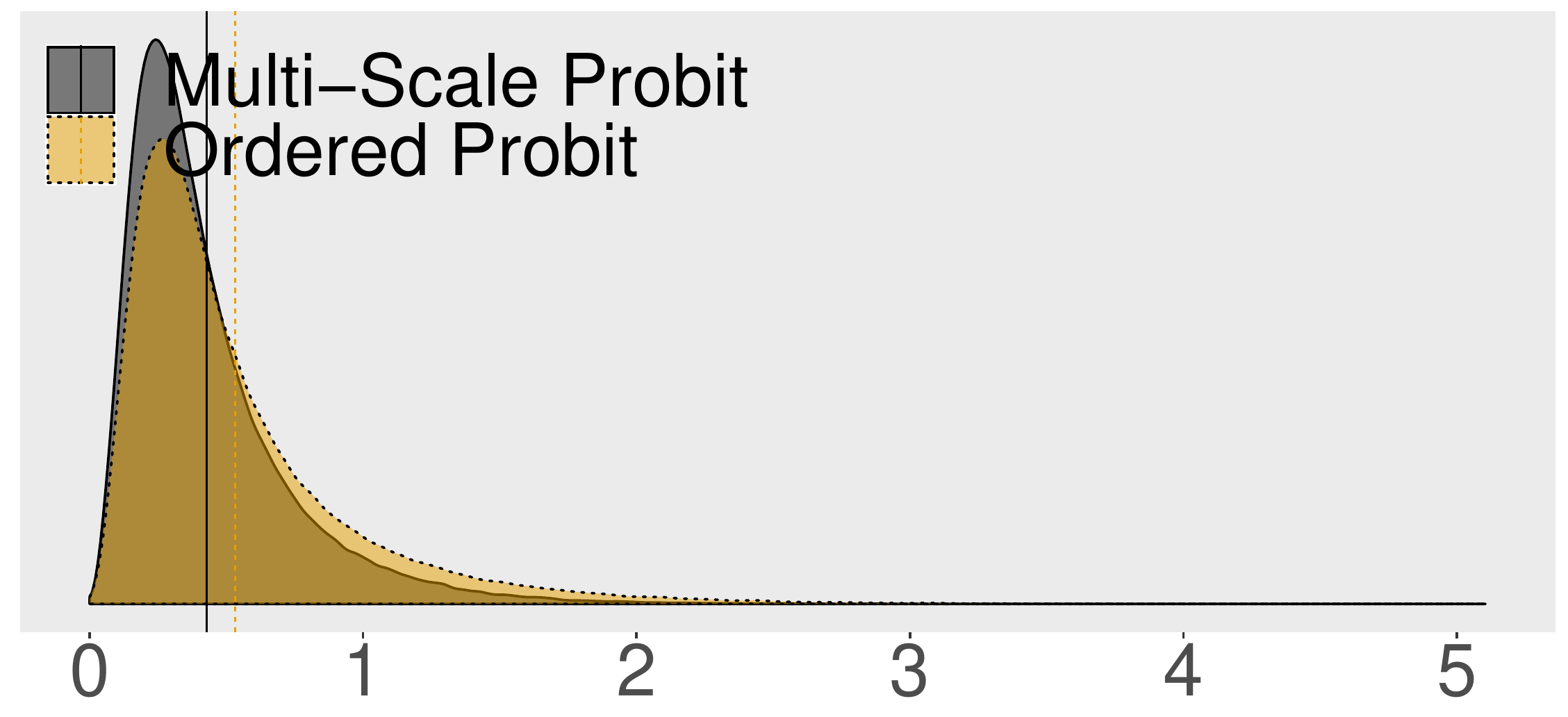} 
    \caption*{$\boldsymbol{\gamma}^{(3)}$} 
  \end{minipage} 
  \caption{The posterior RMSE of $\boldsymbol{\gamma}^{(s)}$ on the three scales for all 500 simulated $p < n$ data sets.
  \label{fig:gamma_rmse600}}
\end{figure}

With regards to root-mean-square error of $\boldsymbol{\gamma}^{(s)}$ ($\boldsymbol{\gamma}^{(s)}_{RMSE}$) we expect a much more modest difference as each $\boldsymbol{\gamma}^{(s)}$ is estimated with the same amount of data in both the single-scale and the Multi-Scale models. This prediction bears in Figure \ref{fig:gamma_rmse600} where the very slight difference between the distributions can probably be explained as an effect of the slightly better estimate of $\boldsymbol{\beta}$ in the Multi-Scale model.

\FloatBarrier

\subsection{Experiment 2: Simulated data, $p > n$}

The same experimental set-up is used as in Experiment 1 but with parameters adapted to simulate a $p > n$ situation. The parameters are chosen to resemble the conditions in the text complexity data. The full set of experimental parameters is listed in Table \ref{tab:simparams60}. In this set-up $p=48$ and $n=40$.

\begin{table}
\begin{center}
\caption{\label{tab:simparams60}Experimental parameters for testing our model under $p>n$ conditions.}
\begin{tabular}[]{l p{10cm}}
\bf{Parameter} & \bf{Value(s)} \\
\hline
No. of repetitions$^1$ & 500 \\
Total number of draws & 250 000 (product of $^1$ and $^2$) \\
\multicolumn{2}{l}{\textbf{Data}}\\
\hline
No. of covariates ($p$) & 48\\
No. of data points & 40 per $\boldsymbol{\gamma}$ vector, with at least 1 instance per class label. \\
Number of class labels per scale & $(1, 3, 3)$ \\
\multicolumn{2}{l}{\textbf{MCMC hyper-parameters}} \\
\hline
Burn in phase & 50000 steps \\
Thinning & 1 step in 100 is stored \\
No. of stored draws$^2$ & 500 \\
$\sigma^2_{\gamma^{(1)}}$ & 5.0\\
$\sigma^2_{\gamma^{(2)}}$ & 1.9\\
$\sigma^2_{\gamma^{(3)}}$ & 1.9\\
& All empirically chosen to get a mean acceptance rate close to 0.234 as suggested by \textcite{Roberts1997WeakAlgorithms} \\
$\mu_0$ & (0, ..., 0) \\
$\Lambda_0$ & $0.1 \times I$ \\
\hline
\end{tabular}
\end{center}
\end{table}

\begin{figure}[h!]
    \begin{minipage}[b]{0.32\linewidth}\captionsetup{width=\textwidth}
    \centering
    \includegraphics[width=\linewidth]{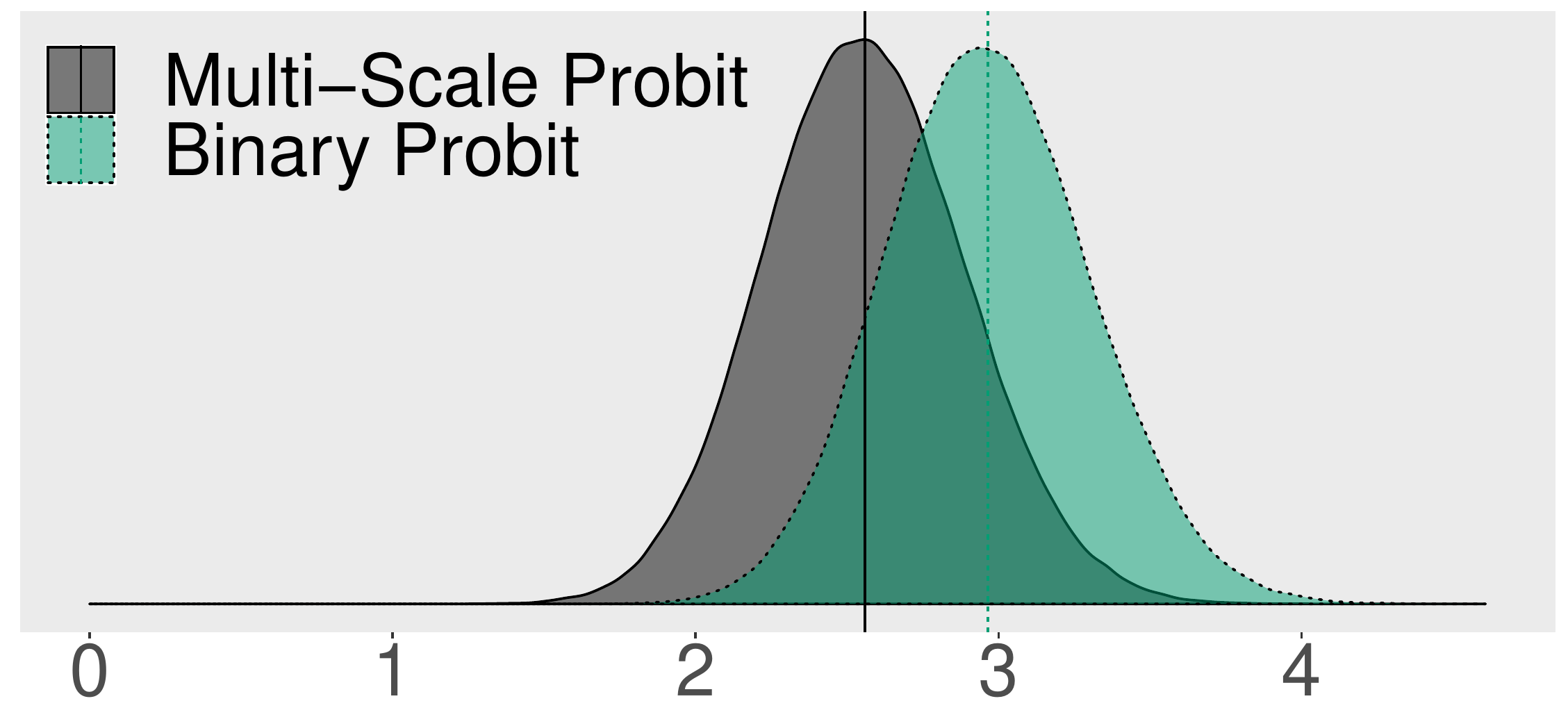} 
    \caption*{1)} 
  \end{minipage} 
  \hfill
  \begin{minipage}[b]{0.32\linewidth}\captionsetup{width=\textwidth}
    \centering
    \includegraphics[width=\linewidth]{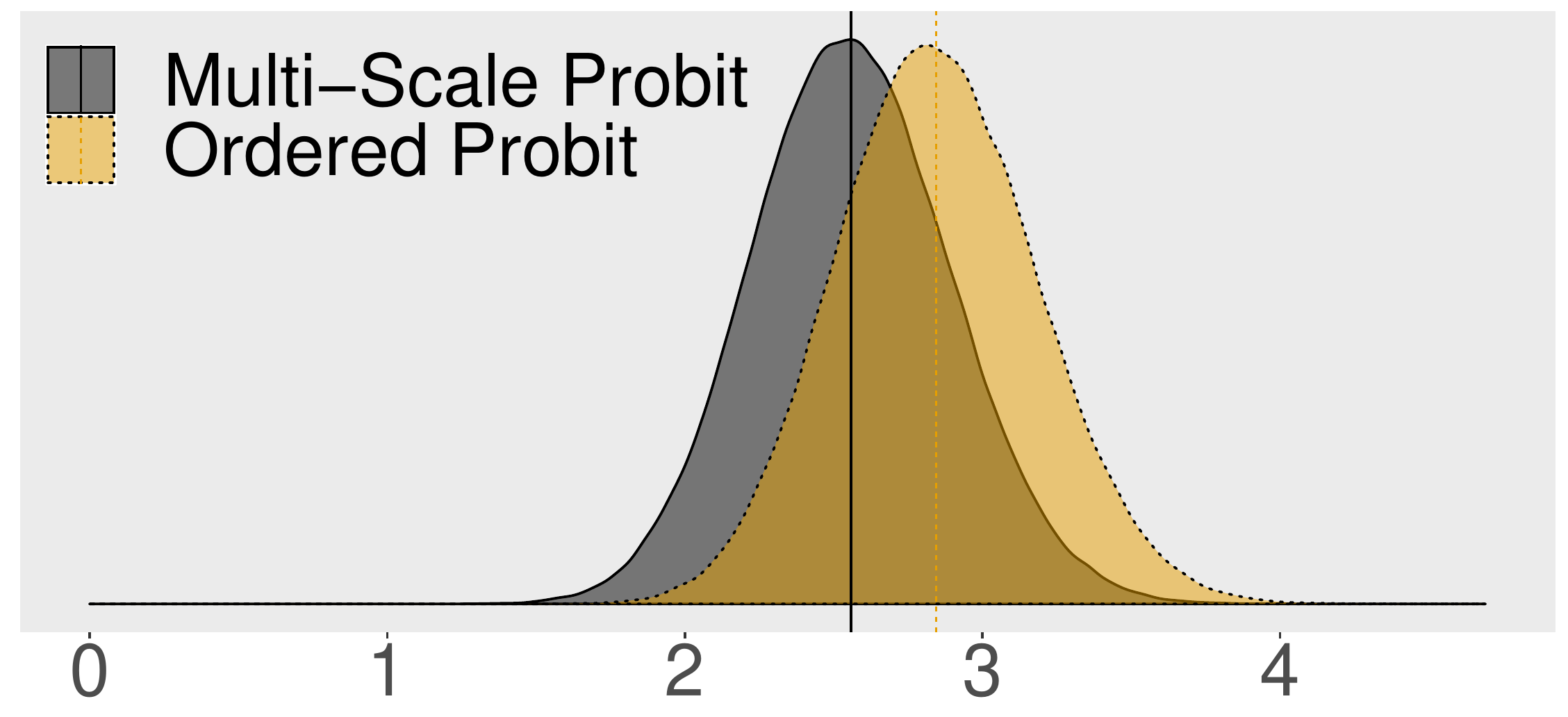} 
    \caption*{2)} 
  \end{minipage}
  \hfill
    \begin{minipage}[b]{0.32\linewidth}\captionsetup{width=\textwidth}
    \centering
    \includegraphics[width=\linewidth]{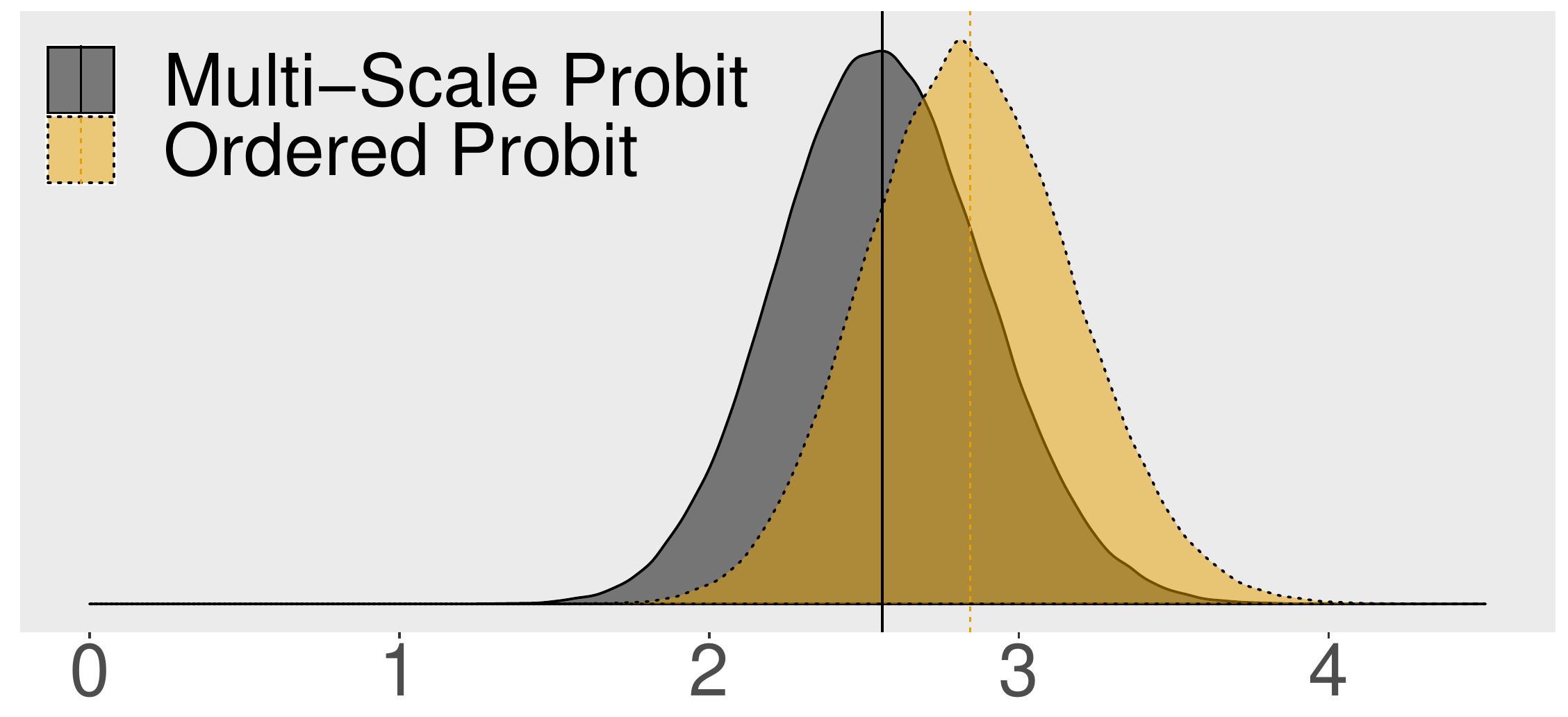} 
    \caption*{3)} 
  \end{minipage} 
  \caption{The posterior distributions of $\boldsymbol{\beta}_{RMSE}$ on the three scales for all 500 simulated $p > n$ data sets.
  \label{fig:beta_rmse60}}
\end{figure}

We then start by inspecting the $\boldsymbol{\beta}_{RMSE}$ for all 250 000 draws and plot the distributions in Figure \ref{fig:beta_rmse60}. The $\boldsymbol{\beta}_{RMSE}$ is much larger than for the $p < n$ case, which is expected with only 1/10 of the amount of training data, but the Multi-Scale model still outperforms the single-scale models.



\begin{figure}[h!]
    \begin{minipage}[b]{0.32\linewidth}\captionsetup{width=\textwidth}
    \centering
    \includegraphics[width=\linewidth]{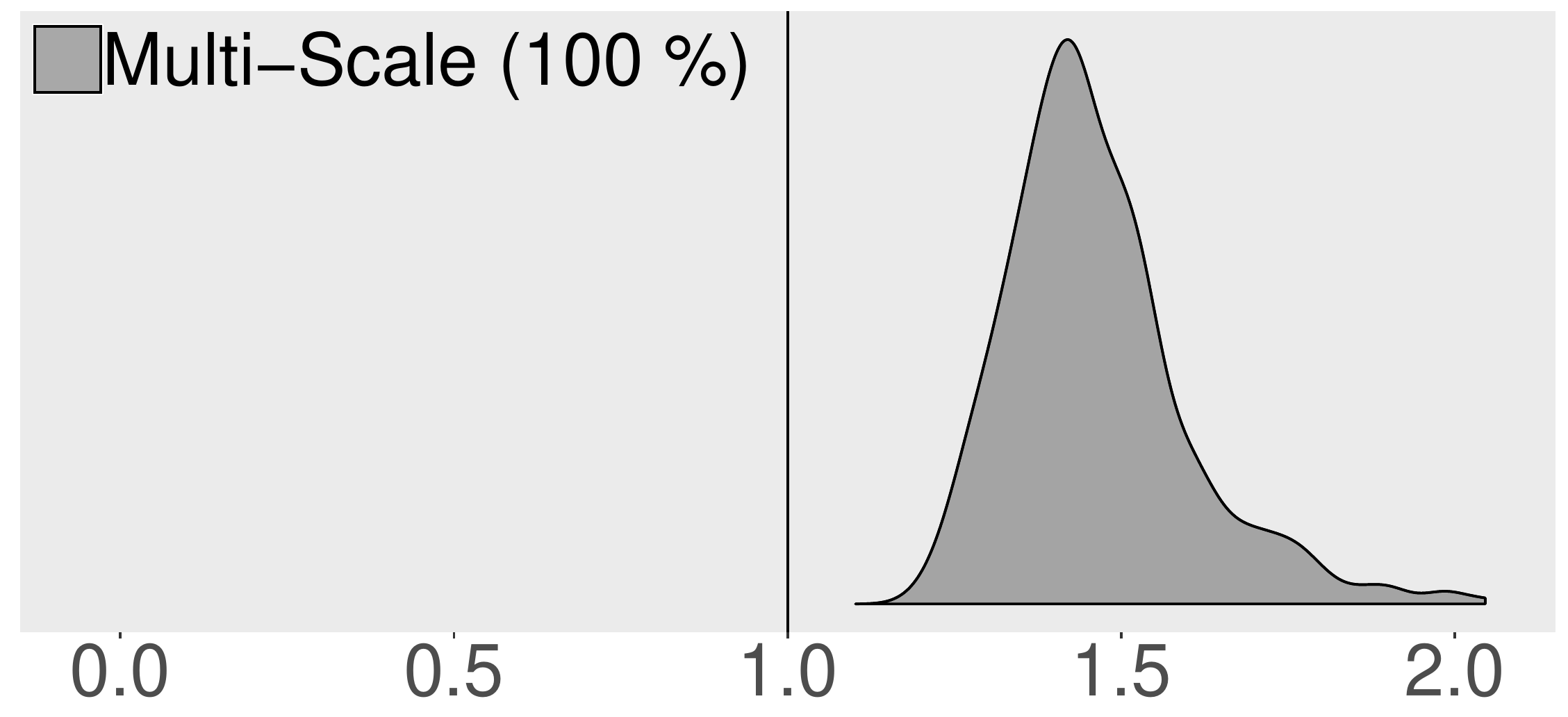} 
    \caption*{1)} 
  \end{minipage} 
  \hfill
  \begin{minipage}[b]{0.32\linewidth}\captionsetup{width=\textwidth}
    \centering
    \includegraphics[width=\linewidth]{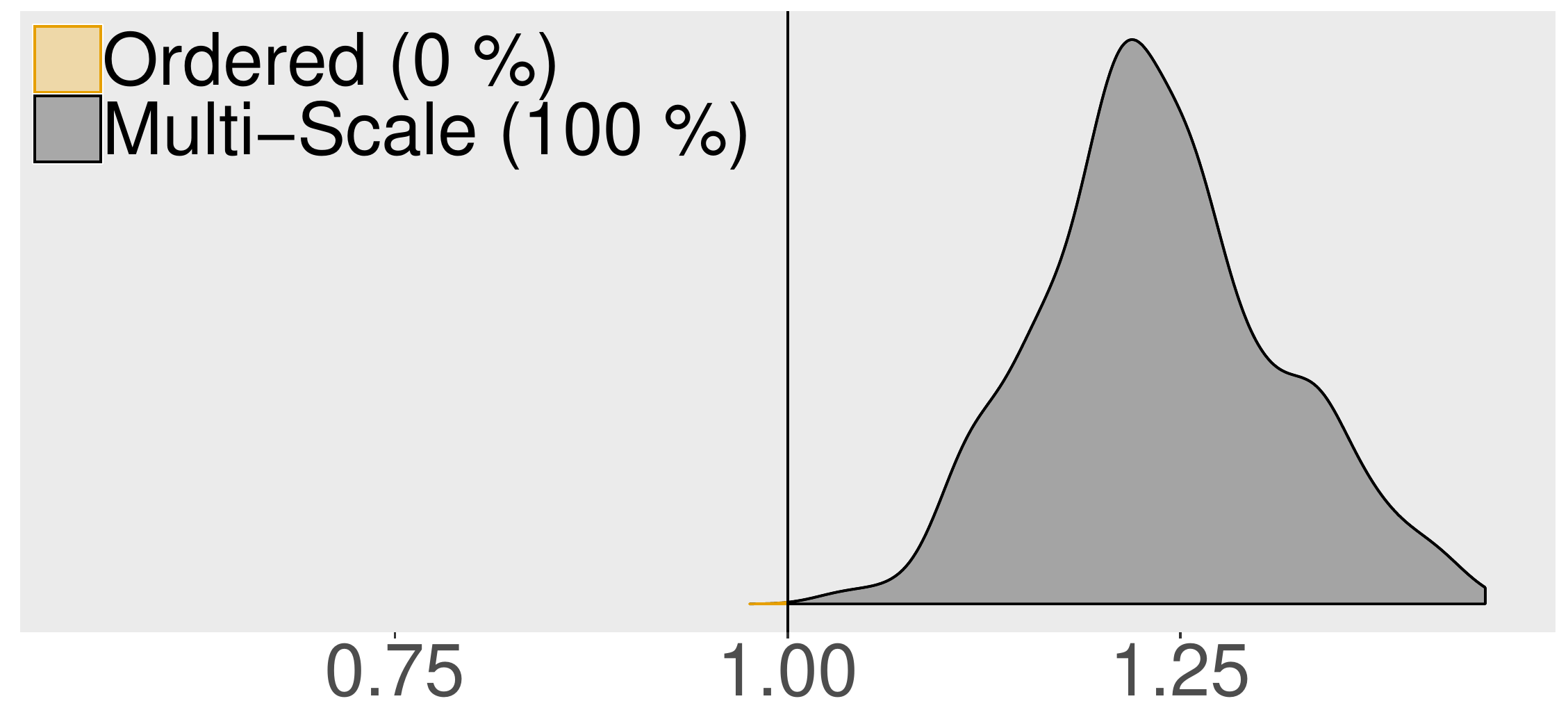}
    \caption*{2)} 
  \end{minipage}
  \hfill
    \begin{minipage}[b]{0.32\linewidth}\captionsetup{width=\textwidth}
    \centering
    \includegraphics[width=\linewidth]{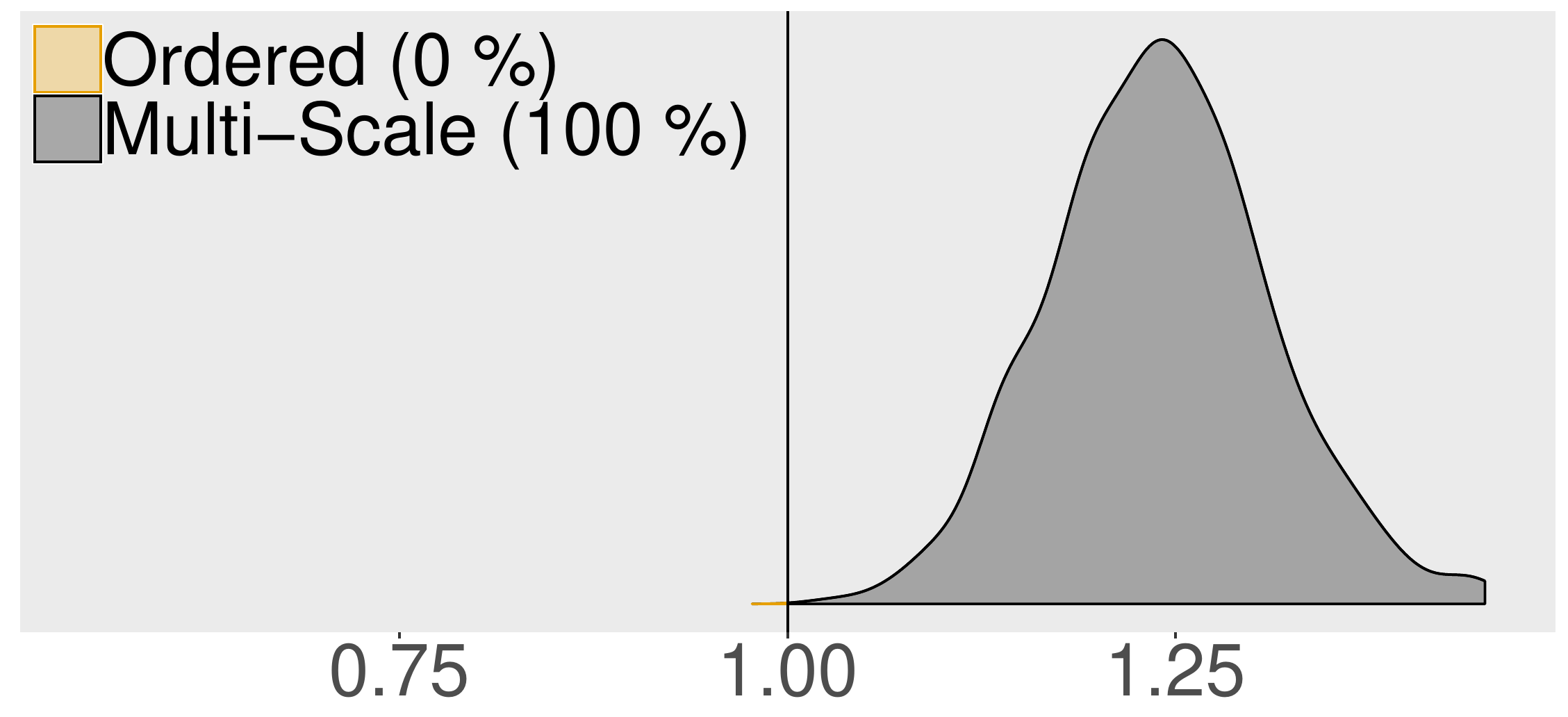}
    \caption*{3)} 
  \end{minipage} 
  \caption{The posterior distributions of mean $\boldsymbol{\beta}_{RMSE}$ ratio between Multi-Scale and single-scale models on each of the three scales for all 500 simulated $p > n$ data sets.
  \label{fig:beta_rmse_ratio60}}
\end{figure}

 As the overlap is quite large we would like to see to whether the Multi-Scale model consistently outperforms the single-scale models on a majority of the simulated data sets. We can do this by computing the posterior mean $\boldsymbol{\beta}_{RMSE}$ for each model and each of the 500 simulated data sets. We then compute the ratio between the posterior mean $\boldsymbol{\beta}_{RMSE}$ for each data set. The result is plotted in Figure \ref{fig:beta_rmse_ratio60} which indicates that the Multi-Scale model consistently outperforms the single-scale models.

\begin{figure}[h!]
    \begin{minipage}[b]{0.32\linewidth}\captionsetup{width=\textwidth}
    \centering
    \includegraphics[width=\linewidth]{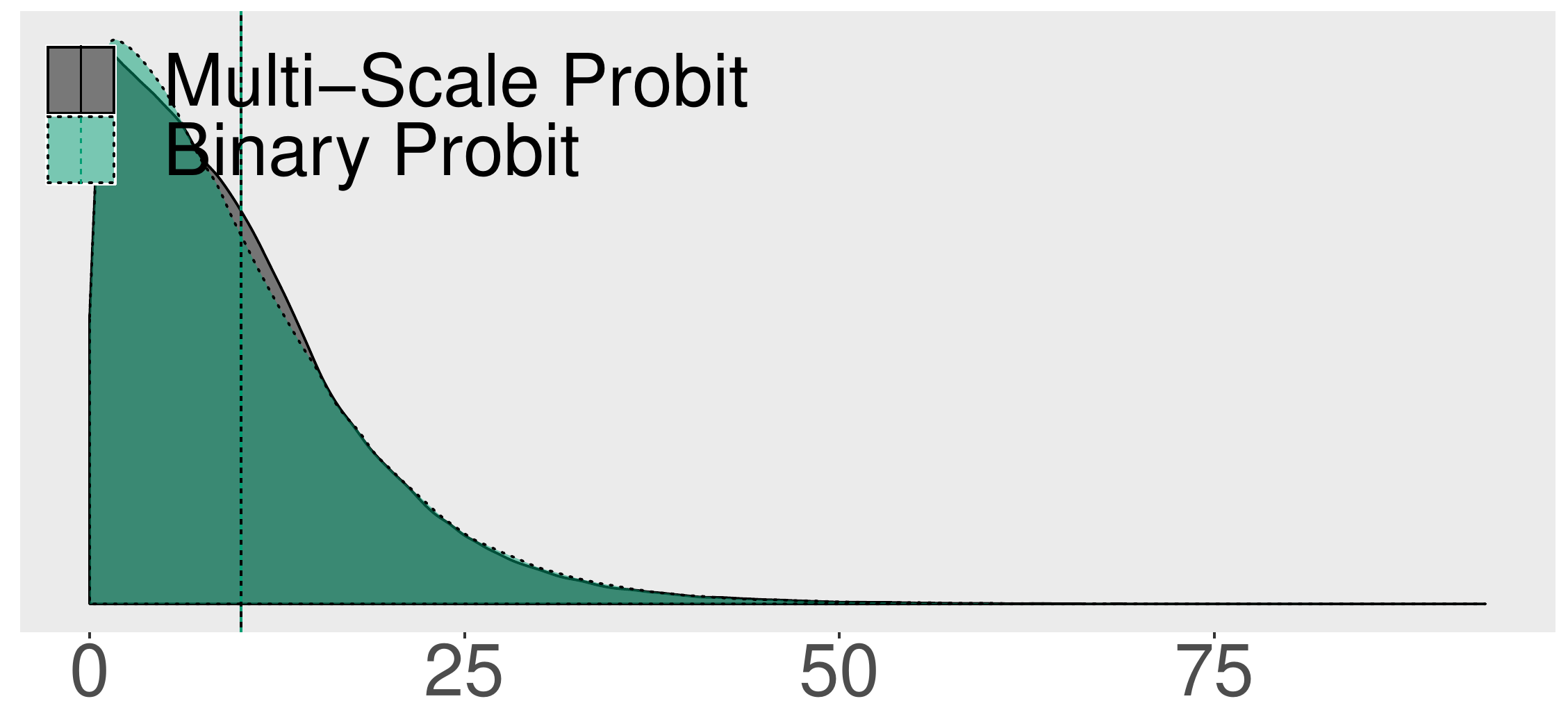} 
    \caption*{$\boldsymbol{\gamma}^{(1)}$} 
  \end{minipage} 
  \hfill
  \begin{minipage}[b]{0.32\linewidth}\captionsetup{width=\textwidth}
    \centering
    \includegraphics[width=\linewidth]{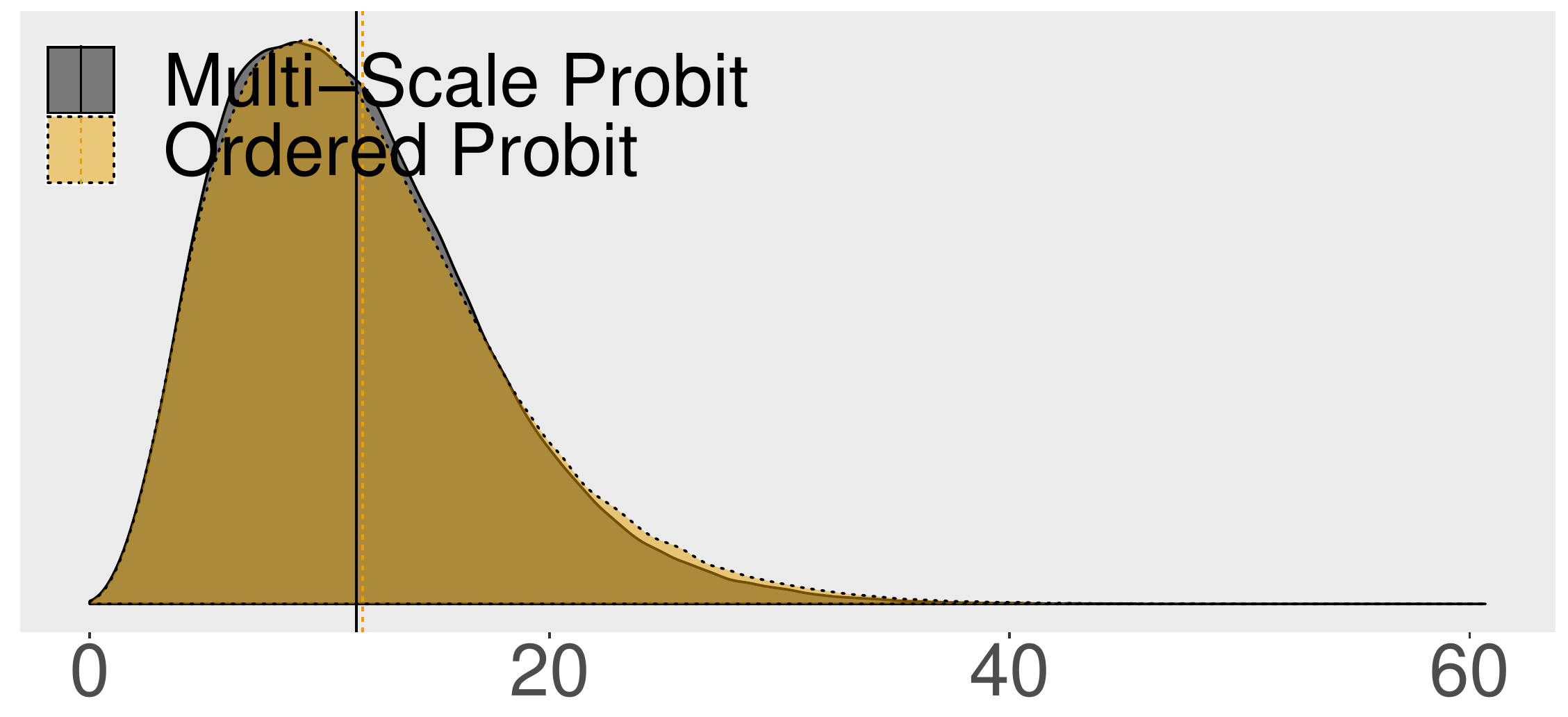} 
    \caption*{$\boldsymbol{\gamma}^{(2)}$} 
  \end{minipage}
  \hfill
    \begin{minipage}[b]{0.32\linewidth}\captionsetup{width=\textwidth}
    \centering
    \includegraphics[width=\linewidth]{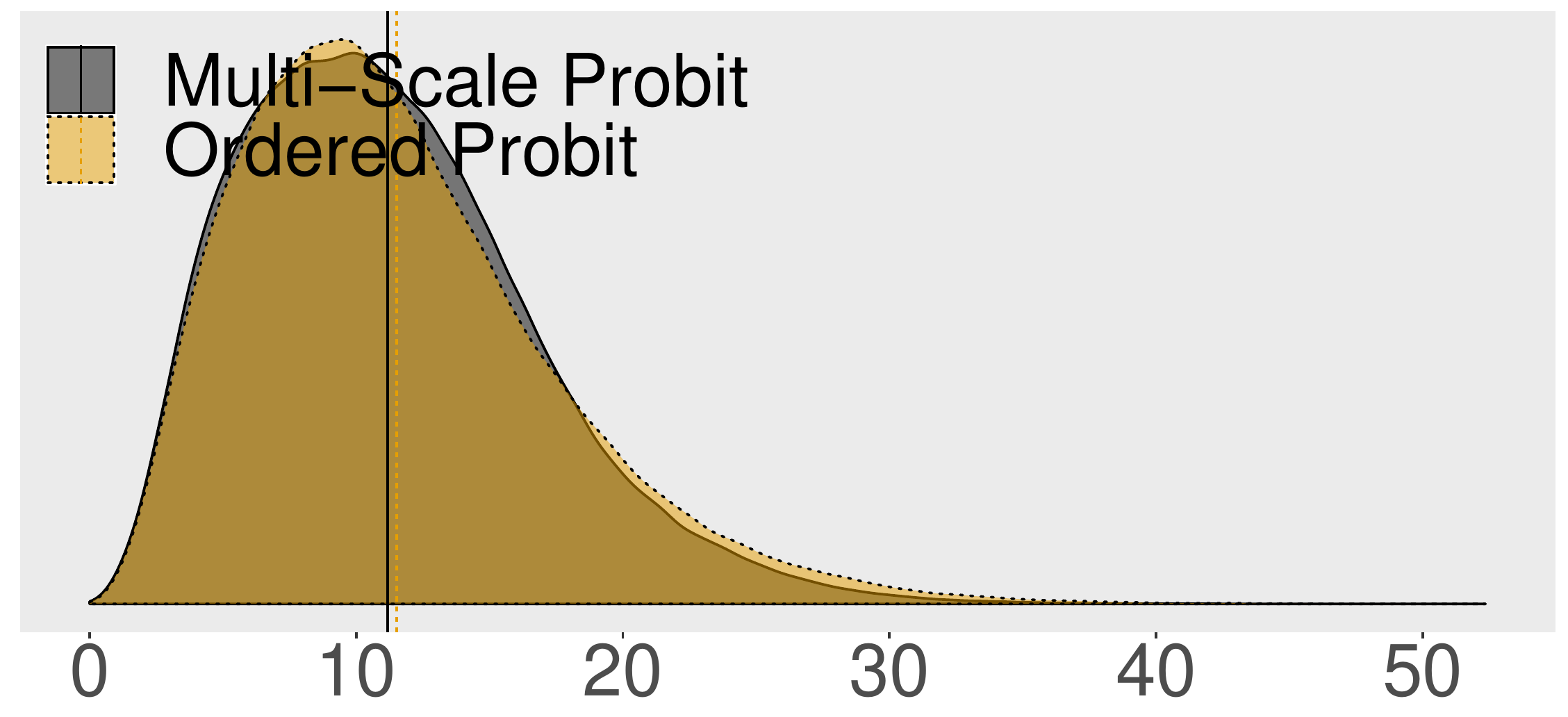} 
    \caption*{$\boldsymbol{\gamma}^{(3)}$} 
  \end{minipage} 
  \caption{The posterior distributions RMSE of $\boldsymbol{\gamma}^{(s)}$ on the 3 scales for all 500 simulated  $p > n$ data sets.
  \label{fig:gamma_rmse60}}
\end{figure}

Again, as we can see in Figure \ref{fig:gamma_rmse60}, there is no noticeable difference in ($\boldsymbol{\gamma}^{(s)}_{RMSE}$). 

\FloatBarrier

\section{Application to text complexity analysis}
\label{sec:ApplTextComplexity}

In this section we will illustrate the workings and predictive performance of the Multi-Class Probit in an application to text complexity analysis, or as it is often referred to, readability analysis. Corpora relevant to text complexity analysis are usually organised by an approximate scale used by a specific publisher, such as a publisher of children's fiction with texts aimed at different age groups. In other cases, a corpus might consist of only easy-to-read (ETR) texts from a single source, such as an easy-to-read newspaper or news aggregator. These can be combined with a corpus containing similar texts but written for a more typical readership, such as a regular newspaper.

Data driven modelling approaches have therefore been restricted to using a single corpus, aggregated corpora by lowest common denominator (e.g. easy to read vs regular text) or a manual re-labelling with existing annotations as support. Our proposed Multi-Scale Probit model is an attempt to allow for using all existing data on potentially different scales in a single model to learn about a single underlying latent readability factor.

It could be argued that the definition of text complexity varies somewhat between genres and domains, and for that reason we have decided to only include data from a single genre, fiction, in this experiment. However, see Section \ref{sec:concfuture} for a proposed approach to integrating multiple domains in our model.

\subsection{Feature set}

We have used a subset of features from the set of 118 features covered in \textcite{Falkenjack2013FeaturesText} by discarding some features with majority zero or constant values in any data set. We also removed features to make sure the Pearson correlation between any pair of features was below $.6$. This cut-off point was selected as it provided a reasonable trade-off between the condition number of the data matrix ($\kappa \approx 3413$) and the number of included features (48). The included features and short descriptions of these are listed in Table \ref{tab:features}.

\subsection{Corpora}

The text data comes from five different sources, three publishers of easy-to-read fiction with different text complexity labelling schemes, and two publishers of general fiction aimed at typical adult readers, Table~\ref{tab:corpora}.

\begin{table}
\begin{center}
\caption{\label{tab:corpora}The corpora used to evaluate the model with regards to text complexity.}
\begin{tabular}[]{l p{10cm}}
\bf{Publisher} & \bf{Number of texts} \\
\hline
Lättlästförlaget & 14 easy-to-read \\ 
Legimus & 11 aimed at 3-9 year olds, 7 aimed at 10-12 year olds, 5 aimed at 13-19 year olds \\
Hegas & 3 very easy, 5 easy, 6 moderately easy\\
Norstedts & 23 aimed at typical adult readers \\
Bonnier & 129 aimed at typical adult readers \\
\hline
\end{tabular}
\end{center}
\end{table}

The data is organised into three sets. One binary set combining the ETR texts from Lättlästförlaget with a sample from Norstedts and Bonnier, one set combining the three levels of Legimus texts with a sample from Norstedts and Bonnier as fourth most complex level, and one following the same strategy with Hegas texts. Each of these three sets represents a different scale of text complexity, two with 4 levels and one with 2 levels. We will refer to these three data sets as LL, Legimus and Hegas.

As in the case with simulated data, we want to estimate the performance of the models given different inputs. To evaluate the performance of the model and the estimation methodology we generate 500 data sets by randomly splitting the data into training sets consisting of 2/3 of the data, and test sets containing the remaining 1/3 as validation set.

\subsection{Predictive performance}

Each figure below contains three sub-figures. Each sub-figure contains either two distributions or a single distribution representing a comparison between two distributions. In the case where two distributions are plotted, one distribution, coloured blue or yellow, represents the performance of a single-scale model, Probit or Ordinal Probit, estimated using training data from a single corpus and evaluated using validation data from the same corpus. The other distribution, coloured grey, represents the performance of the Multi-Scale Probit estimated using all three corpora but evaluated only using validation data from a single corpus. In the case where only a single distribution is plotted, the colours represent which model performs better in that part of the distribution.

\begin{figure}[h!]
  \begin{center}
  \begin{minipage}[b]{0.32\linewidth}
    \centering
    \includegraphics[width=\linewidth]{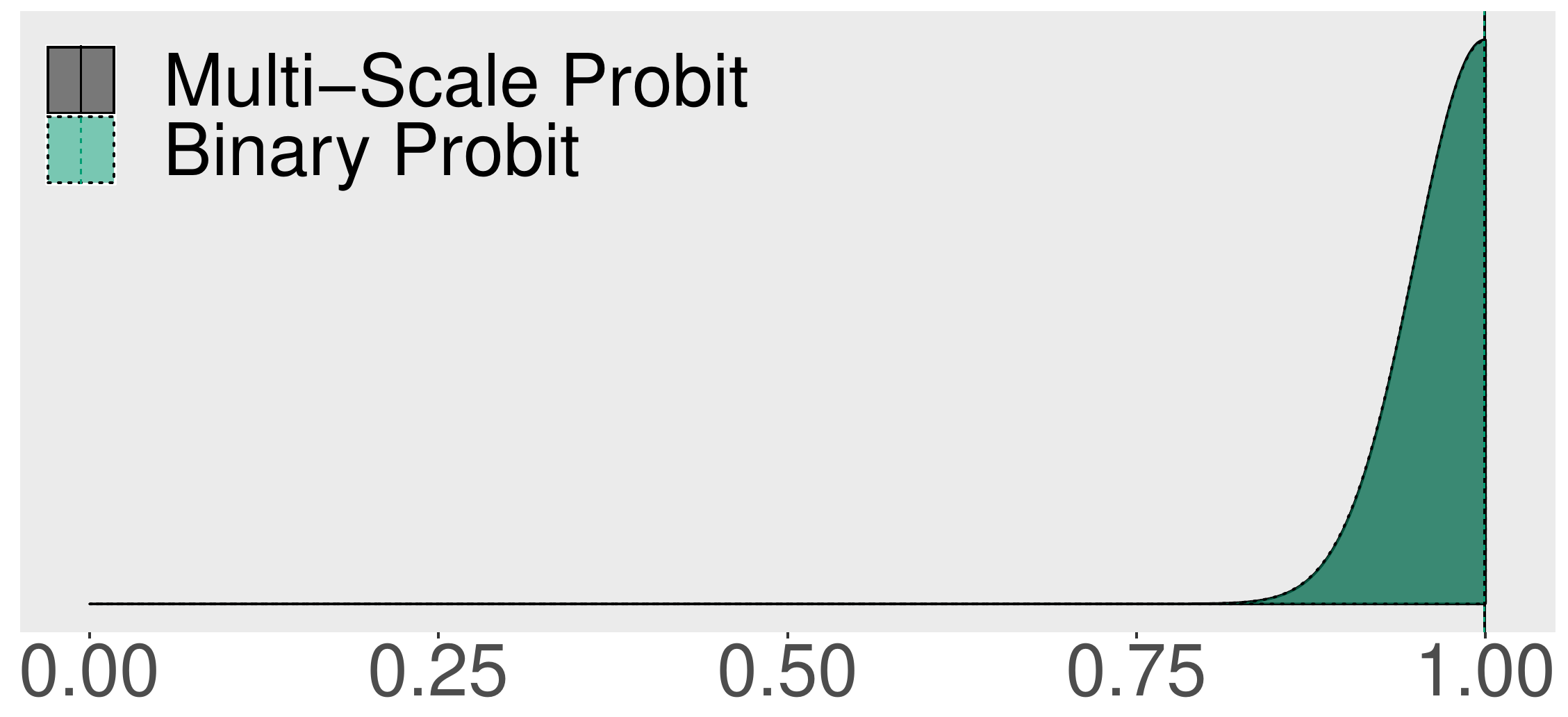}
    \caption*{LL} 
  \end{minipage} 
  \hfill
  \begin{minipage}[b]{0.32\linewidth}
    \centering
    \includegraphics[width=\linewidth]{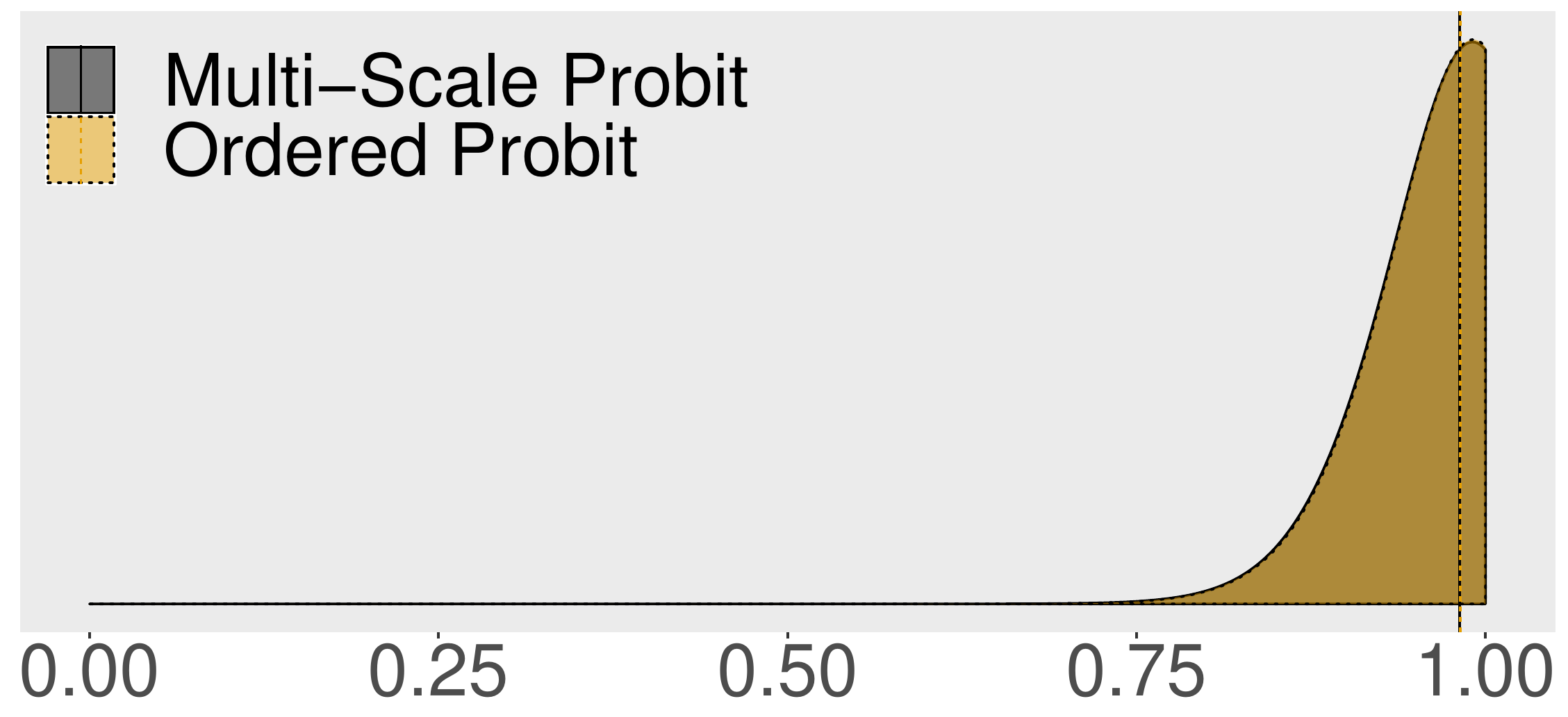} 
    \caption*{Legimus} 
  \end{minipage}
  \hfill
  \begin{minipage}[b]{0.32\linewidth}
    \centering
    \includegraphics[width=\linewidth]{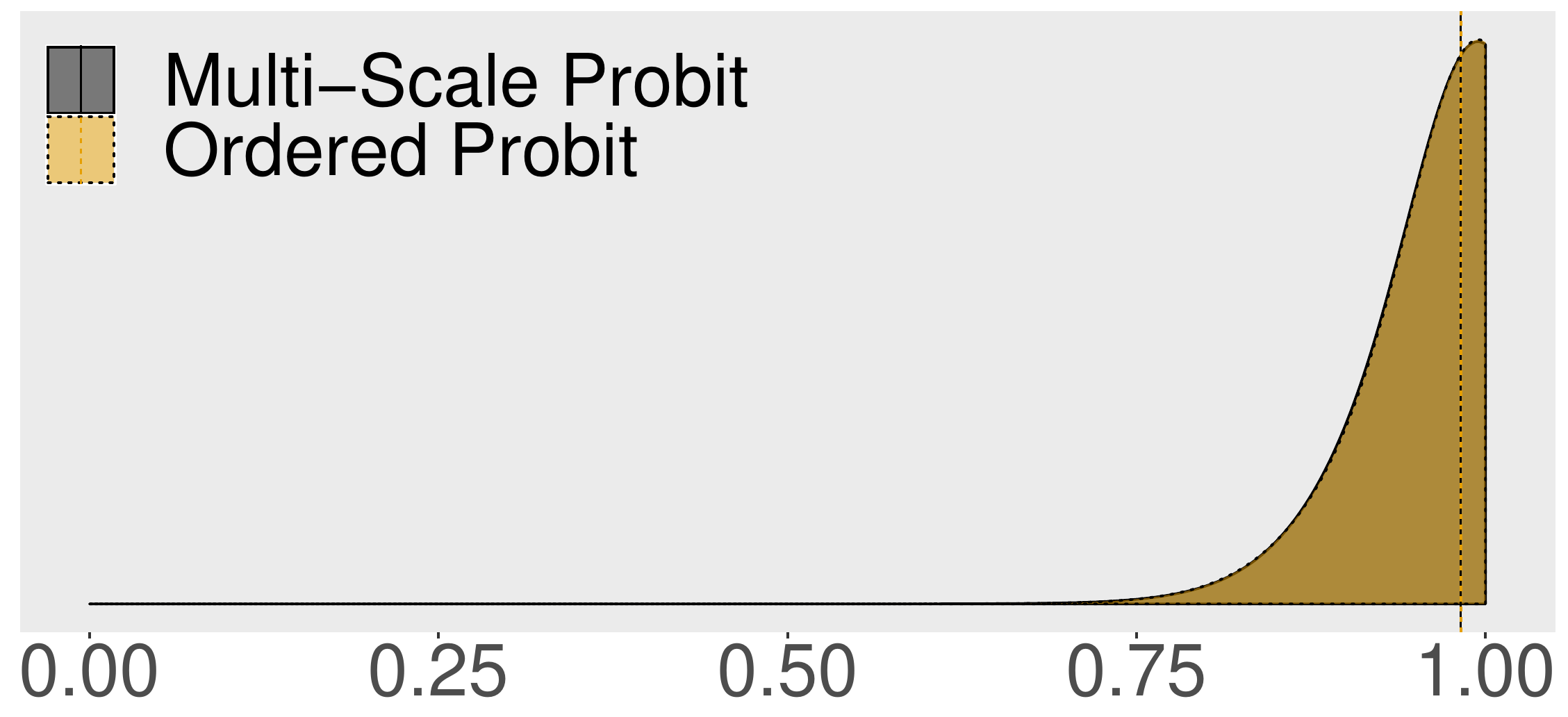} 
    \caption*{Hegas} 
  \end{minipage} 
  \end{center}
  \caption{The posterior distributions for in-sample $F_1$ scores of the text data, plotted per measurement scale.}
  \label{fig:F1_text_train}
\end{figure}

In Figure \ref{fig:F1_text_train} we can see that as with the simulated data, in-sample performance does not differ noticeably between single-scale Probits and the Multi-Scale Probit. As discussed in Section \ref{sec:experiments}, this is the expected behaviour, and we will not further plot in-sample performances for any metrics.

\begin{figure}[h!]
  \begin{center}
  \begin{minipage}[b]{0.32\linewidth}
    \centering
    \includegraphics[width=\linewidth]{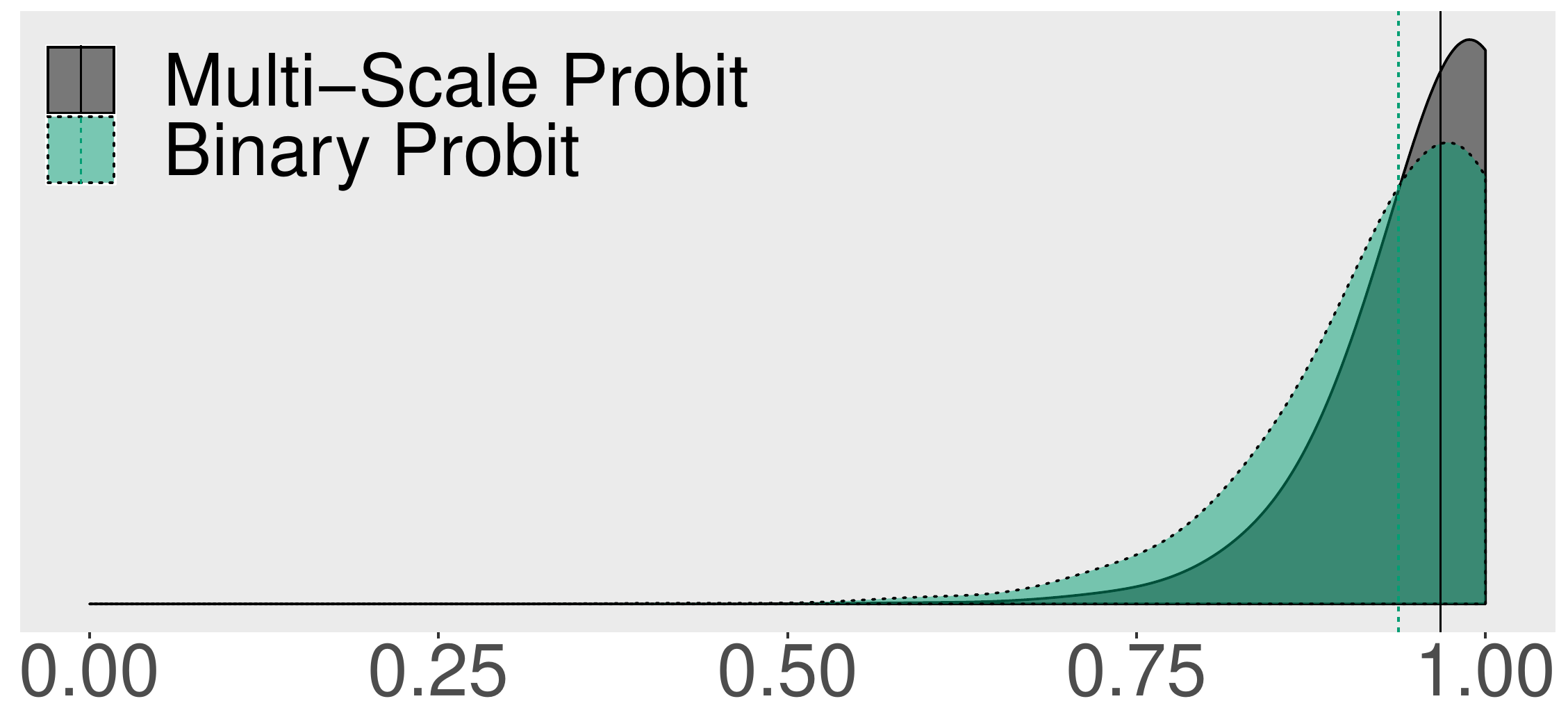} 
    \caption*{LL} 
  \end{minipage} 
  \hfill
  \begin{minipage}[b]{0.32\linewidth}
    \centering
    \includegraphics[width=\linewidth]{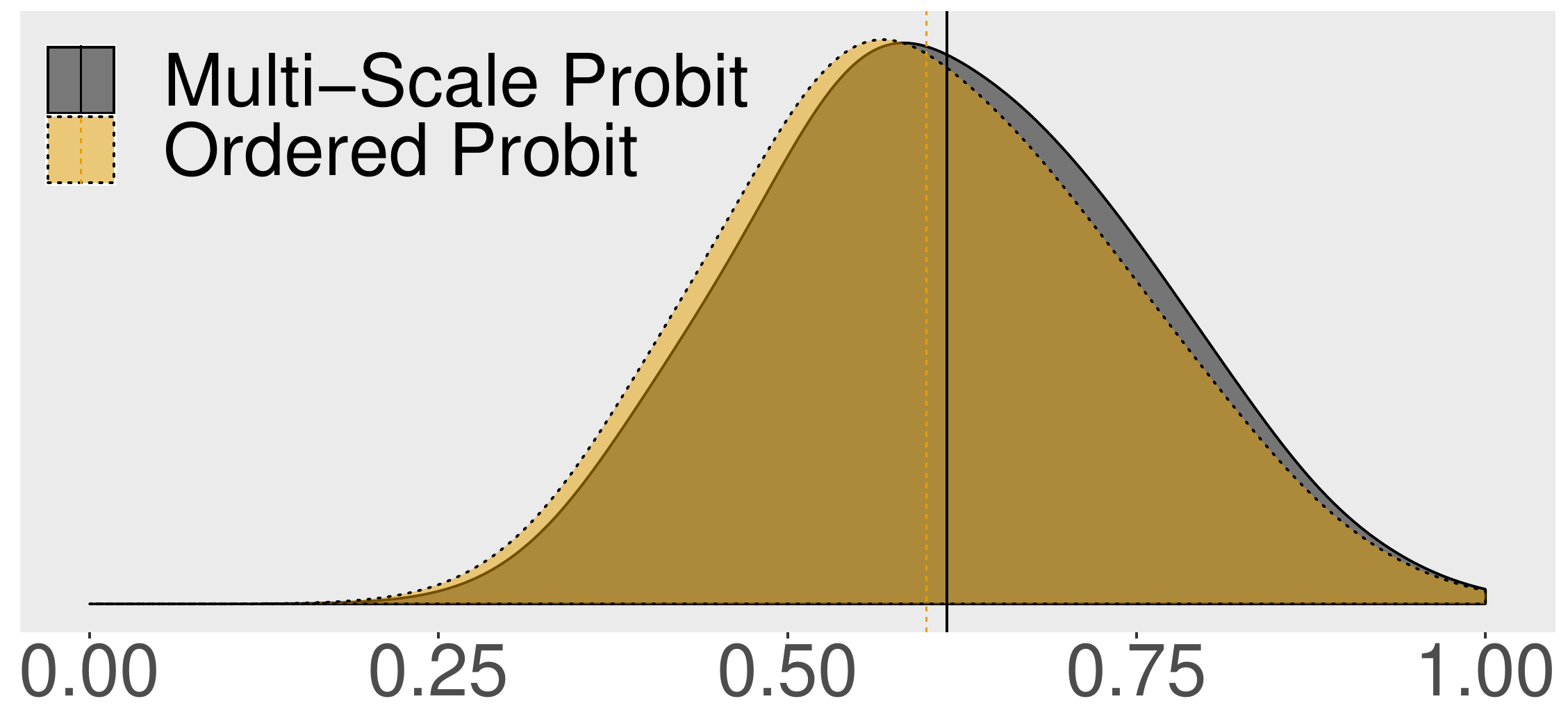} 
    \caption*{Legimus} 
  \end{minipage}
  \hfill
  \begin{minipage}[b]{0.32\linewidth}
    \centering
    \includegraphics[width=\linewidth]{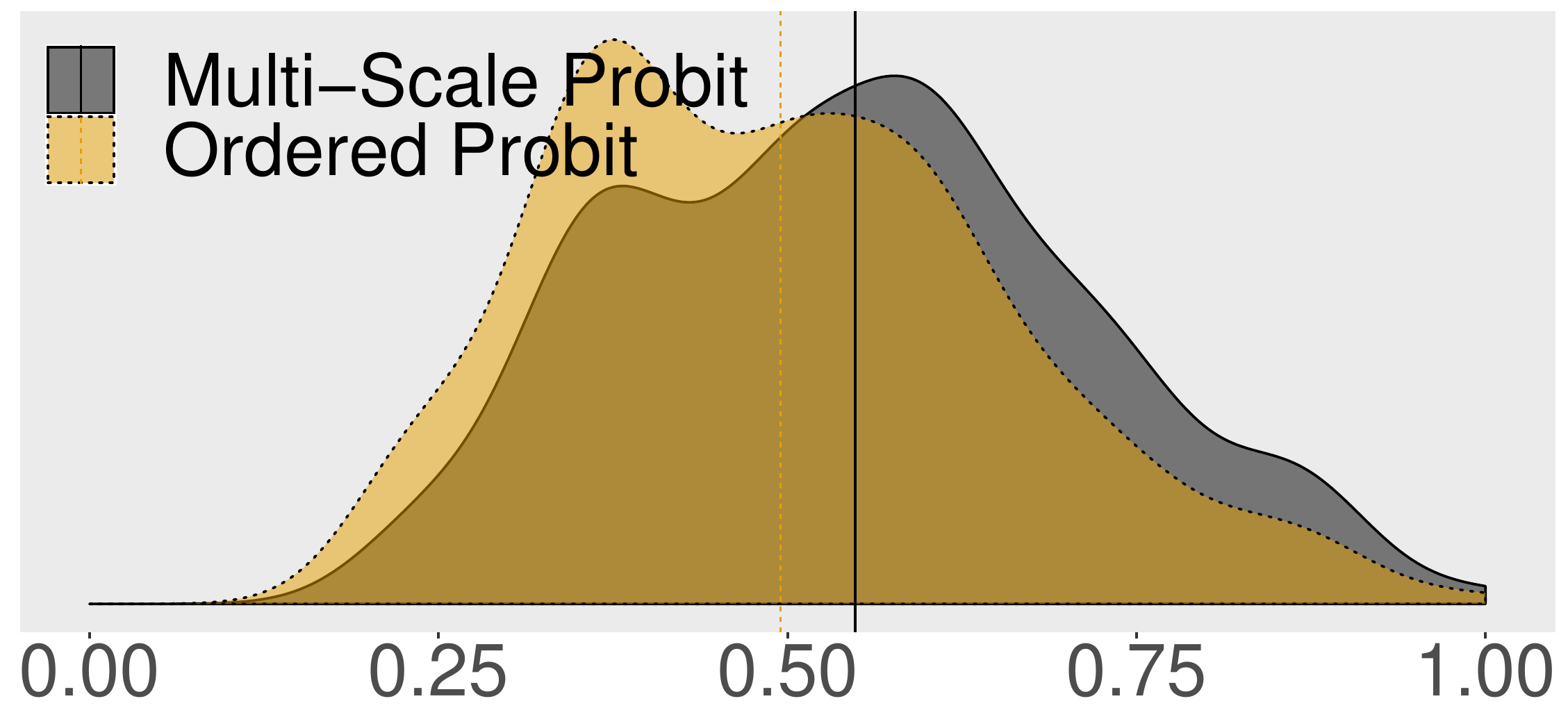} 
    \caption*{Hegas} 
  \end{minipage} 
  \end{center}
  \caption{The posterior distributions for out-of-sample $F_1$ scores of the text data, plotted per measurement scale.}
  \label{fig:F1_text_test}
\end{figure}

Looking at out-of-sample classification performance in Figure \ref{fig:F1_text_test}, we see that the Multi-Scale Probit outperforms the single-scale models, albeit to a smaller extent than in the simulation experiments in Section \ref{sec:experiments}. There is however a large variability in $F_1$ scores over the 500 generated test data sets, which makes it hard to accurately compare models based only on Figure \ref{fig:F1_text_test}. 

\begin{figure}[h!]
  \begin{center}
  \begin{minipage}[b]{0.32\linewidth}
    \centering
    \includegraphics[width=\linewidth]{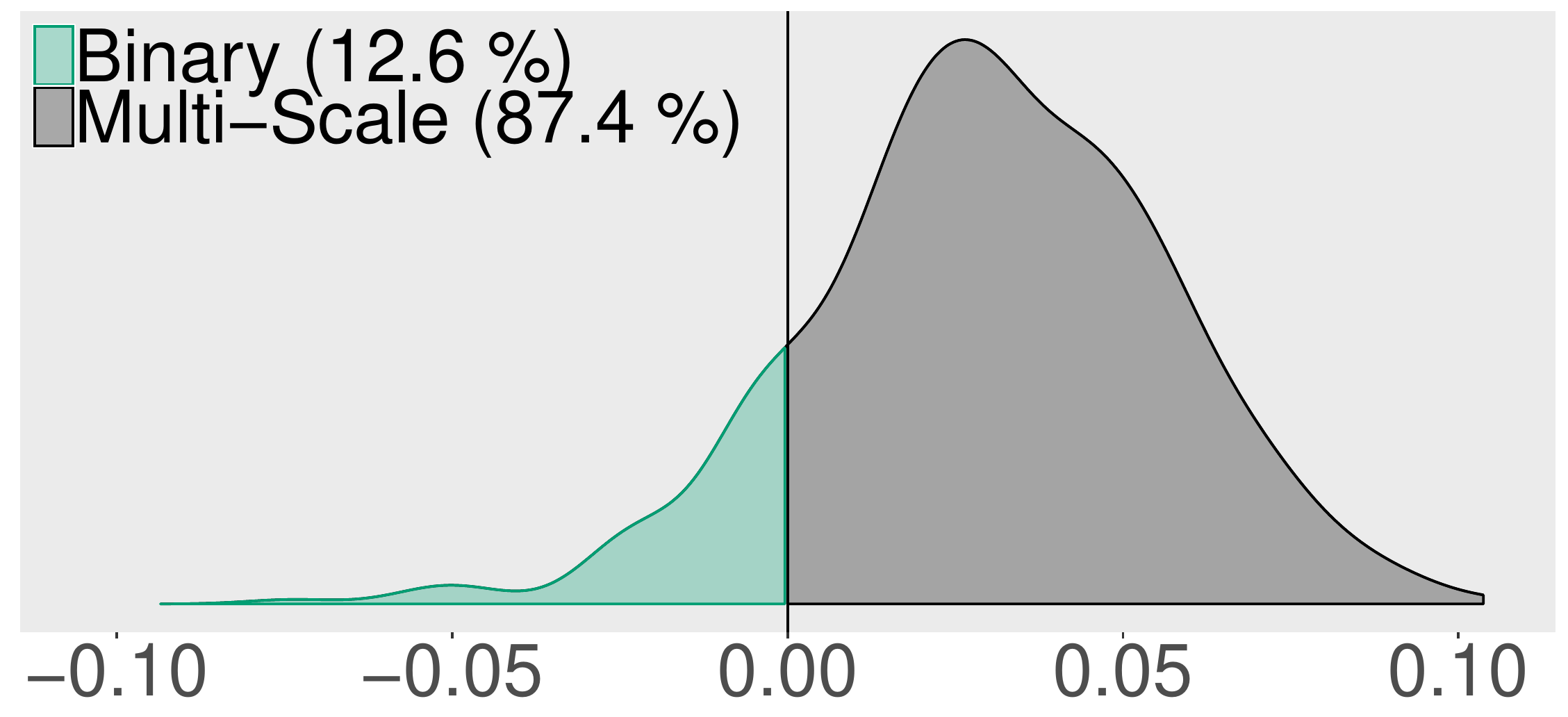} 
    \caption*{LL} 
  \end{minipage} 
  \hfill
  \begin{minipage}[b]{0.32\linewidth}
    \centering
    \includegraphics[width=\linewidth]{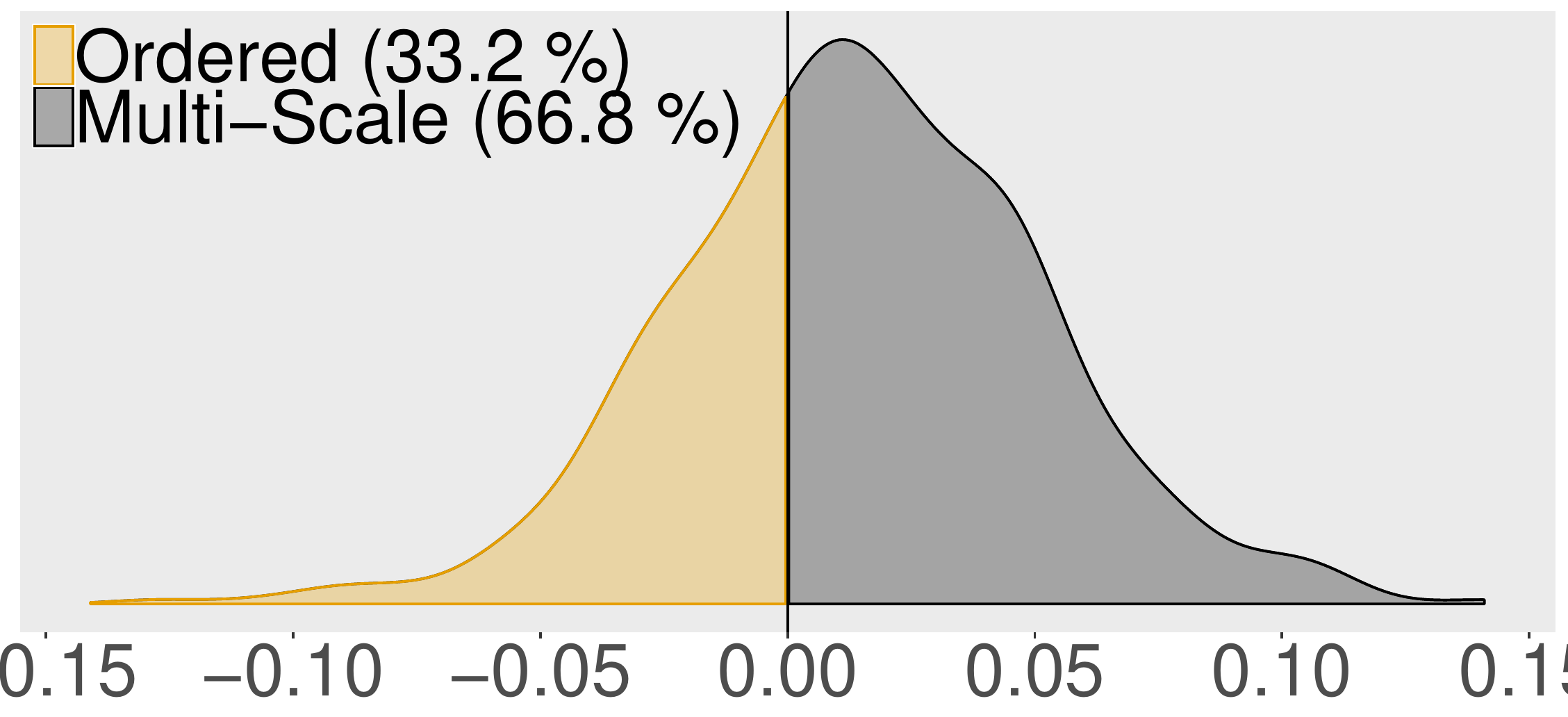} 
    \caption*{Legimus} 
  \end{minipage}
  \hfill
  \begin{minipage}[b]{0.32\linewidth}
    \centering
    \includegraphics[width=\linewidth]{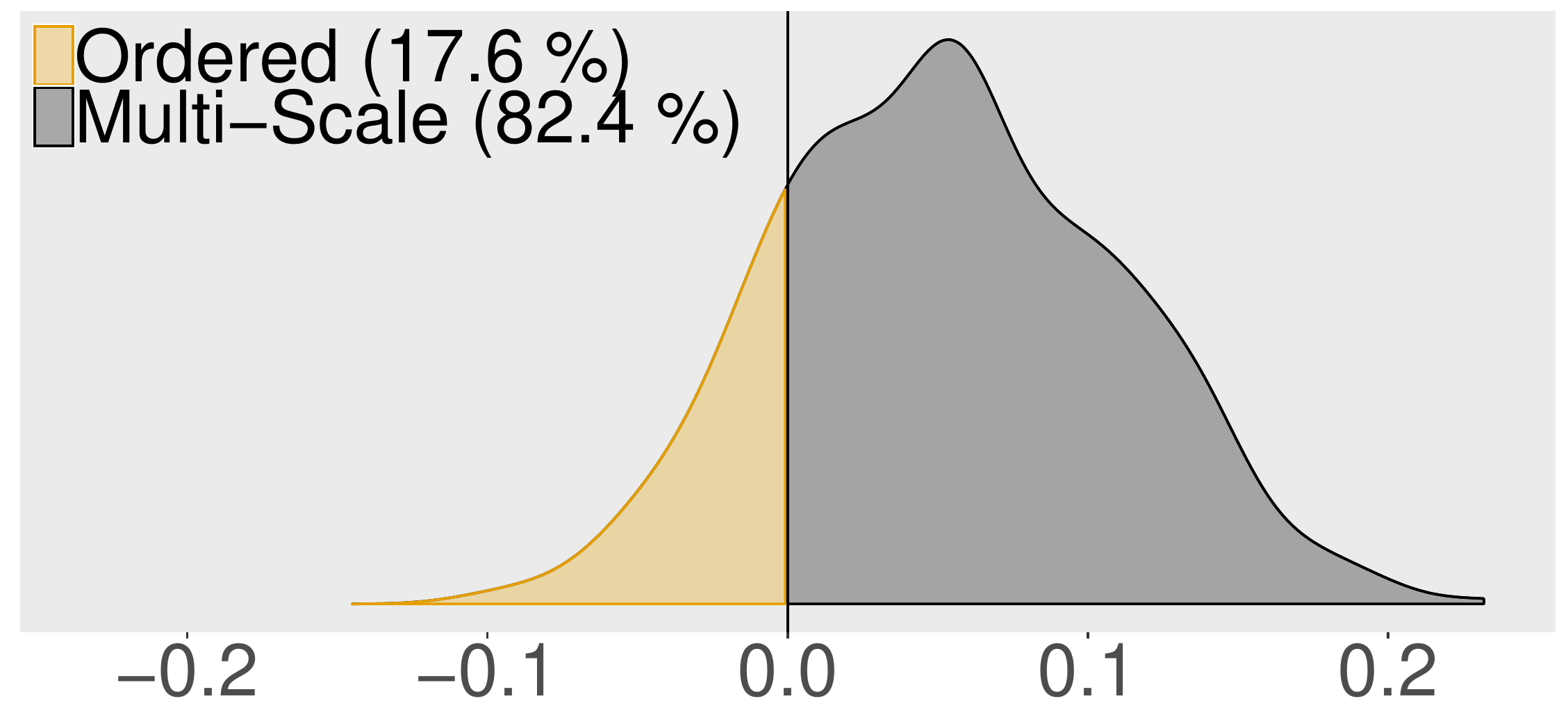} 
    \caption*{Hegas} 
  \end{minipage} 
  \end{center}
  \caption{The posterior distributions for the difference in out-of-sample $F_1$ scores between single-scale and Multi-Scale models on the text data for the 500 different training sets, plotted per measurement scale.}
  \label{fig:F1_text_test_diff}
\end{figure}

Figure \ref{fig:F1_text_test_diff} instead depicts densities of the posterior mean differences between models, that is, the difference between the mean $F_1$ scores for the models for each of the 500 training sets. This assesses whether one model consistently out-performs the other across all generated data sets. Figure \ref{fig:F1_text_test_diff} shows that the Multi-Scale model tends to outperform its single-scale counterpart on a majority of the data sets, in particular for the LL corpus where it is better on 87 \% of the data sets.

\FloatBarrier

\begin{figure}[h!]
  \begin{center}
  \begin{minipage}[b]{0.32\linewidth}
    \centering
    \includegraphics[width=\linewidth]{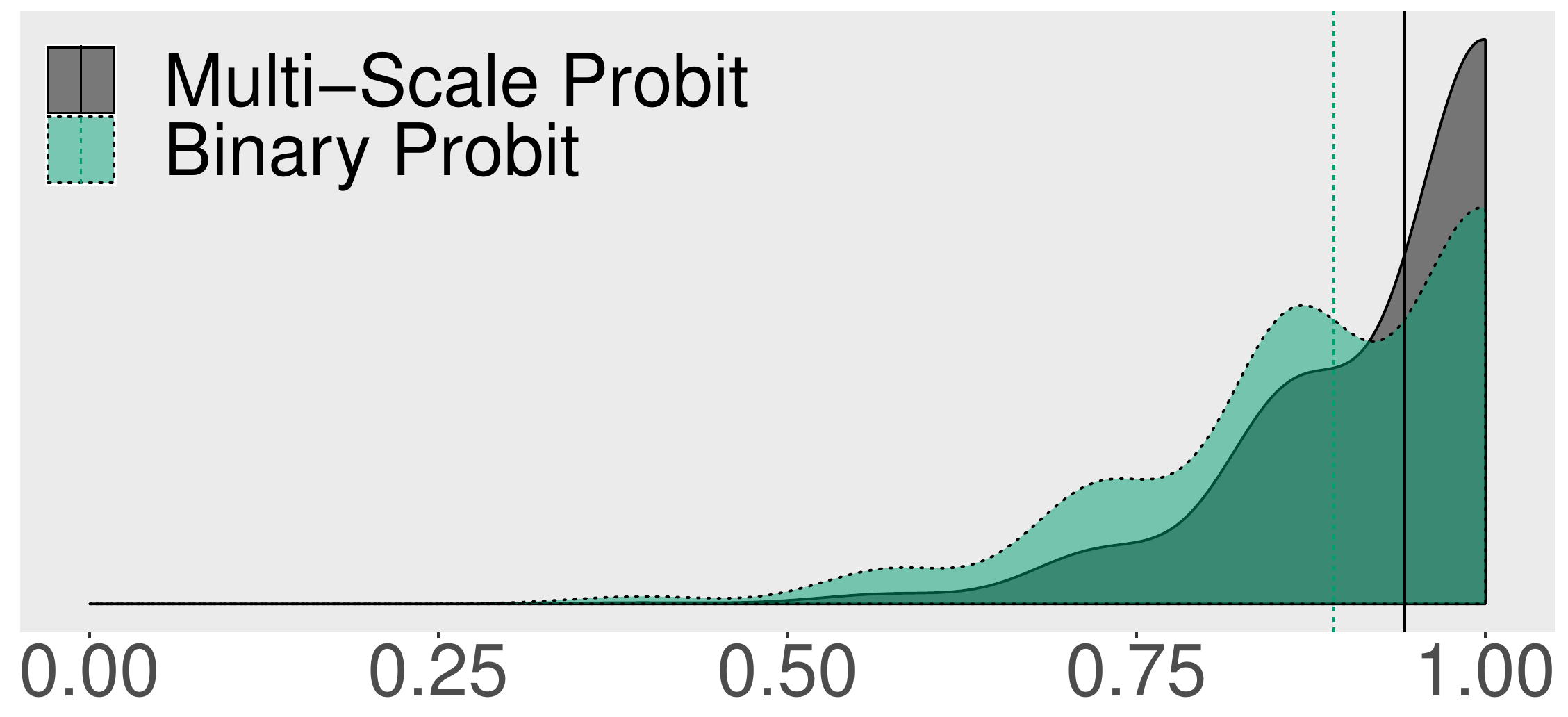} 
    \caption*{LL} 
  \end{minipage} 
  \hfill
  \begin{minipage}[b]{0.32\linewidth}
    \centering
    \includegraphics[width=\linewidth]{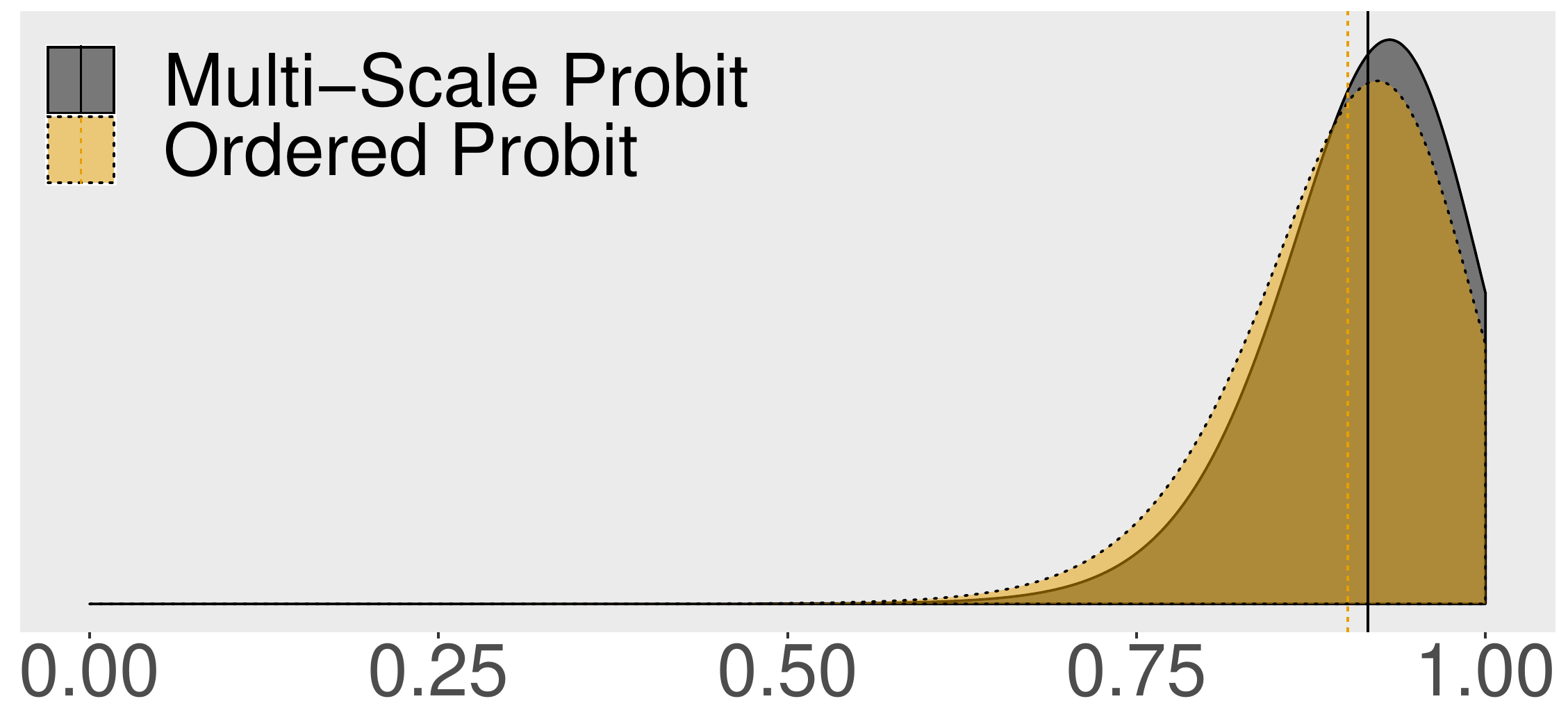} 
    \caption*{Legimus} 
  \end{minipage}
  \hfill
  \begin{minipage}[b]{0.32\linewidth}
    \centering
    \includegraphics[width=\linewidth]{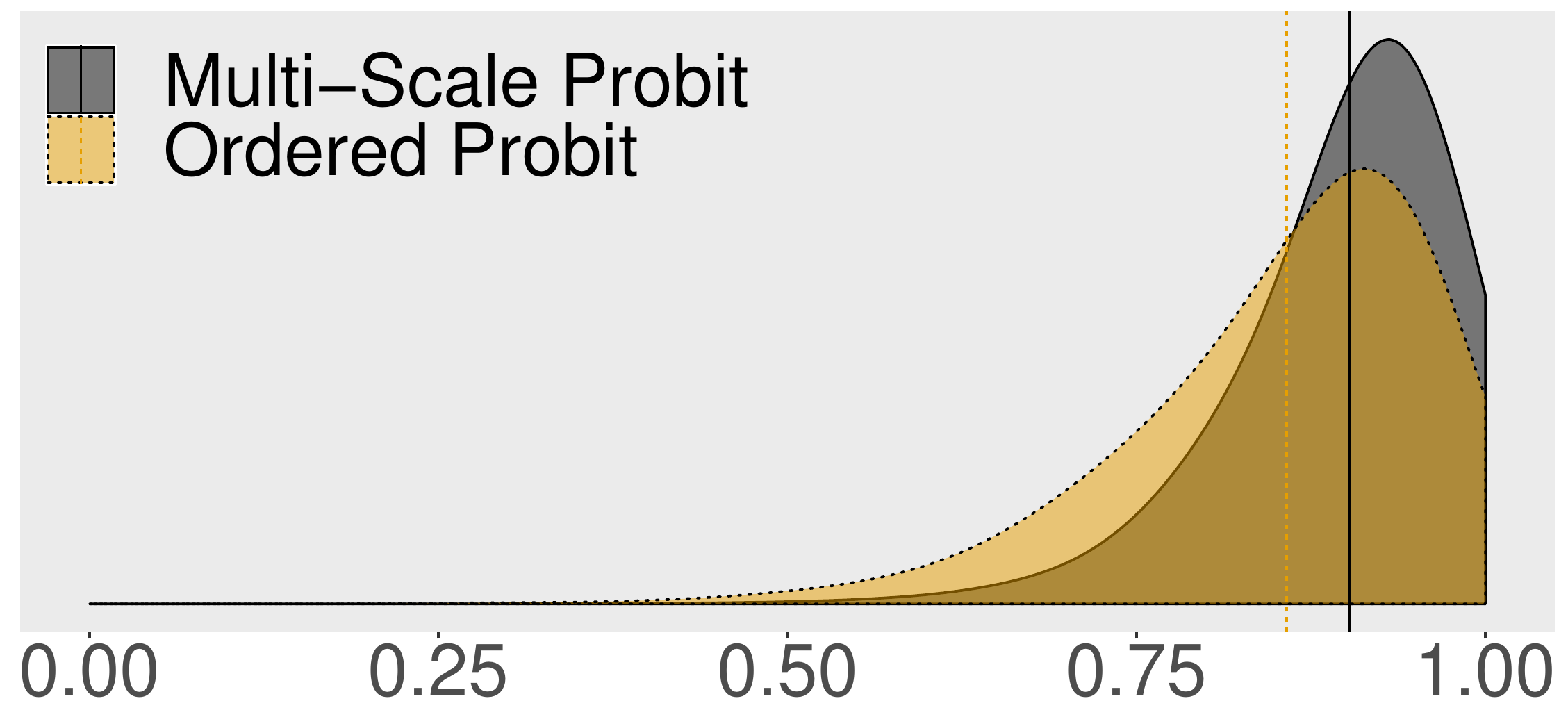} 
    \caption*{Hegas} 
  \end{minipage} 
  \end{center}
  \caption{The posterior distributions for out-of-sample \textit{Kendall} $\tau_B$ correlations of the text data, plotted per measurement scale.}
  \label{fig:kendall_text_test}
\end{figure}

\begin{figure}[h!]
  \begin{center}
  \begin{minipage}[b]{0.32\linewidth}
    \centering
    \includegraphics[width=\linewidth]{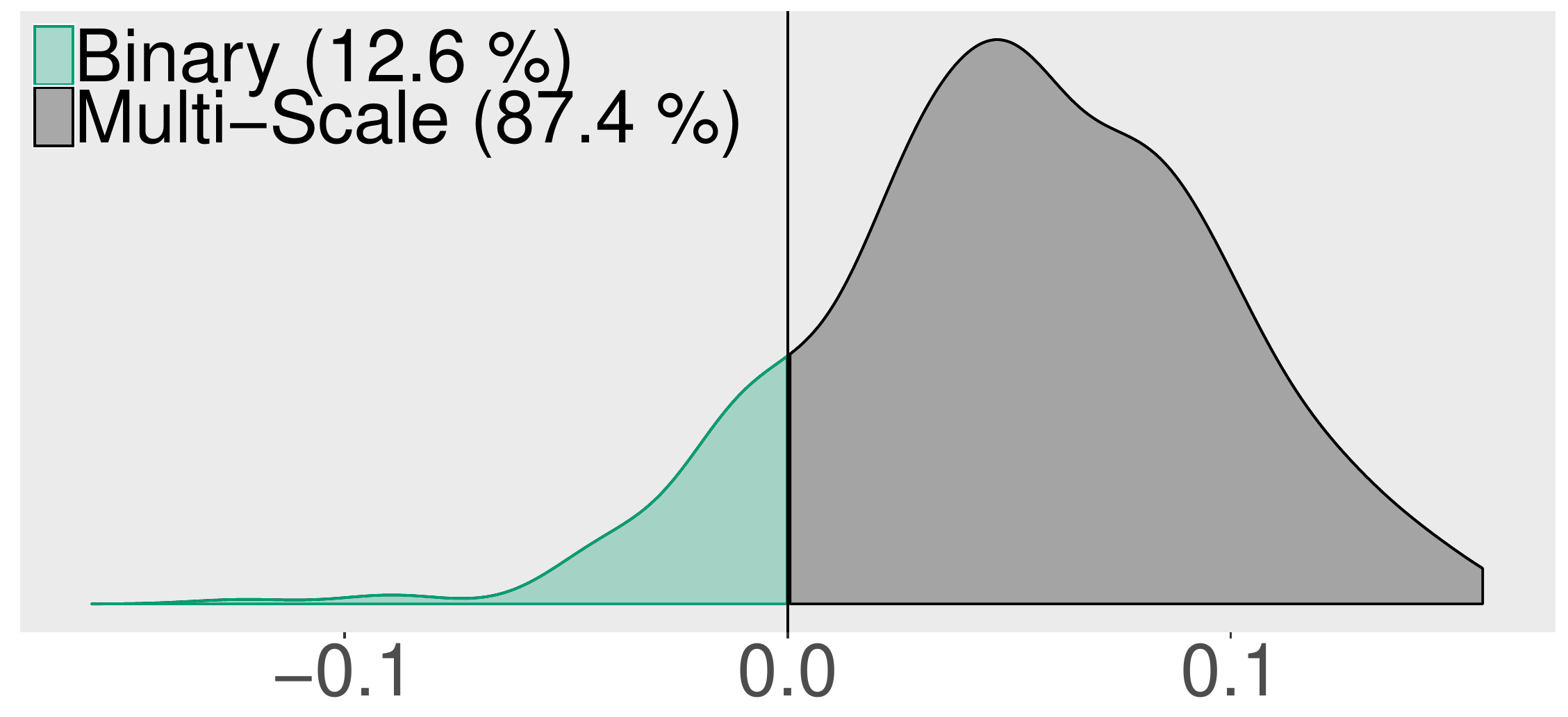} 
    \caption*{LL} 
  \end{minipage} 
  \hfill
  \begin{minipage}[b]{0.32\linewidth}
    \centering
    \includegraphics[width=\linewidth]{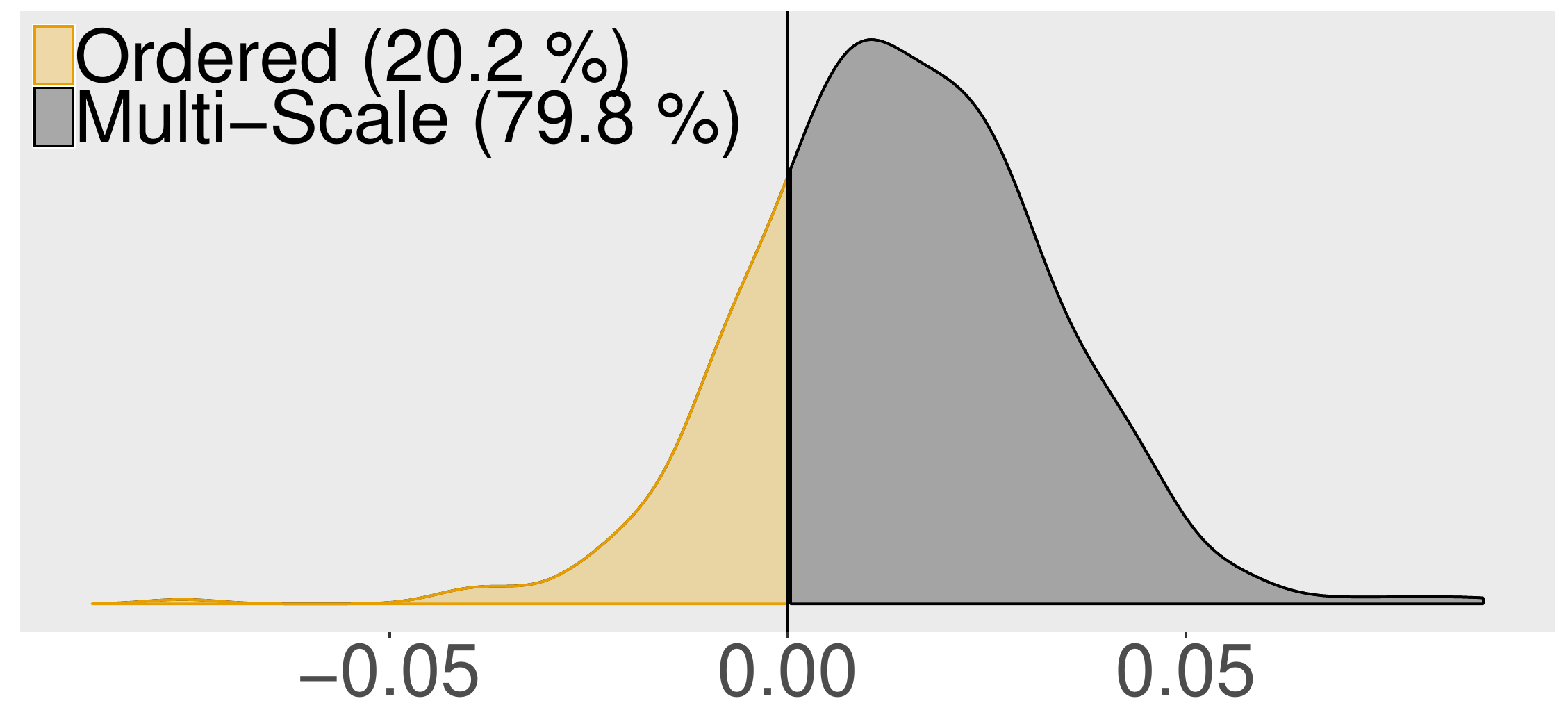} 
    \caption*{Legimus} 
  \end{minipage}
  \hfill
  \begin{minipage}[b]{0.32\linewidth}
    \centering
    \includegraphics[width=\linewidth]{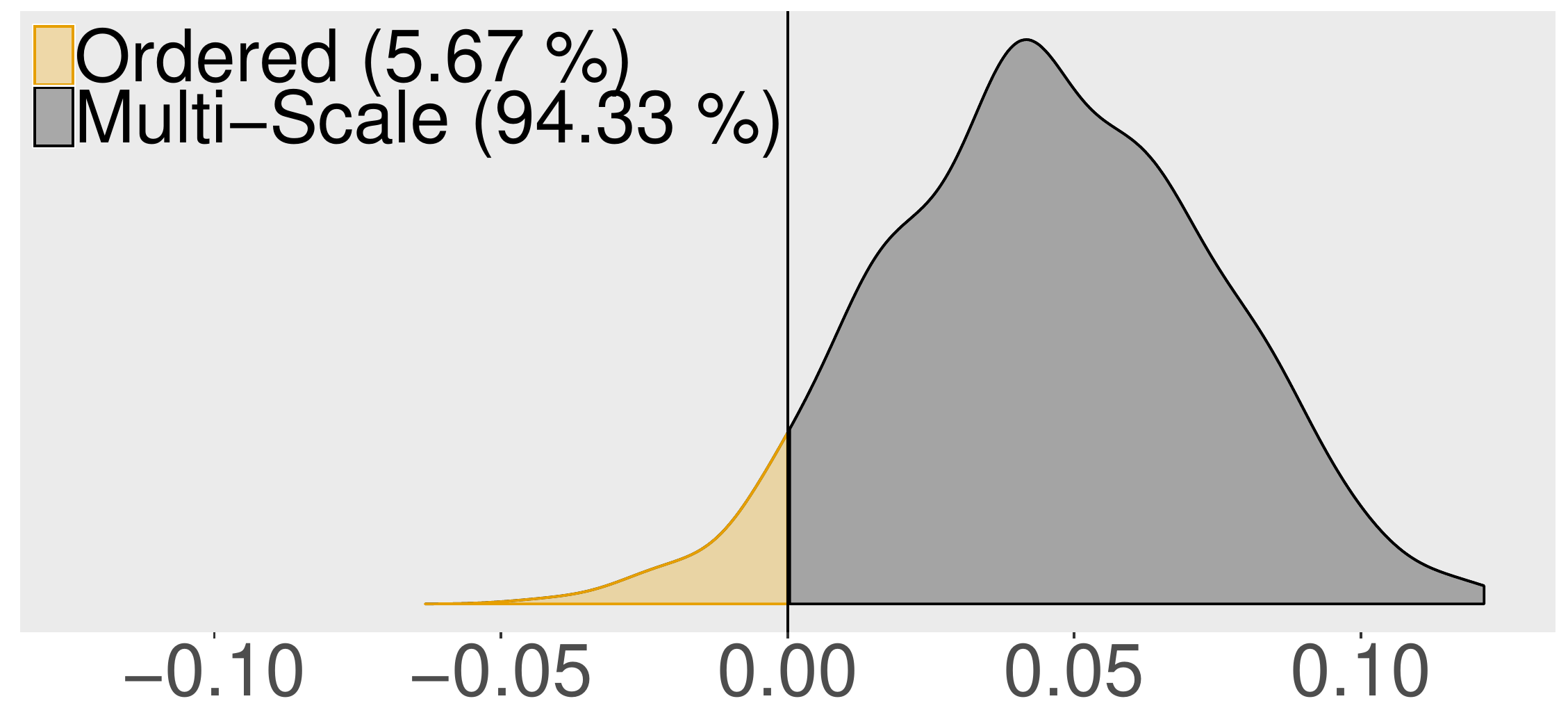} 
    \caption*{Hegas} 
  \end{minipage} 
  \end{center}
  \caption{The posterior distributions for the difference in out-of-sample \textit{Kendall} $\tau_B$ correlation between single-scale and Multi-Scale models on the text data for the 500 different training sets, plotted per measurement scale.}
  \label{fig:kendall_text_test_diff}
\end{figure}

Figure \ref{fig:kendall_text_test} and Figure \ref{fig:kendall_text_test_diff} show that the rankings from the Multi-Scale Probit clearly improves upon the rankings from the single-scale models. In particular, Figure \ref{fig:kendall_text_test_diff} shows that the \textit{Kendall} $\tau_B$ correlations from the Multi-Scale Probit are closer to one than the single-scale models in a clear majority of the 500 generated test data sets.

\FloatBarrier

\begin{figure}[h!]
  \begin{center}
  \begin{minipage}[b]{0.32\linewidth}
    \centering
    \includegraphics[width=\linewidth]{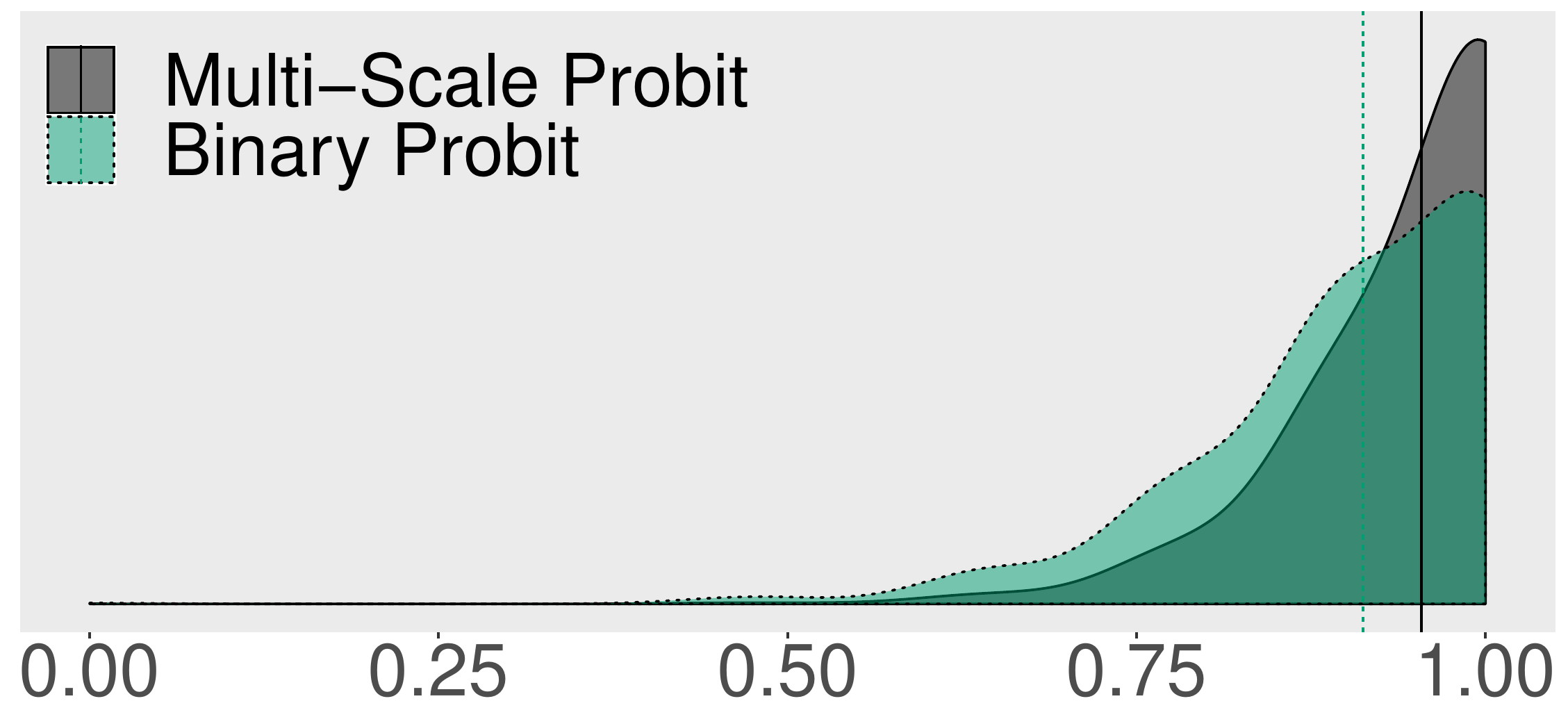} 
    \caption*{LL} 
  \end{minipage} 
  \hfill
  \begin{minipage}[b]{0.32\linewidth}
    \centering
    \includegraphics[width=\linewidth]{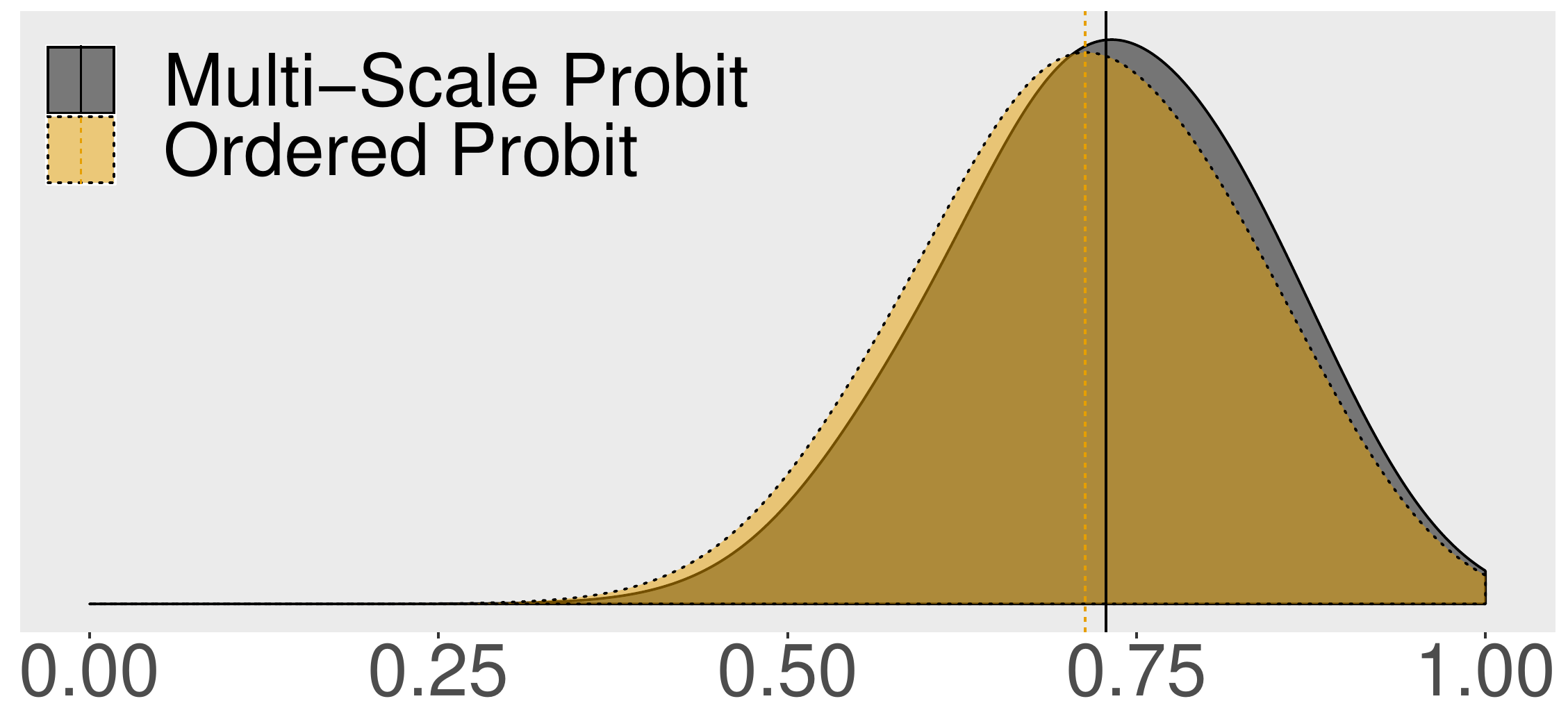} 
    \caption*{Legimus} 
  \end{minipage}
  \hfill
  \begin{minipage}[b]{0.32\linewidth}
    \centering
    \includegraphics[width=\linewidth]{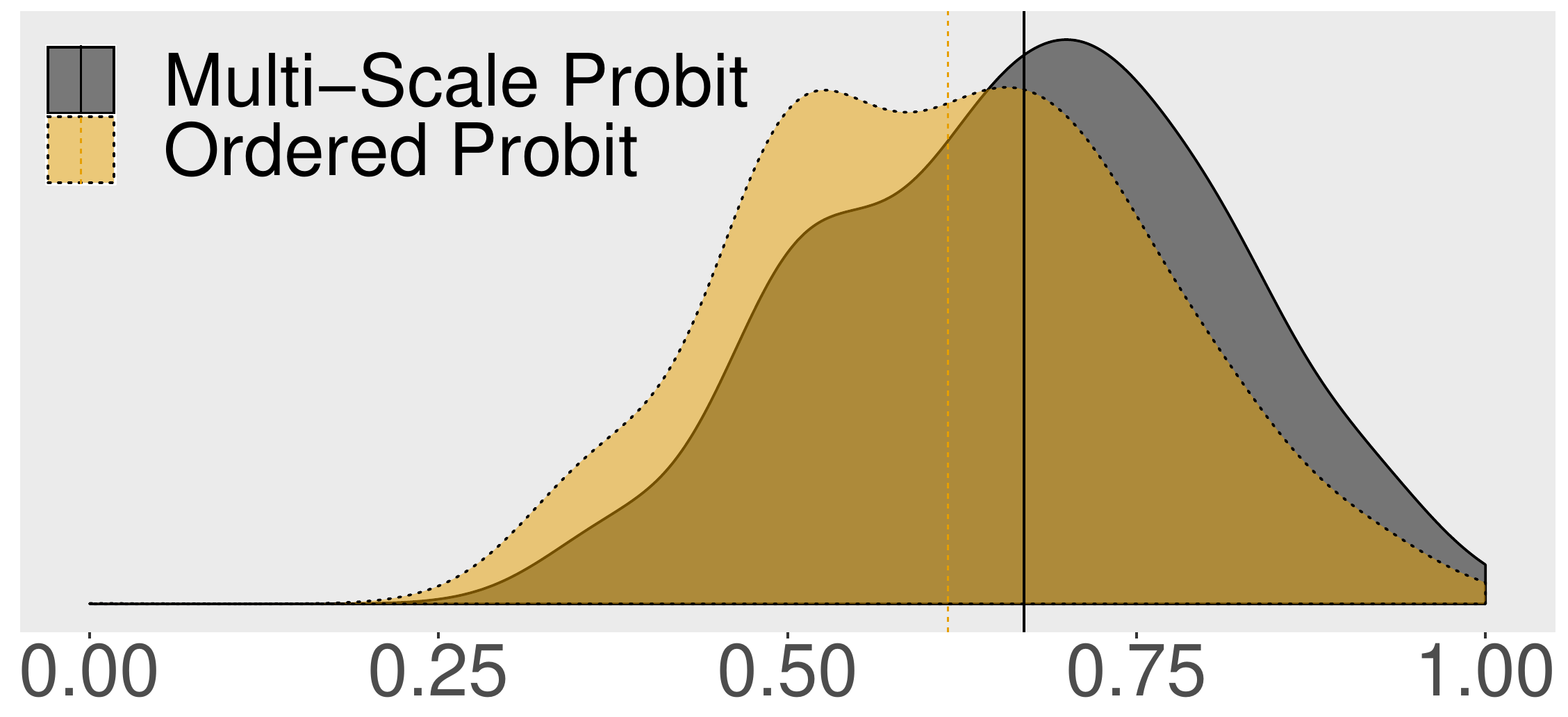} 
    \caption*{Hegas} 
  \end{minipage} 
  \end{center}
  \caption{The posterior distributions for out-of-sample harmonic mean of$F_1$ and \textit{Kendall} $\tau_B$ correlation on the text data, plotted per measurement scale.}
  \label{fig:harm_text_test}
\end{figure}

\begin{figure}[h!]
  \begin{center}
  \begin{minipage}[b]{0.32\linewidth}
    \centering
    \includegraphics[width=\linewidth]{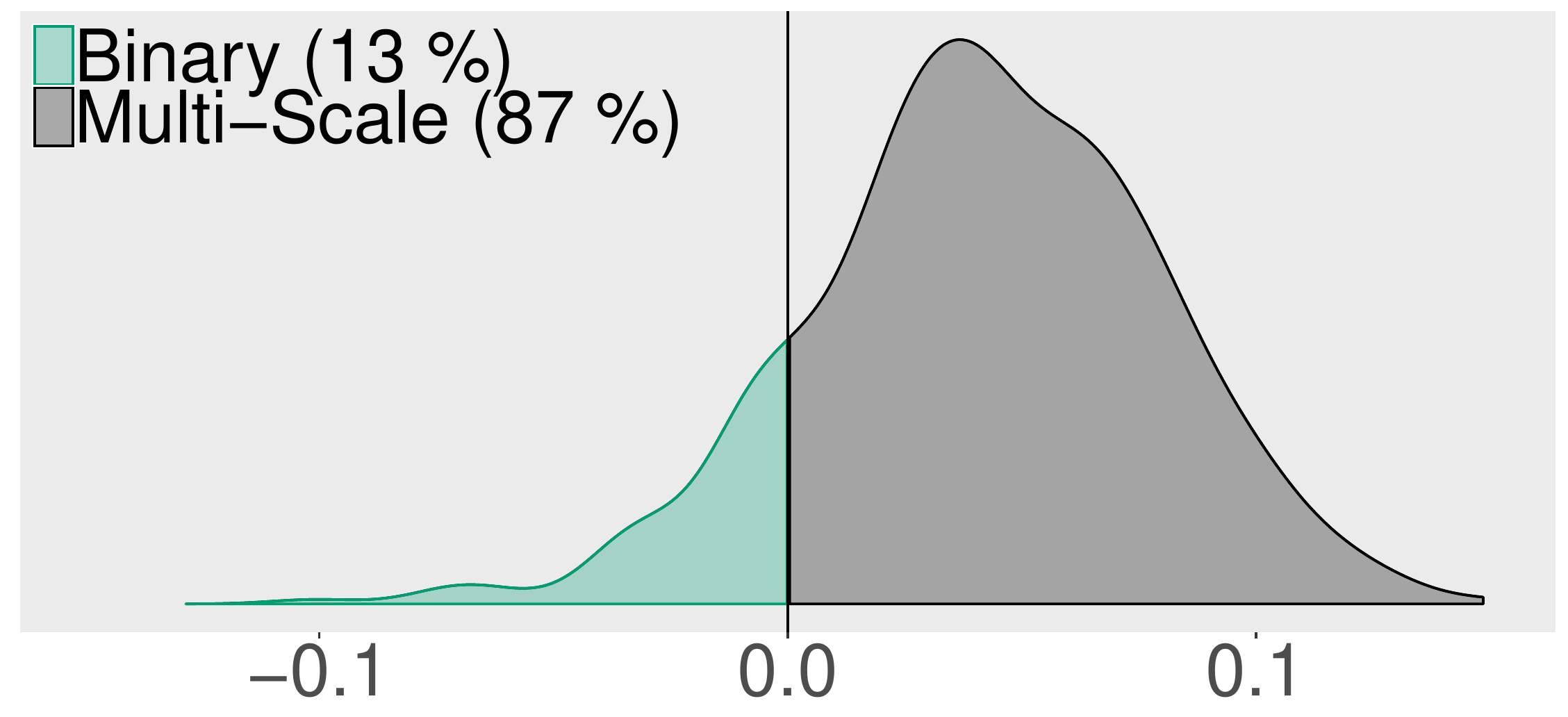} 
    \caption*{LL} 
  \end{minipage} 
  \hfill
  \begin{minipage}[b]{0.32\linewidth}
    \centering
    \includegraphics[width=\linewidth]{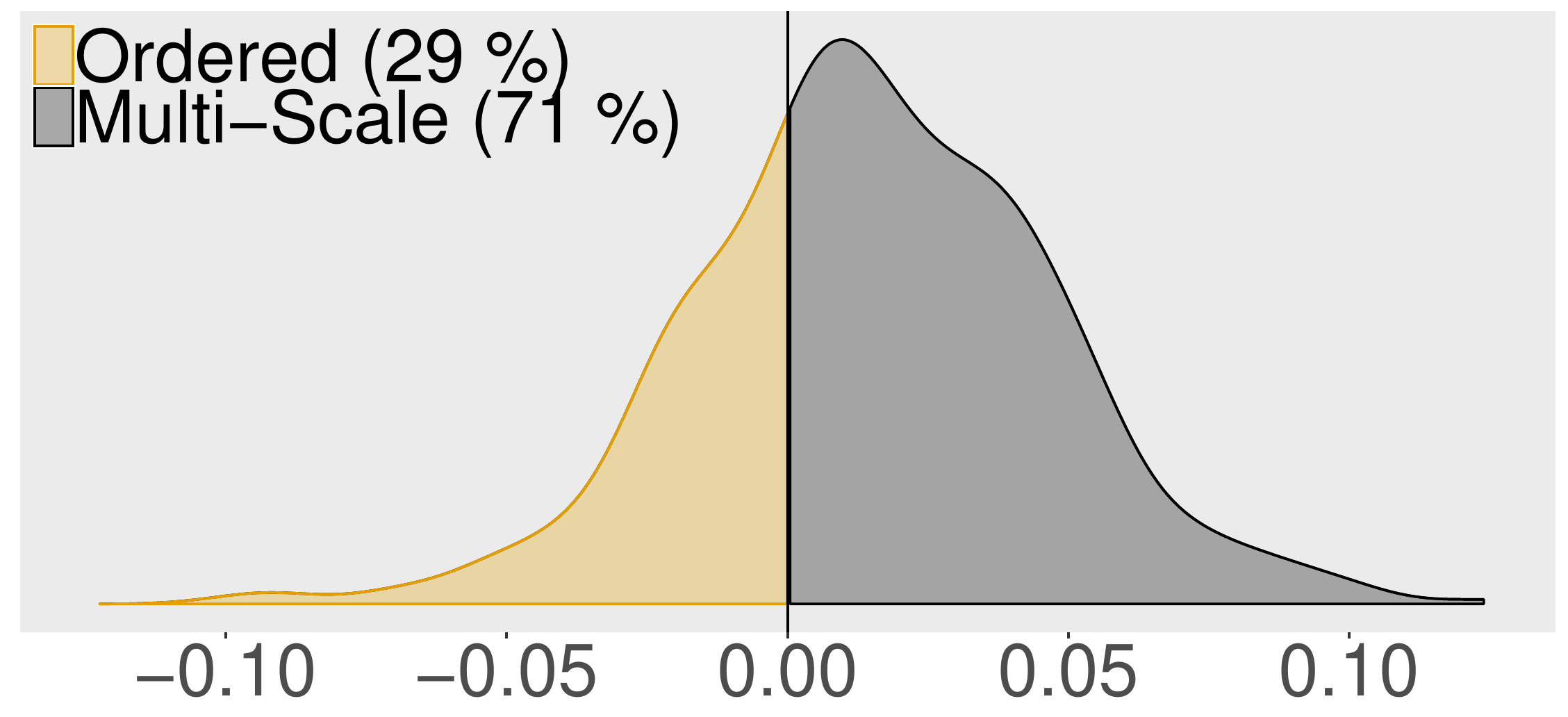} 
    \caption*{Legimus} 
  \end{minipage}
  \hfill
  \begin{minipage}[b]{0.32\linewidth}
    \centering
    \includegraphics[width=\linewidth]{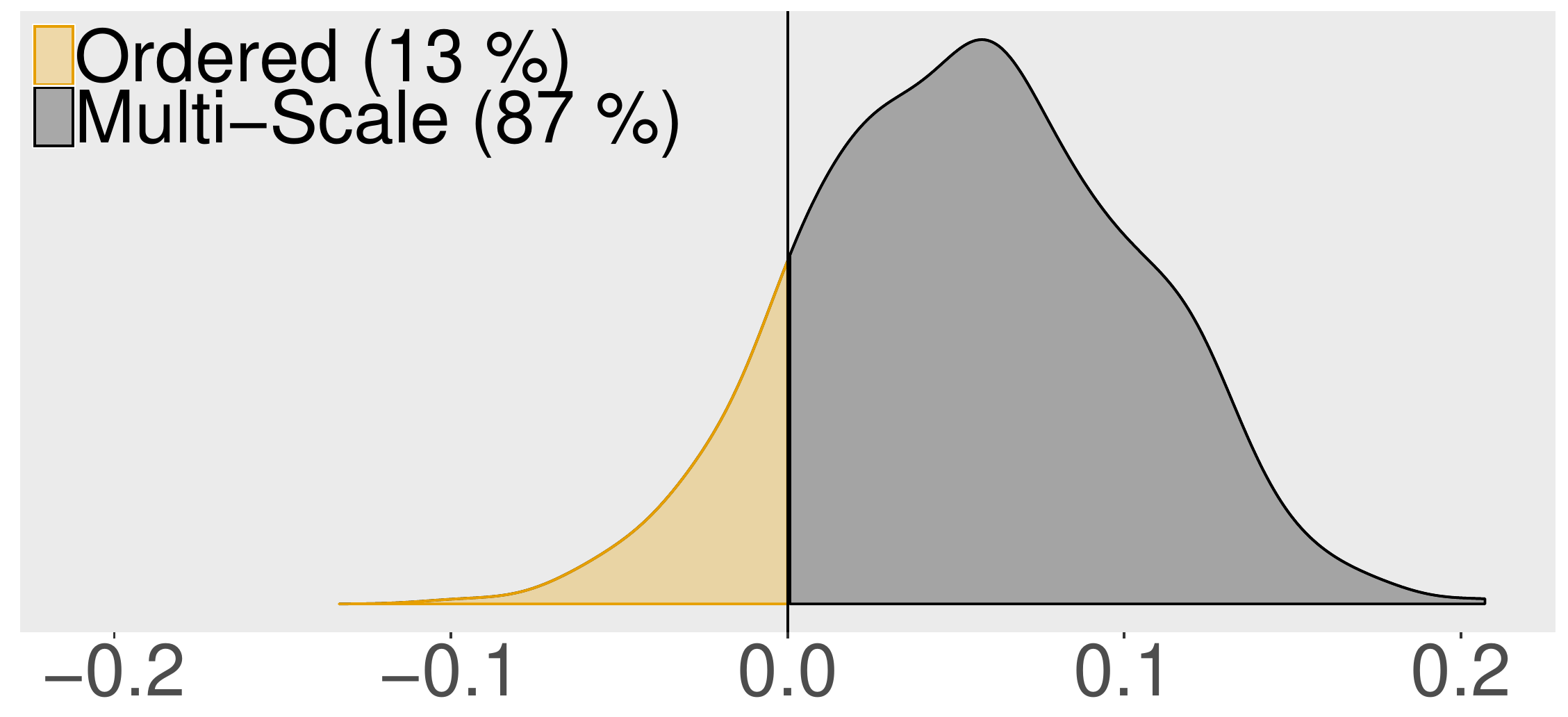} 
    \caption*{Hegas} 
  \end{minipage} 
  \end{center}
  \caption{The posterior distributions for the difference in out-of-sample harmonic mean of $F_1$ and \textit{Kendall} $\tau_B$ correlation between single-scale and Multi-Scale models on the text data for the 500 different training sets, plotted per measurement scale.}
  \label{fig:harm_text_test_diff}
\end{figure}

Figures \ref{fig:harm_text_test} and \ref{fig:harm_text_test_diff} display the posterior distributions of the harmonic mean between $F_1$ scores and \textit{Kendall} $\tau_B$ correlation.

\begin{figure}[h!]
  \begin{center}
  \begin{minipage}[b]{0.32\linewidth}
    \centering
    \includegraphics[width=\linewidth]{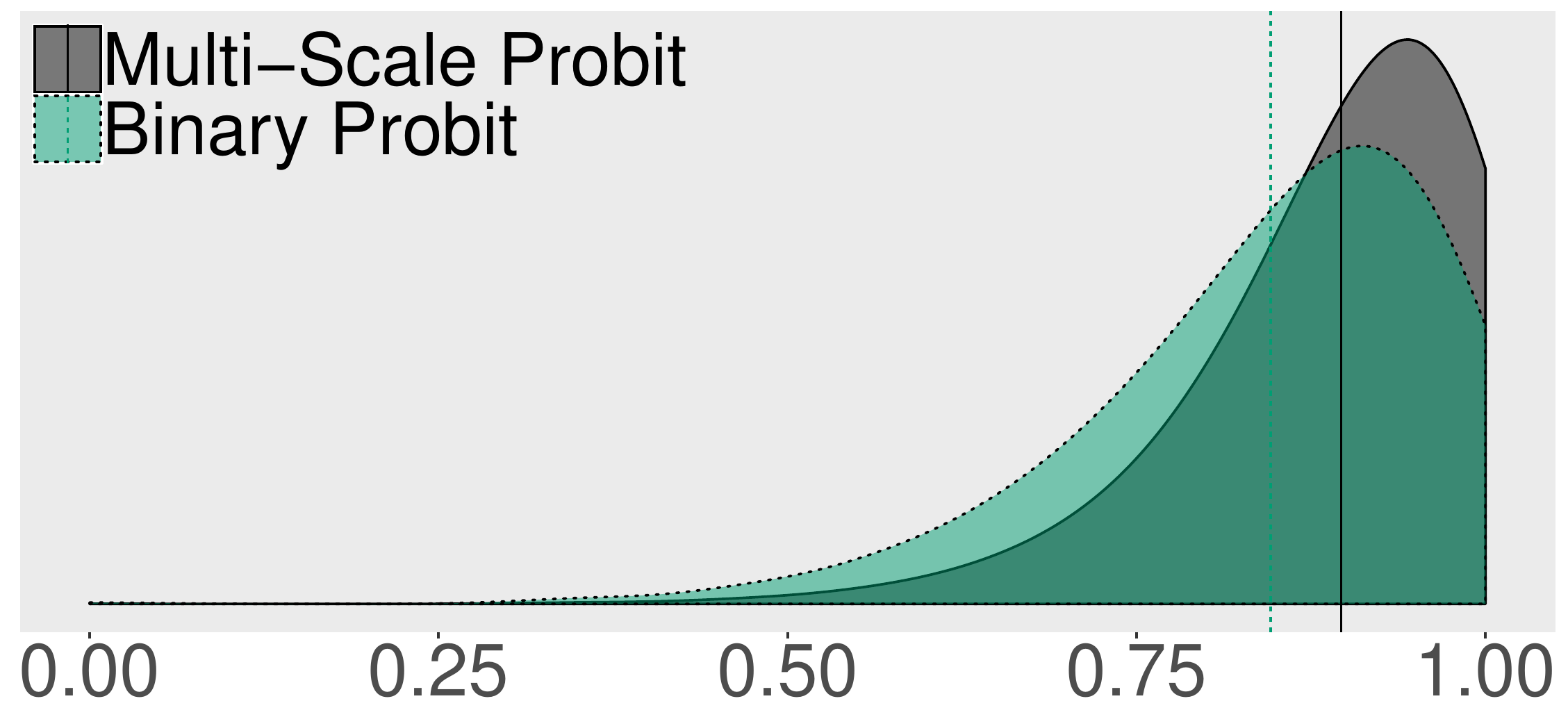} 
    \caption*{LL} 
  \end{minipage} 
  \hfill
  \begin{minipage}[b]{0.32\linewidth}
    \centering
    \includegraphics[width=\linewidth]{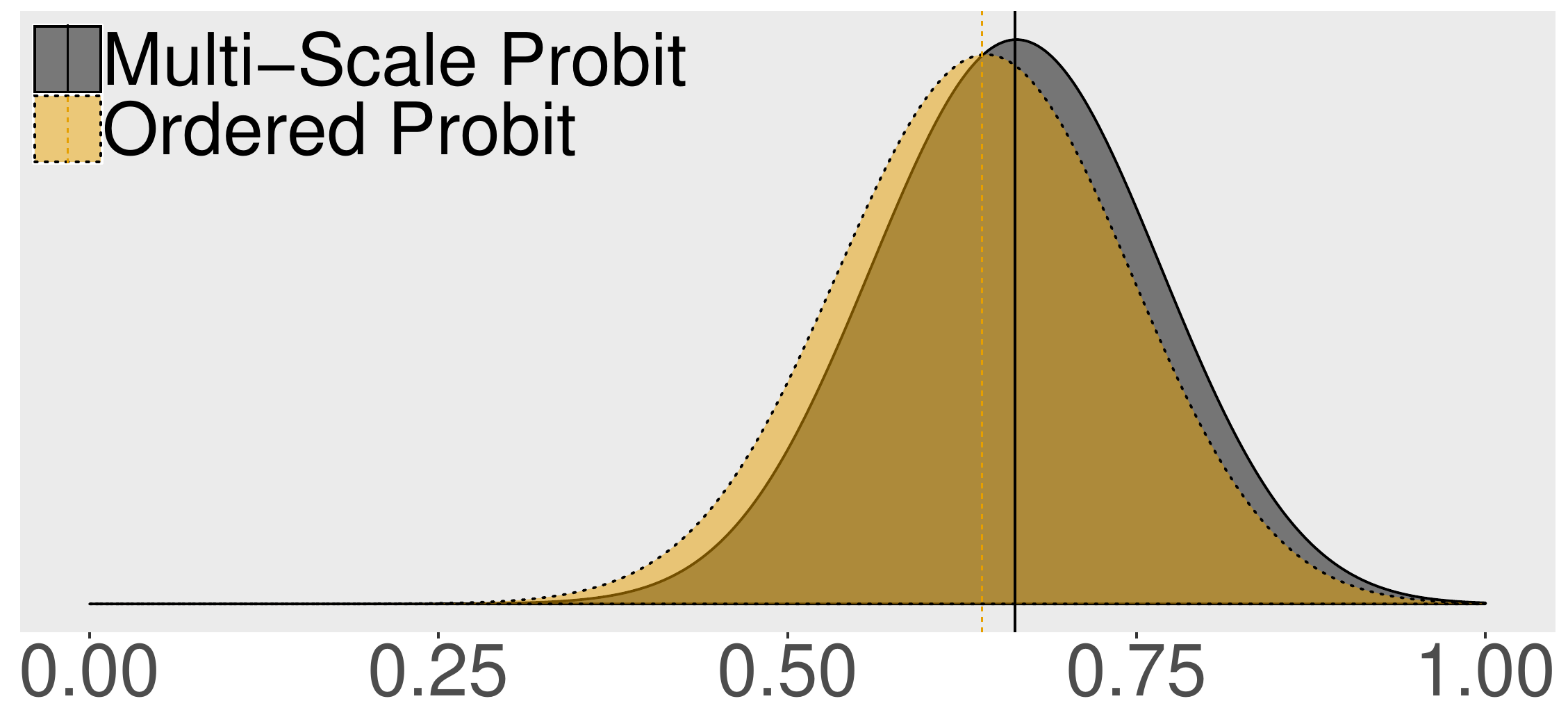} 
    \caption*{Legimus} 
  \end{minipage}
  \hfill
  \begin{minipage}[b]{0.32\linewidth}
    \centering
    \includegraphics[width=\linewidth]{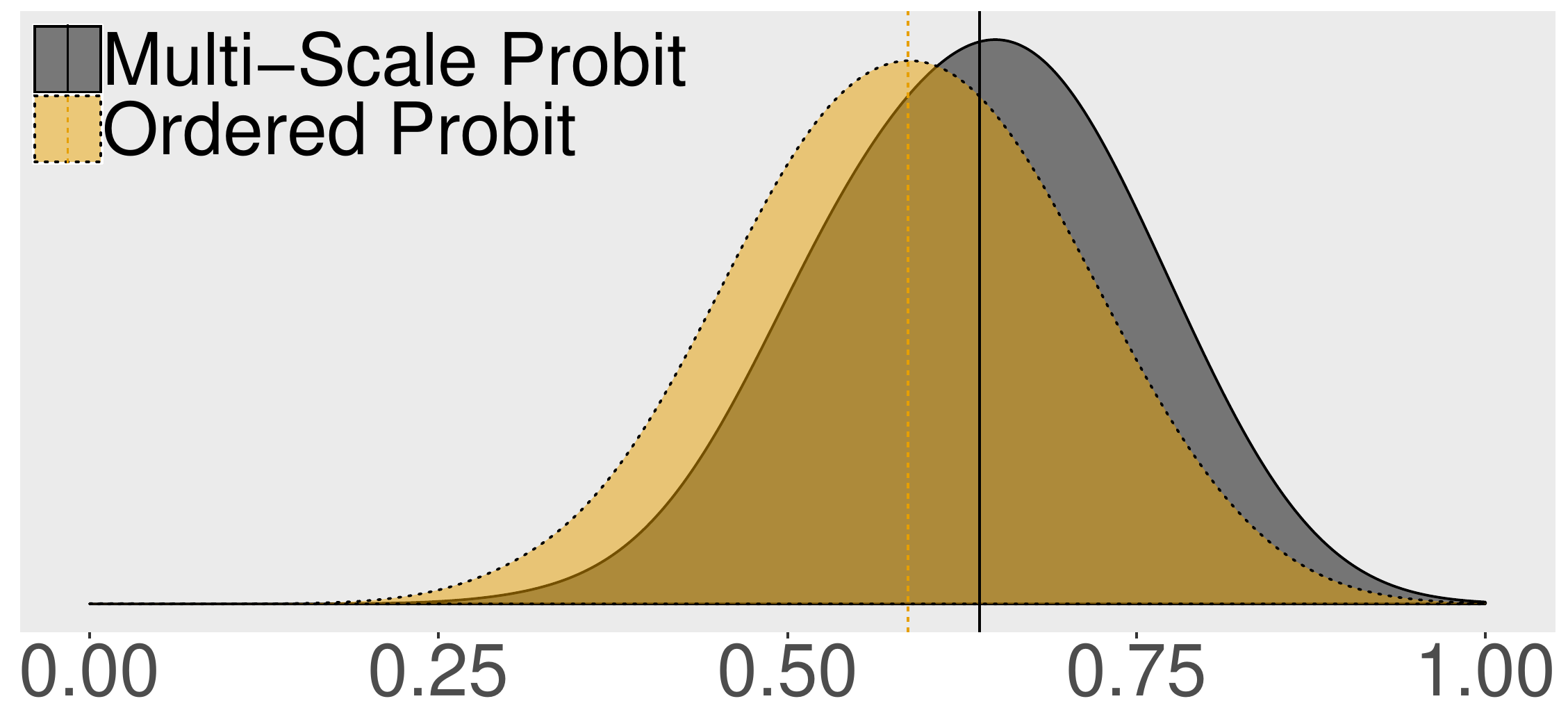} 
    \caption*{Hegas} 
  \end{minipage} 
  \end{center}
  \caption{The posterior distributions for out-of-sample harmonic mean of $F_1$ and \textit{Kendall} $\tau_B$ correlation on the text data, plotted per measurement scale, using only 1/3 of the data for training.}
  \label{fig:harm_text_test_1third}
\end{figure}

\begin{figure}[h!]
  \begin{center}
  \begin{minipage}[b]{0.32\linewidth}
    \centering
    \includegraphics[width=\linewidth]{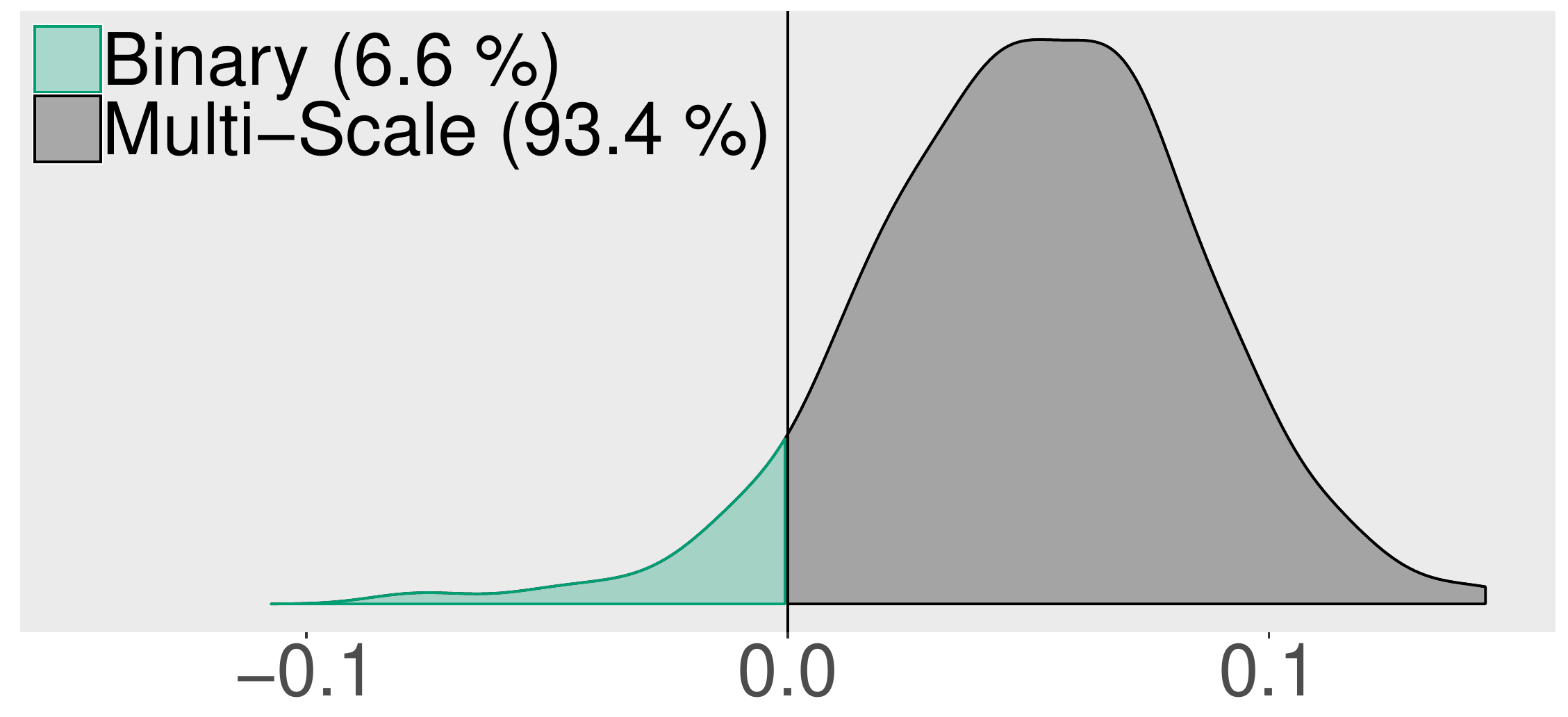} 
    \caption*{LL} 
  \end{minipage} 
  \hfill
  \begin{minipage}[b]{0.32\linewidth}
    \centering
    \includegraphics[width=\linewidth]{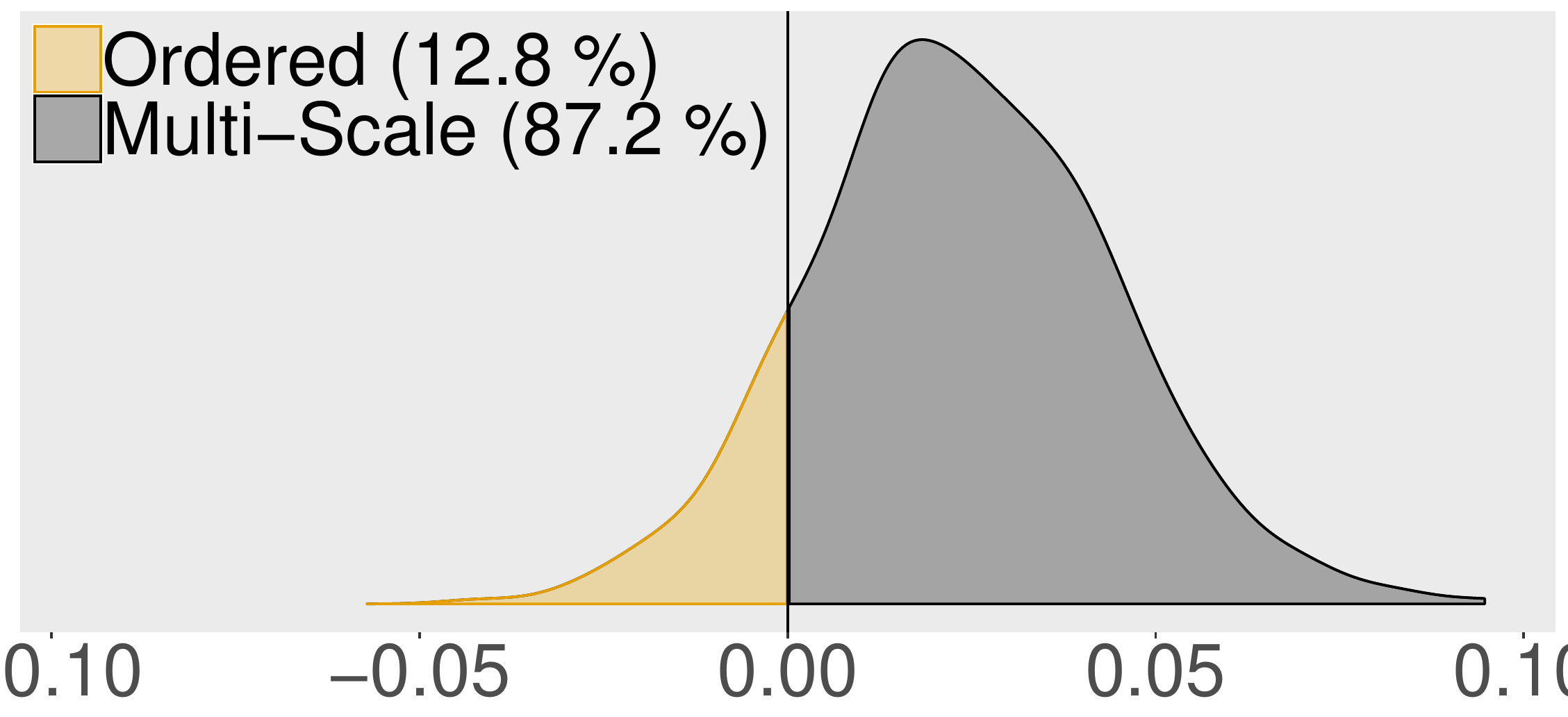} 
    \caption*{Legimus} 
  \end{minipage}
  \hfill
  \begin{minipage}[b]{0.32\linewidth}
    \centering
    \includegraphics[width=\linewidth]{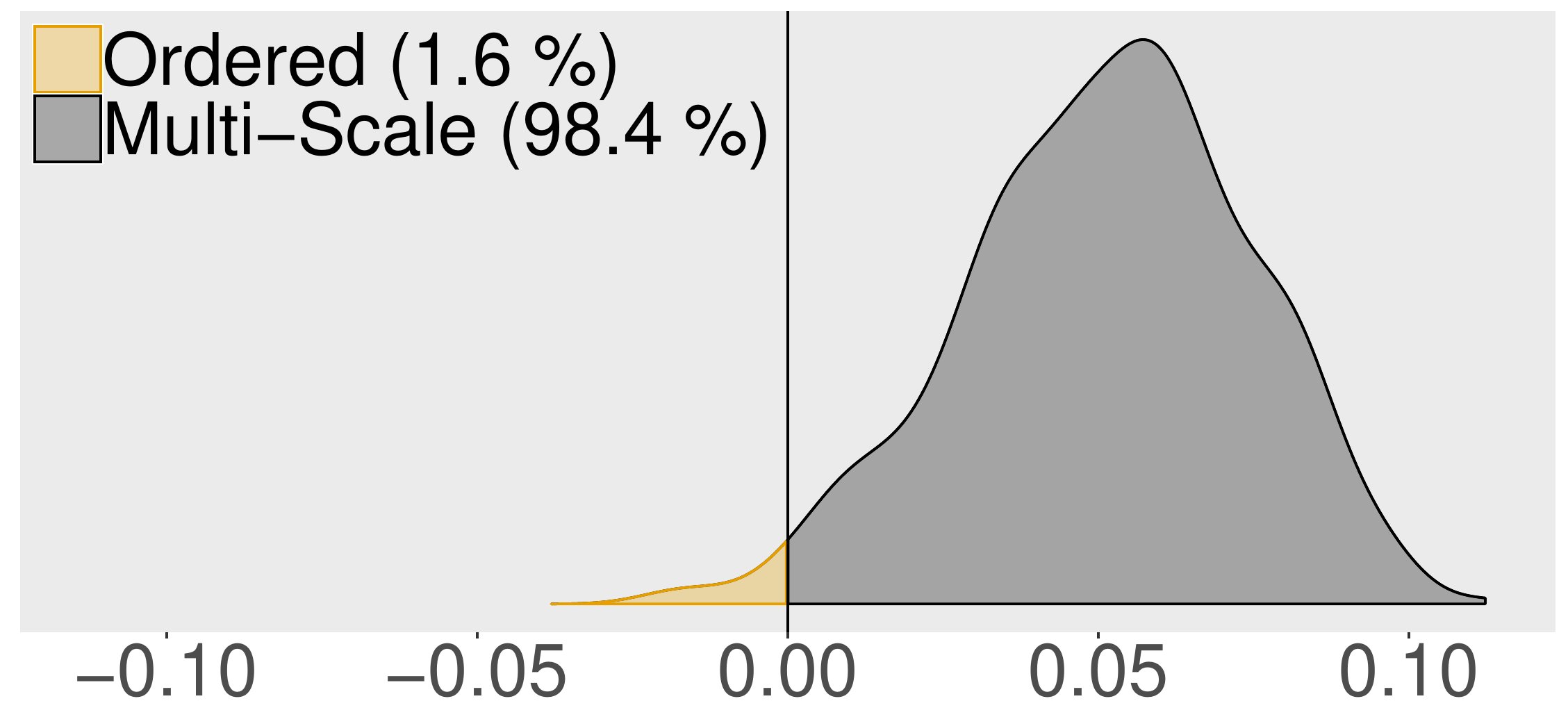} 
    \caption*{Hegas} 
  \end{minipage} 
  \end{center}
  \caption{The posterior distributions for the difference in out-of-sample harmonic mean of $F_1$ and \textit{Kendall} $\tau_B$ correlation between single-scale and Multi-Scale models on the text data for the 500 different training sets, plotted per measurement scale, using only 1/3 of the data for training.}
  \label{fig:harm_text_test_diff_1third}
\end{figure}

Finally, we explore how the models perform with less training data by repeating the above experiments, but this time using only 1/3 of the data for training and evaluating on the remaining 2/3. The training-test split is again repeated 500 times. Figure \ref{fig:harm_text_test_1third} displays the posterior distributions of harmonic means of $F_1$ score and \textit{Kendall} $\tau_B$ correlation for 500 different training sets using this set-up, and Figure \ref{fig:harm_text_test_diff_1third} shows the differences between the models for each data set. It is clear that the advantage of the Multi-Scale Probit increases with smaller training data sets.

There are two opposing factors that determine the relative success of the Multi-Scale model: the advantage of pooling the data over multiple corpora against the restriction to a single latent variable driving all corpora. In highly informative data sets with many data points and low-dimensional feature sets the benefits from data pooling may not outweigh the disadvantage of the single latent variable restriction, assuming that the corpora do not fully satisfy the restriction. 

\subsection{Posterior analysis}

In order to get the best possible posterior estimate, we ran 8 chains of the Gibbs sampler on all data from the three corpora and combined the resulting samples. 

\subsubsection{Posterior for $\boldsymbol{\beta}$}

As with all Bayesian regression models we can inspect the posterior distribution for each coefficient $\beta_k$ in order to reason about its influence on the latent variable. In this context, this is equivalent to reasoning about the influence of a linguistic feature on text complexity. In this case there are 48 covariates and we have selected a few illustrative examples. 

\begin{figure}[h!]
  \begin{center}
  \begin{minipage}[b]{0.32\linewidth}
    \centering
    \includegraphics[width=\linewidth]{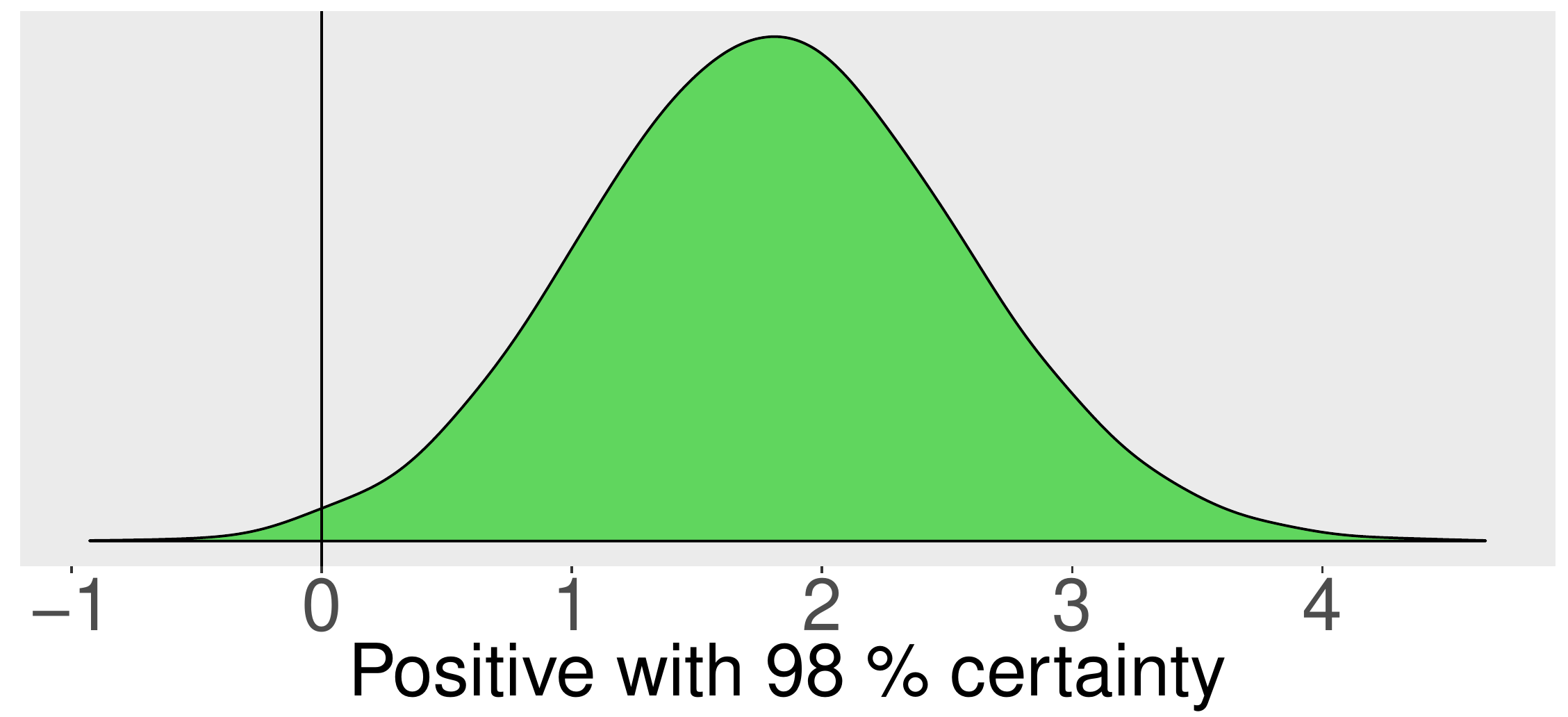}
    \caption*{pos\_HP} 
  \end{minipage} 
  \hfill
  \begin{minipage}[b]{0.32\linewidth}
    \centering
    \includegraphics[width=\linewidth]{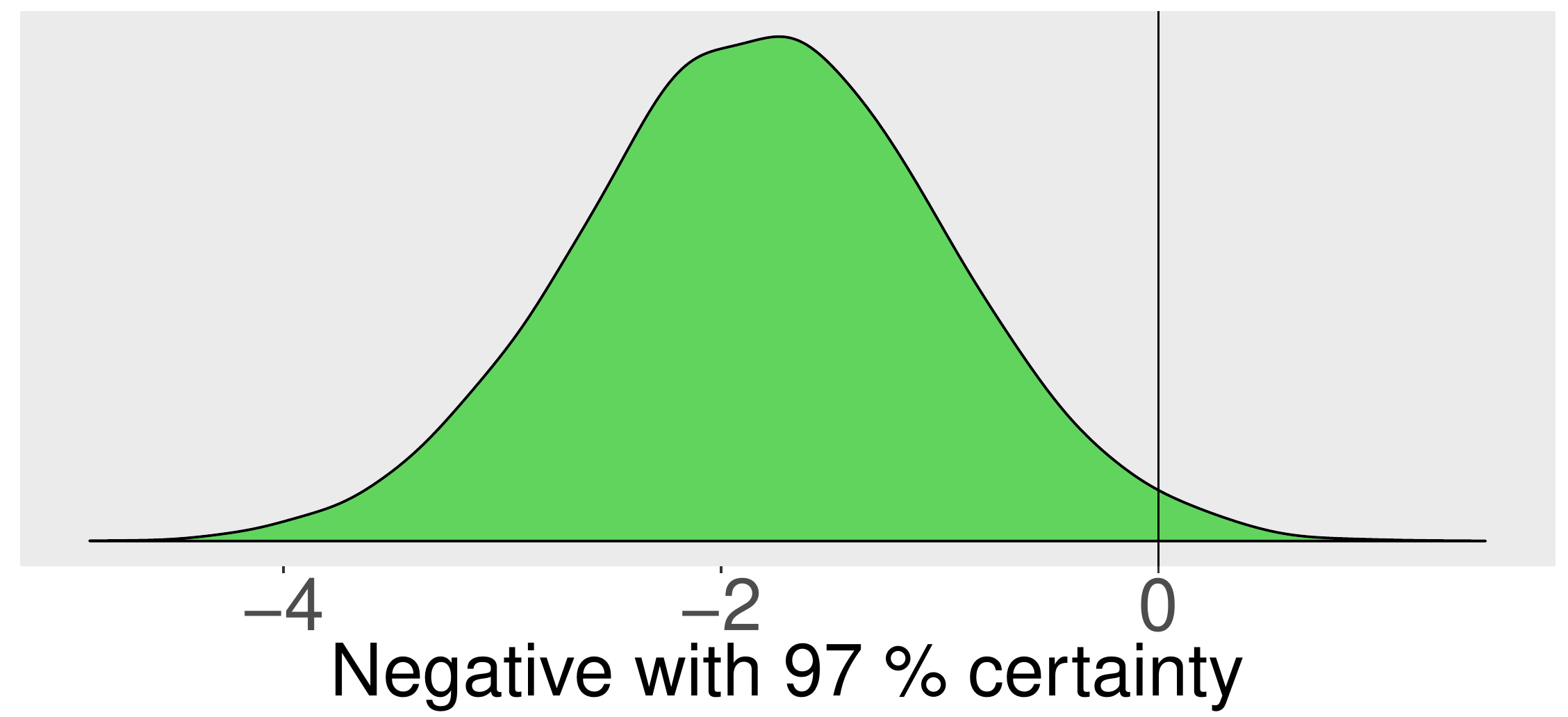}
    \caption*{ratioSweVocTotal} 
  \end{minipage}
  \hfill
  \begin{minipage}[b]{0.32\linewidth}
    \centering
    \includegraphics[width=\linewidth]{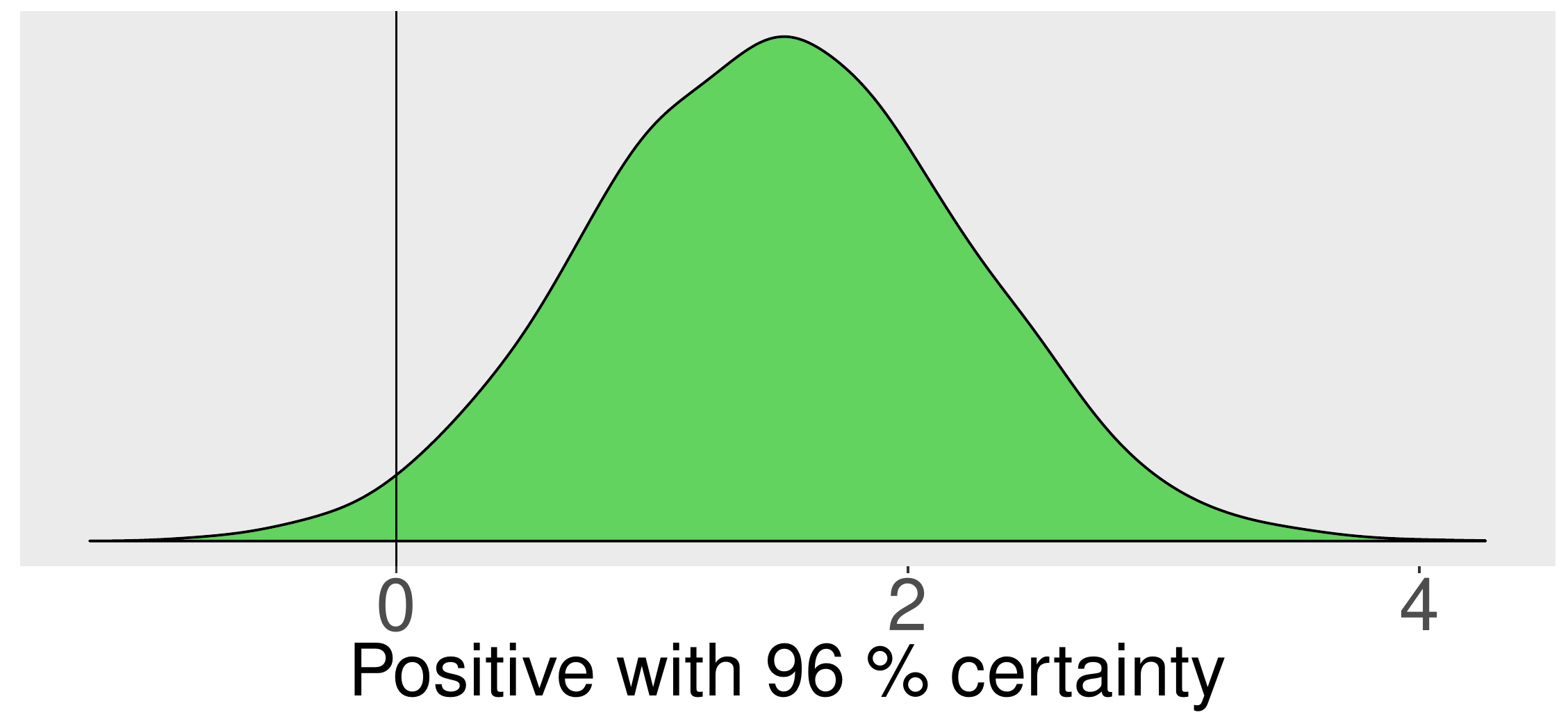}
    \caption*{ratioRightDeps} 
  \end{minipage} 
  \end{center}
  \caption{The three marginal posteriors for coefficients in $\boldsymbol{\beta}$ with the least uncertainty.}
  \label{fig:beta_posterior_certain}
\end{figure}

The three covariates with the least uncertainty in the posterior are frequency of relative/interrogative pronouns (pos\_HP), for example \textit{vem} (\textit{who}), \textit{vad} (\textit{what}), and \textit{vilket} (\textit{that}), the ratio of words existing in any category in the SweVoc lexicon (ratioSweVocTotal), and the ratio of grammatical dependency relations where the dependent occurs after its head word (ratioRightDeps). The marginal posteriors for each of these are plotted in Figure \ref{fig:beta_posterior_certain}.

The frequency of relative/interrogative pronouns has a rather certain \textit{positive} influence on text complexity. In this context \textit{positive} means that a higher frequency of relative/interrogative pronouns indicate a \textit{more} complex text. The feature ratioSweVocTotal instead has a relatively strong \textit{negative} influence on text complexity. That is, a larger proportion of words in the text which belong to a lexicon of common and "simple" words result in a lower text complexity value, that is, a less complex text.

\begin{figure}[h!]
  \begin{center}
  \begin{minipage}[b]{0.32\linewidth}
    \centering
    \includegraphics[width=\linewidth]{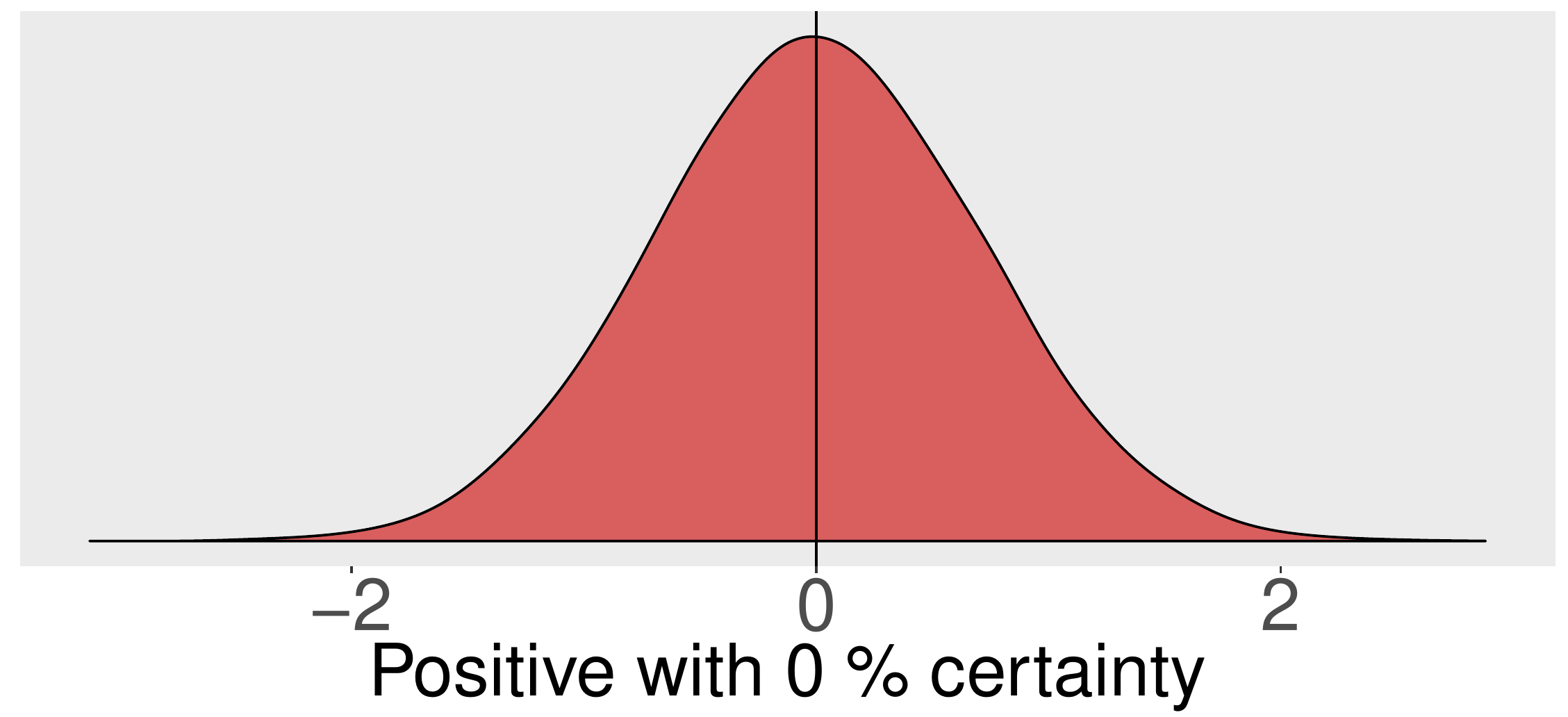}
    \caption*{dep\_VO} 
  \end{minipage} 
  \hfill
  \begin{minipage}[b]{0.32\linewidth}
    \centering
    \includegraphics[width=\linewidth]{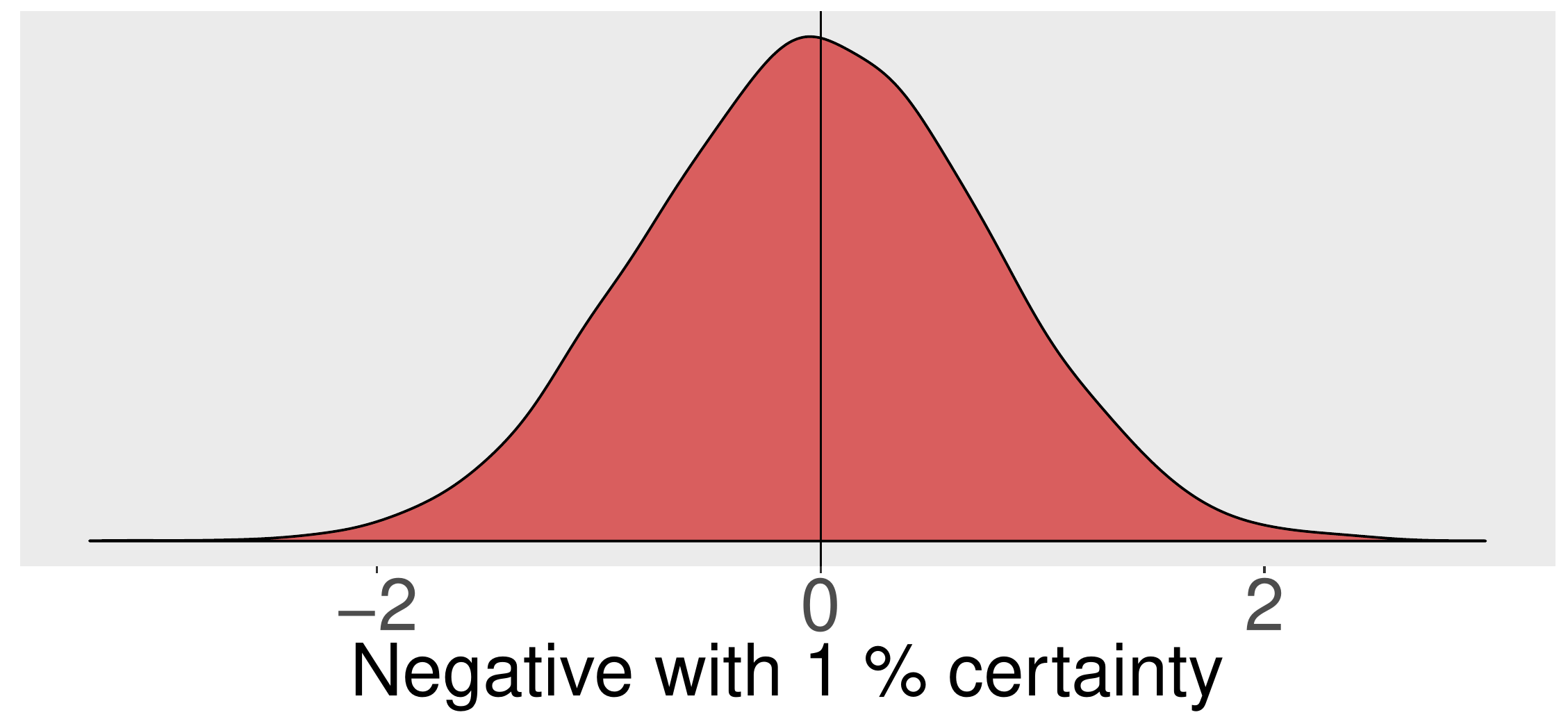}
    \caption*{dep\_MA} 
  \end{minipage}
  \hfill
  \begin{minipage}[b]{0.32\linewidth}
    \centering
    \includegraphics[width=\linewidth]{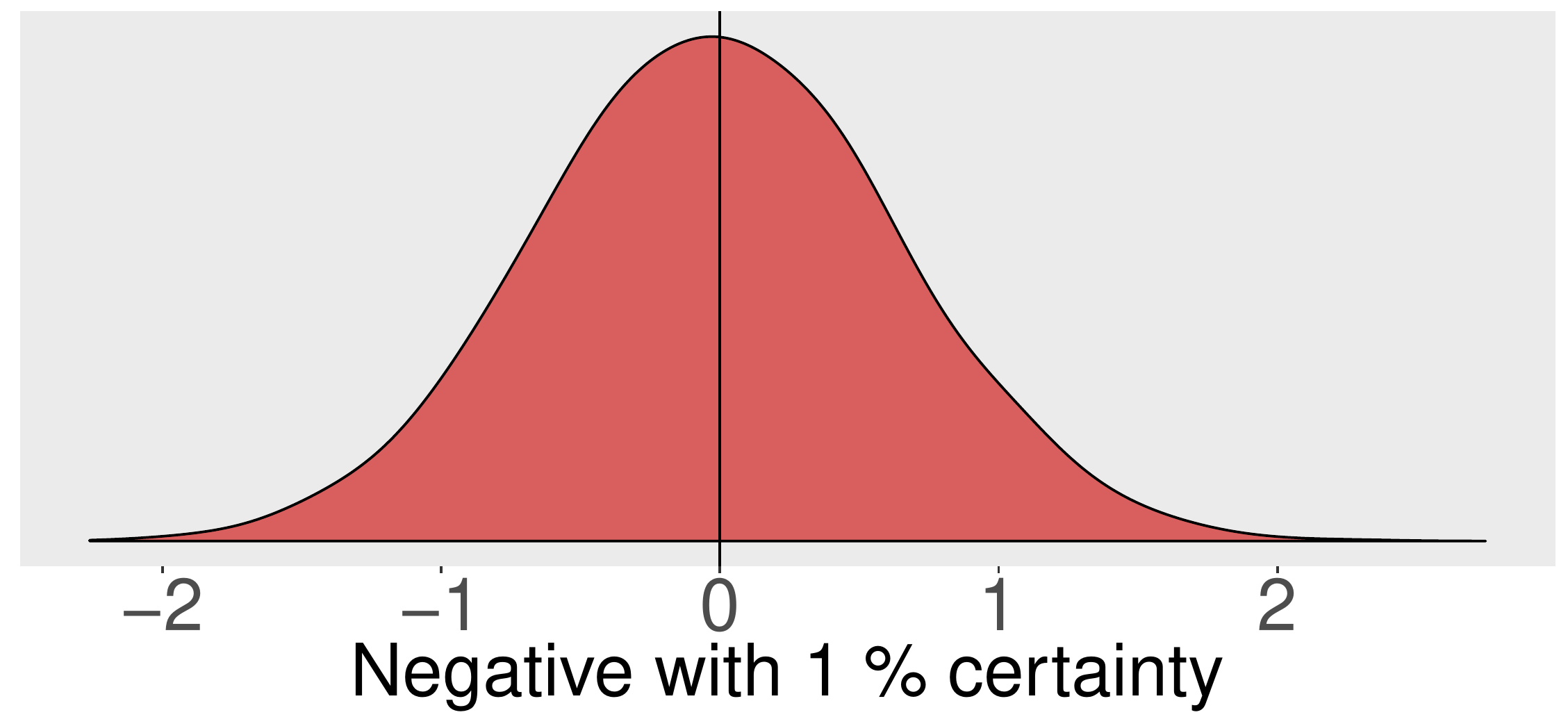}
    \caption*{verbArity5} 
  \end{minipage} 
  \end{center}
  \caption{The three marginal posteriors for coefficients in $\boldsymbol{\beta}$ with the most uncertainty.}
  \label{fig:beta_posterior_uncertain}
\end{figure}

These marginal $\beta_k$ posteriors can be contrasted to the three most uncertain marginal $\beta_k$ posteriors. These are the frequency of the infinitive object complement grammatical construct (dep\_VO), the frequency of attitude adverbials (dep\_MA), and the frequency of verbs with exactly 5 dependants (verbArity5). The marginal posteriors for each of these are plotted in Figure \ref{fig:beta_posterior_uncertain}. These features are hence likely to not be informative about the complexity of the texts.

Contrasting these results quickly to previous research on the classification performance of linguistic features in the context of Support Vector Machines \parencite{Falkenjack2013FeaturesText, Falkenjack2014ClassifyingParsing} we can see that our results are quite different. \textcite{Falkenjack2014ClassifyingParsing} found that the ratio of relative/interrogative pronouns performed barely better than chance on the task of classifying mixed-genre easy-to-read texts. The ratio of SweVoc words and the ratio of rightward dependencies were clearly better than chance but were not among the strongest predictors. It should be noted however that our feature set is a subset of the feature set used by \textcite{Falkenjack2013FeaturesText} and that we are also comparing very different types of analyses using different data sets. Our results do agree with \textcite{Falkenjack2013FeaturesText} regarding rate of infinitive object complements and attitude adverbials not being particularly strong predictors of text complexity. 

\subsubsection{Posterior for $\boldsymbol{\gamma}$}

\begin{figure}[h!]
  \begin{center}
    \includegraphics[width=\linewidth]{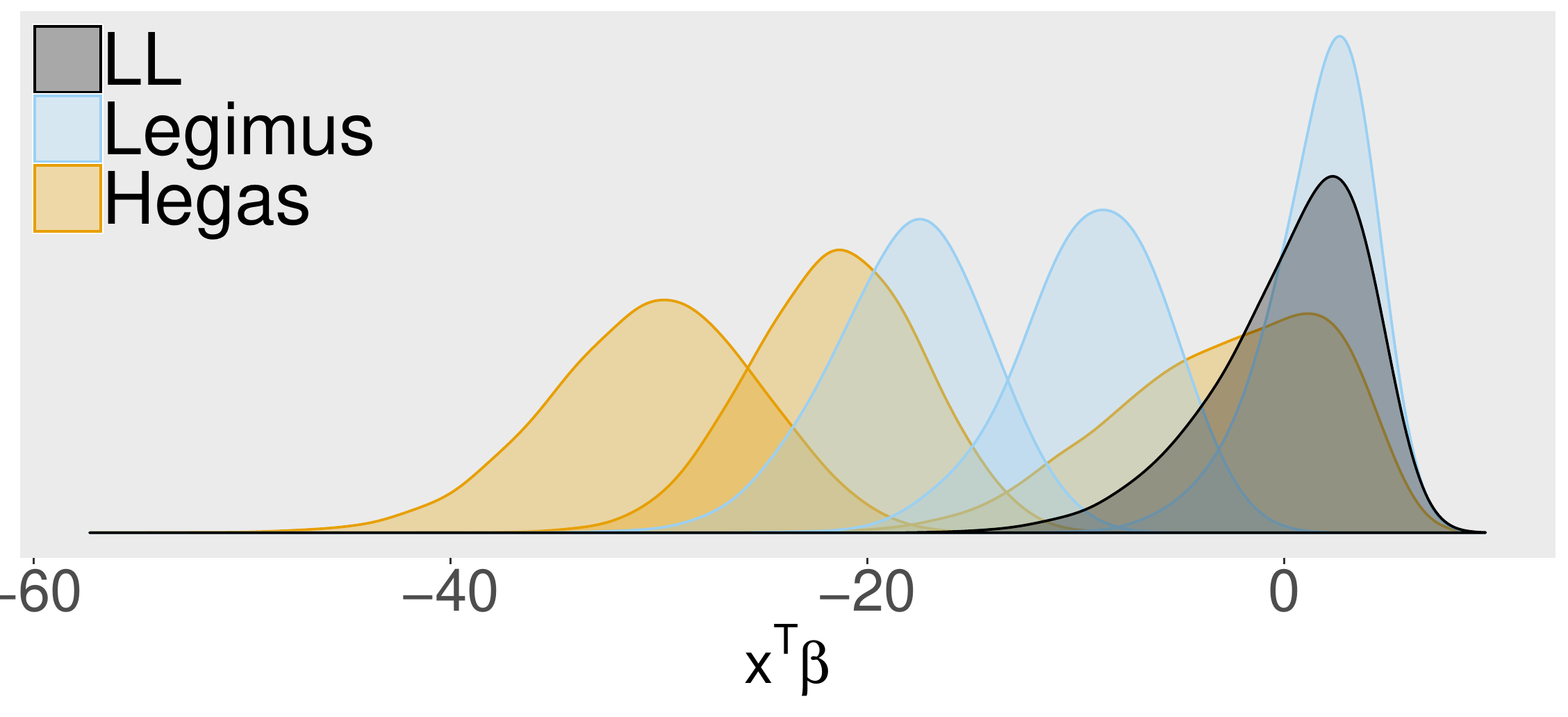}
  \end{center}
  \caption{The marginal posterior distributions for all $\gamma$-values from the Multi-Scale Probit model.}
  \label{fig:gamma_posteriors_combined}
\end{figure}

One of the strongest arguments for the Multi-Scale Probit model compared to single-scale Probit models is the ability to compare scales to each other. In Figure \ref{fig:gamma_posteriors_combined} we can see the marginal posteriors for $\boldsymbol{\gamma}$ for all three scales, as estimated by the Multi-Scale Probit, plotted together. 

We can see from the figure that the posterior modes of the highest threshold are similar for all three data sets. This is to be expected as the texts constituting the most complex category for each scale all come from the corpus made up by combining the Norstedts and Bonnier corpora (see Table \ref{tab:corpora}). This can be interpreted as creating a shared ceiling for the three scales. However, the thresholds are unevenly distributed on the parts of the scale estimated using different corpora. For instance, even though the Legimus and Hegas corpora each contain three categories (four when the Norstedts/Bonnier texts are added) the Legimus scale seems more fine grained on the interval $[-20,0]$ while the Hegas scale seems more fine grained on the interval $[-40,-20]$. This visualisation also illustrates how we can compute the probability distribution for, for example, the suitable Hegas-category of a text from the LL corpus.

\begin{figure}[h!]
  \begin{center}
    \includegraphics[width=\linewidth]{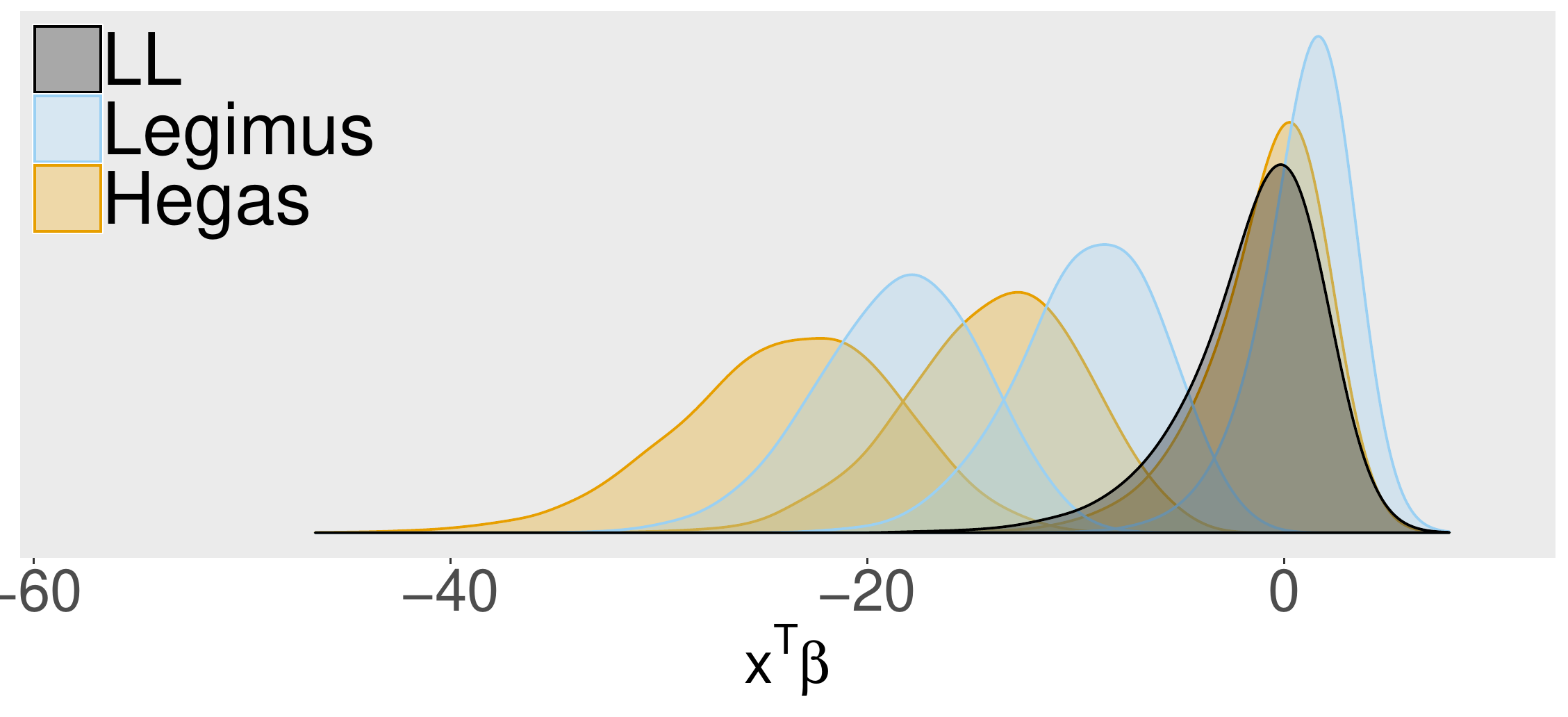}
  \end{center}
  \caption{The marginal posterior distributions for all $\gamma$-values from separate single-scale Probit and Ordered Probit models.}
  \label{fig:gamma_posteriors_separate}
\end{figure}

We can contrast this with each posterior estimated using separate-scale models, which we plot in Figure \ref{fig:gamma_posteriors_separate}. The posterior modes of the highest threshold for each scale no longer line up as well, i.e. there no longer seems to be a shared ceiling. The thresholds along the lower parts of the scale seem more evenly distributed but we can no longer see that the scales are more fine grained on different intervals and that the most complex Hegas category encompasses the two most complex Legimus categories, and that the least complex Legimus category is split among the two least complex Hegas categories.

\FloatBarrier

\section{Conclusion and Future Work}
\label{sec:concfuture}

We have shown that the Multi-Scale Probit can be fitted to data with a shared latent variable measured on different scales and that this new Probit outperforms the traditional binary Probit and the Ordered Probit in the majority of cases when data is sparse but multiple previously incompatible data sets are available. The model performs better than established Probit models with regards to both classification and ranking.

The multi-scale assumption of a single latent variable driving all corpora imposes a restriction which will have to be weighed against the advantage of pooling data. In situations where data is less scarce and the predictive accuracy on specific scales is important the Multi-Scale Probit might not measure up to a single-scale model. On the other hand, in the typical situation in practical work when data are scarce and many features are used, the advantages of data pooling are obvious. In applications such as ours, where we are explicitly modeling a generalisation of nominally equivalent scales, the slight averaging effect from pooling might even be viewed as an advantage. We also note that the assumption of a single latent readability factor makes the model highly interpretable, which is in itself a strong point for the proposed model.

All in all we find these results very promising. Below are some suggestions for issues for future research.


\subsection{The $p > n$ problem}
\label{sec:p>n}

We fixed the prior precision for $\boldsymbol{\beta}$, $\Lambda_0$, for reasons of simplicity, but it is straightforward to treat the shrinkage parameter as an unknown parameter with a Gamma prior. The full conditional posterior of the shrinkage parameter then follows an inverse Gamma distribution, which is easy to sample from in a separate Gibbs update step.

Another approach to the $p > n$-problem would be to use Bayesian variable selection in order to lower the number of covariates. \textcite{George1993VariableSampling} indicate how variable selection could be integrated into a Gibbs sampler for Bayesian linear regression. Since the update step for $\beta$ in the Multi-Scale Probit is a simple linear regression update, it is straightforward to implement Bayesian variable selection and to sample a binary variable selection indicator for each feature jointly with $\beta$ in the Gibbs sampler \parencite{Smith1996NonparametricSelection}.

\subsection{The generality/specificity trade-off}
\label{sec:tradeoff}

The version of the Multi-Scale Probit model presented here makes the assumption that the latent variable is exactly the same for each data set. However, it is not difficult to imagine ways to model scale specific deviations from a mostly shared latent variable. For instance, scale specific variable selection could be introduced into the model where each coefficient of the latent variable is split into a shared and a scale specific part. A prior would then be used to put as much of the effect as possible into the shared latent variable and only the small deviations into the scale specific parts. This can be combined with variable selection to learn if a single latent variable is needed for each corpus, see \textcite{Villani2012GeneralizedMixtures} for a similar approach in a different context.

\subsection{Linguistic application}
\label{sec:application}

Our application to text complexity in Section \ref{sec:ApplTextComplexity} can certainly be extended by linguists in a number of interesting ways, and it will be interesting to see the model applied to other corpora or other situations with classification problems using data sets with different ordinal scales.

\printbibliography

\FloatBarrier

\appendix

\section{Features}

This appendix contains short descriptions of the features used in Section \ref{sec:application} as well as plots for the marginal posteriors for all coefficients in $\boldsymbol{\beta}$.

\subsection*{Feature descriptions}

\tablecaption{\label{tab:features} The set of text based covariates.}
\tablefirsthead{

\bf{Feature} & \bf{Description} \\
\hline}
\tablehead{%
\multicolumn{2}{ c }%
{{\bfseries  Continued from previous page}} \\

\bf{Feature} & \bf{Description} \\
\hline}
\tabletail{%
\hline
\multicolumn{2}{ c }{{\bfseries Continued on next page}} \\
}
\tablelasttail{%
\hline
\multicolumn{2}{ c }{{\bfseries Concluded}} \\ }
\begin{center}

\begin{supertabular}{ l  p{10cm} }
ratioSweVocTotal & Total ratio of words from the SweVoc lexicon \\

ratioSweVocD & Ratio of words from the SweVoc D category (words for everyday use)\\

ratioSweVocH & Ratio of words from the SweVoc H category (other highly frequent words) \\

\multicolumn{2}{ l }{{\bfseries Part-of-Speech tag frequencies}} \\
\hline
pos\_RG & Cardinal number \\

pos\_HP & Interrogative/Relative Pronoun \\

pos\_RO & Ordinal number \\

pos\_MID &  \\

pos\_HD & Interrogative/Relative Determiner \\

pos\_KN & Conjunction \\

pos\_HA & Interrogative/Relative Adverb \\

pos\_PM & Proper Noun \\

pos\_PS & Possessive \\

lexicalDensity & Ratio of nouns, verbs, adjectives and adverbs to all words \\

\multicolumn{2}{ l }{{\bfseries Dependency type tag frequencies}} \\
\hline
dep\_VS & Infinitive subject complement \\

dep\_VO & Infinitive object complement \\

dep\_I. & Question mark \\

dep\_RA & Place adverbial \\

dep\_IF & Infinitive verb phrase minus infinitive marker \\

dep\_MA & Attitude adverbial \\

dep\_.F & Coordination at main clause level \\

dep\_XX & Unclassifiable grammatical function \\

dep\_IO & Indirect object \\

dep\_IQ & Colon \\

dep\_.A & Conjunctional adverbial \\

dep\_IU & Exclamation mark \\

dep\_AA & Other adverbial \\

dep\_AG & Agent \\

dep\_.. & Coordinating conjunction \\

dep\_CA & Contrastive adverbial \\

dep\_FS & Dummy subject \\

dep\_KA & Comparative adverbial \\

dep\_XF & Fundament phrase \\

dep\_FP & Free subjective predicative complement \\

dep\_OA & Object adverbial \\

dep\_TA & Time adverbial \\

dep\_HD & Head \\

dep\_DB & Doubled function \\

dep\_SP & Subjective predicative complement \\

dep\_OP & Object predicative \\

dep\_OO & Direct object \\

dep\_PL & Verb particle \\

\multicolumn{2}{ l }{{\bfseries Dependency structure features}} \\
\hline
ratioRightDeps & The ratio of dependency relations where the head word occurs after the dependent \\

verbArity0 & The frequency of verbs with no dependents \\

verbArity1 & The frequency of verbs with 1 dependent \\

verbArity2 & \hspace{53pt} \textquotedbl \hspace{53pt} 2 dependents \\

verbArity3 & \hspace{53pt} \textquotedbl \hspace{53pt} 3 dependents  \\

verbArity5 & \hspace{53pt} \textquotedbl \hspace{53pt} 5 dependents  \\

verbArity6 & \hspace{53pt} \textquotedbl \hspace{53pt} 6 dependents  \\

\end{supertabular}%
\end{center}

\subsection*{Marginal posteriors for $\boldsymbol{\beta}$}

\begin{figure}[h!]
  \begin{center}
      \begin{minipage}[b]{0.32\linewidth}
    \centering
    \includegraphics[width=\linewidth]{images/beta_posterior_0_dep_VO.pdf}
    \caption*{dep\_VO} 
  \end{minipage}
  \hfill
  \begin{minipage}[b]{0.32\linewidth}
    \centering
    \includegraphics[width=\linewidth]{images/beta_posterior_1_verbArity5.pdf}
    \caption*{verbArity5} 
  \end{minipage}
  \hfill
  \begin{minipage}[b]{0.32\linewidth}
    \centering
    \includegraphics[width=\linewidth]{images/beta_posterior_1_dep_MA.pdf}
    \caption*{dep\_MA} 
  \end{minipage}
  \hfill
  \begin{minipage}[b]{0.32\linewidth}
    \centering
    \includegraphics[width=\linewidth]{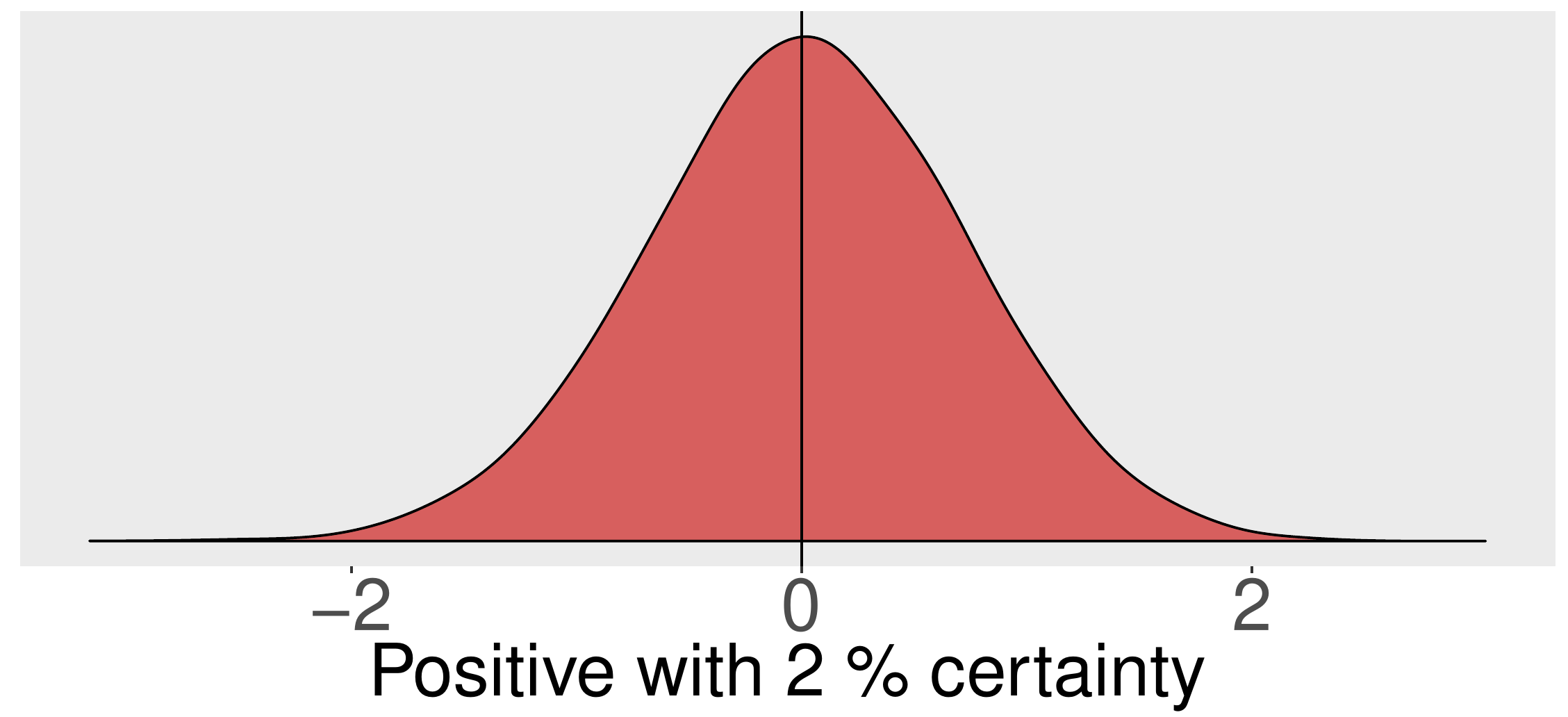}
    \caption*{pos\_RG} 
  \end{minipage}
  \hfill
  \begin{minipage}[b]{0.32\linewidth}
    \centering
    \includegraphics[width=\linewidth]{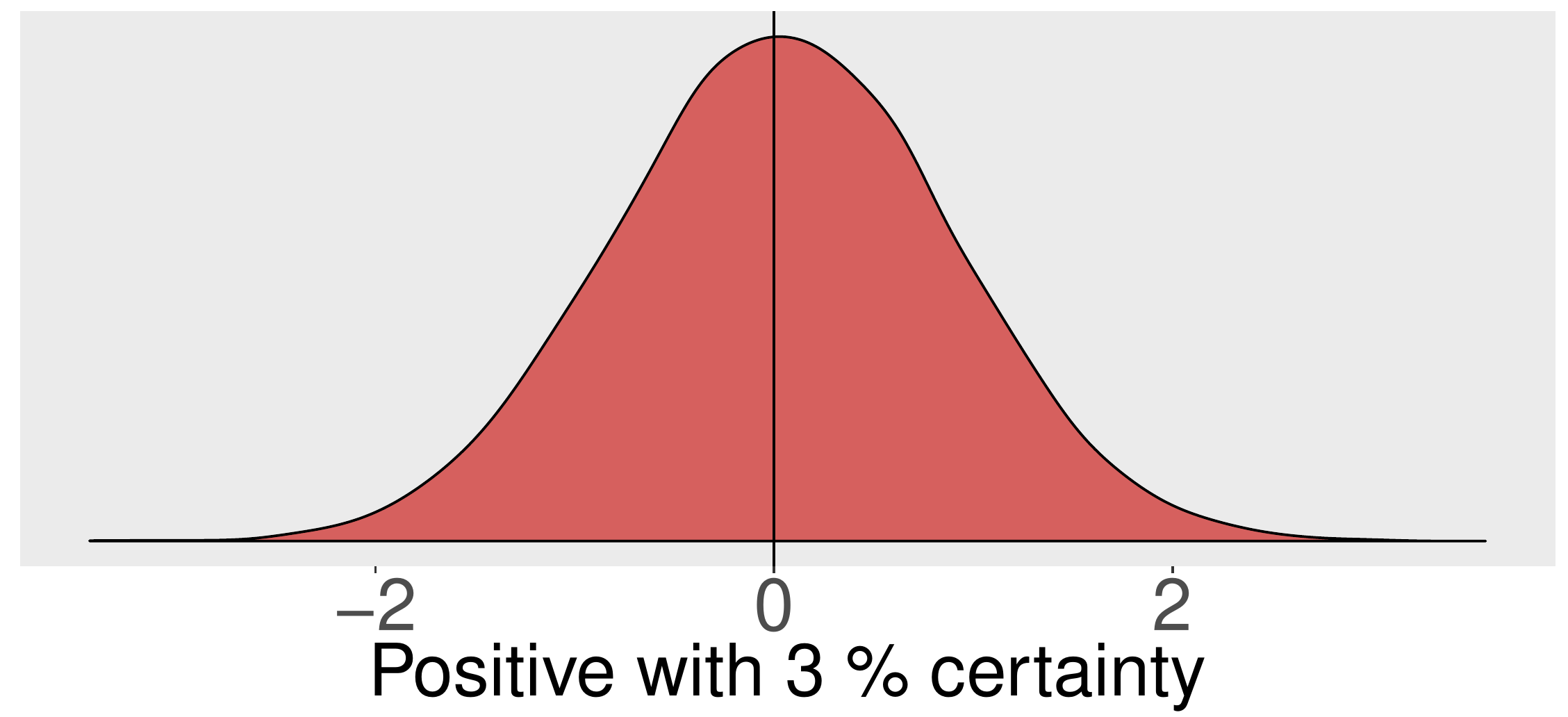}
    \caption*{verbArity3} 
  \end{minipage}
  \hfill
  \begin{minipage}[b]{0.32\linewidth}
    \centering
    \includegraphics[width=\linewidth]{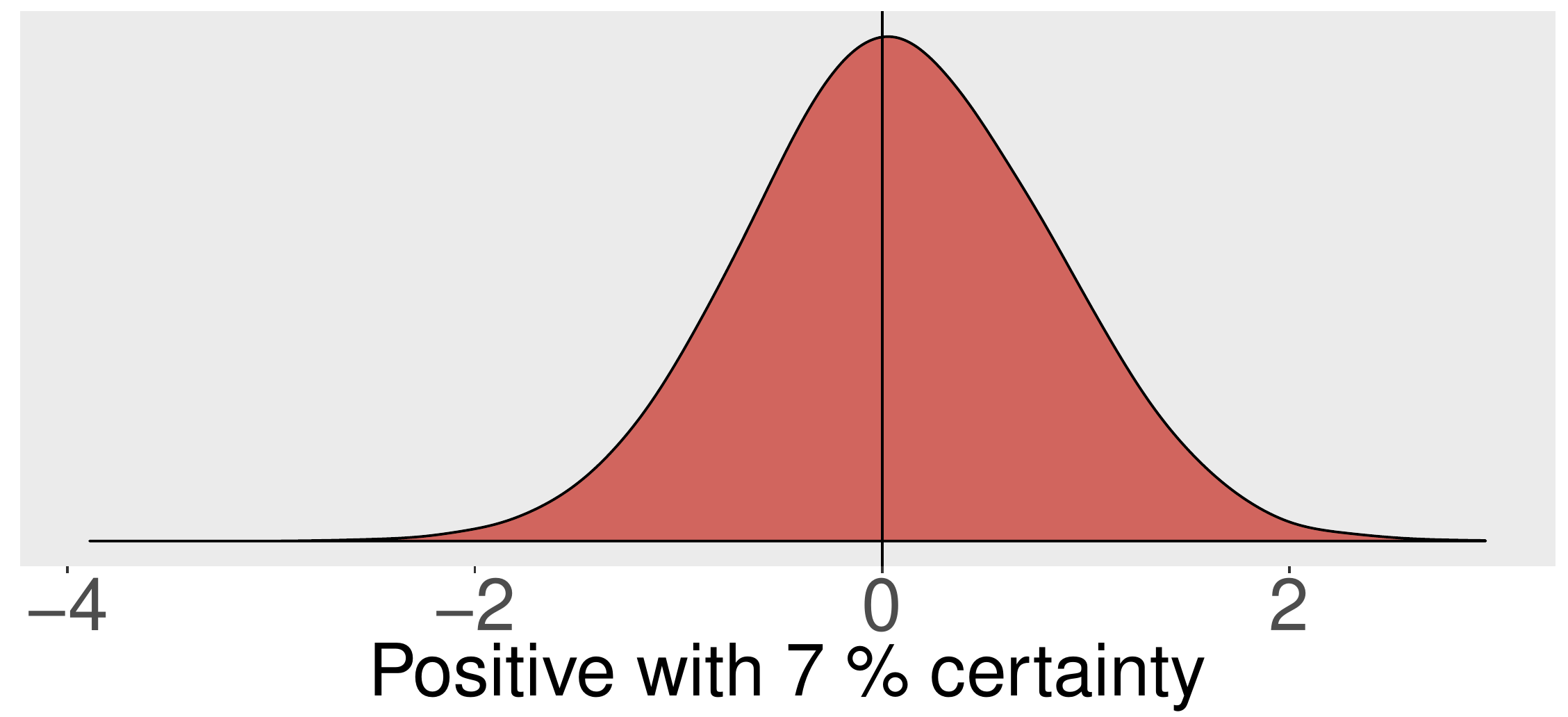}
    \caption*{dep\_OO} 
  \end{minipage}
  \hfill
  \begin{minipage}[b]{0.32\linewidth}
    \centering
    \includegraphics[width=\linewidth]{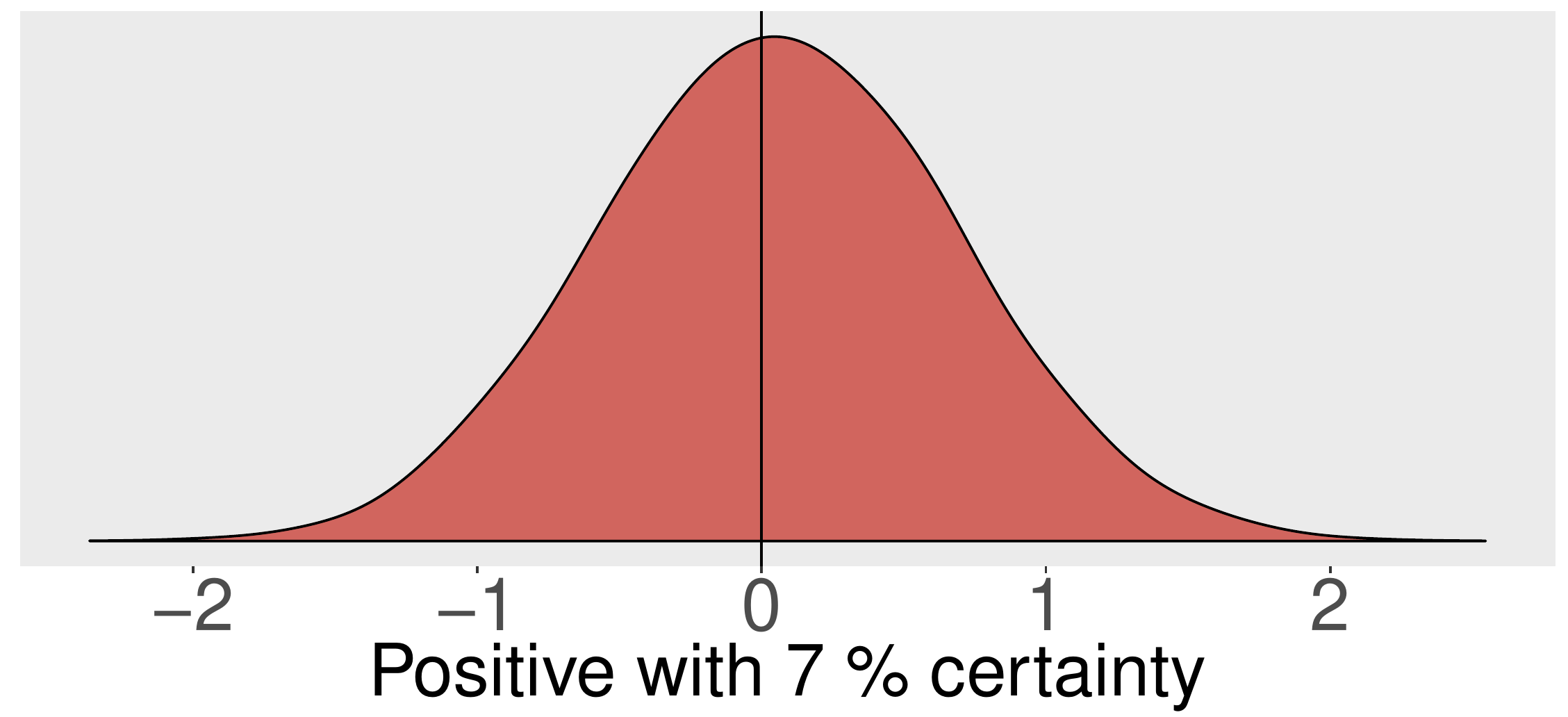}
    \caption*{dep\_KA} 
  \end{minipage}
  \hfill
  \begin{minipage}[b]{0.32\linewidth}
    \centering
    \includegraphics[width=\linewidth]{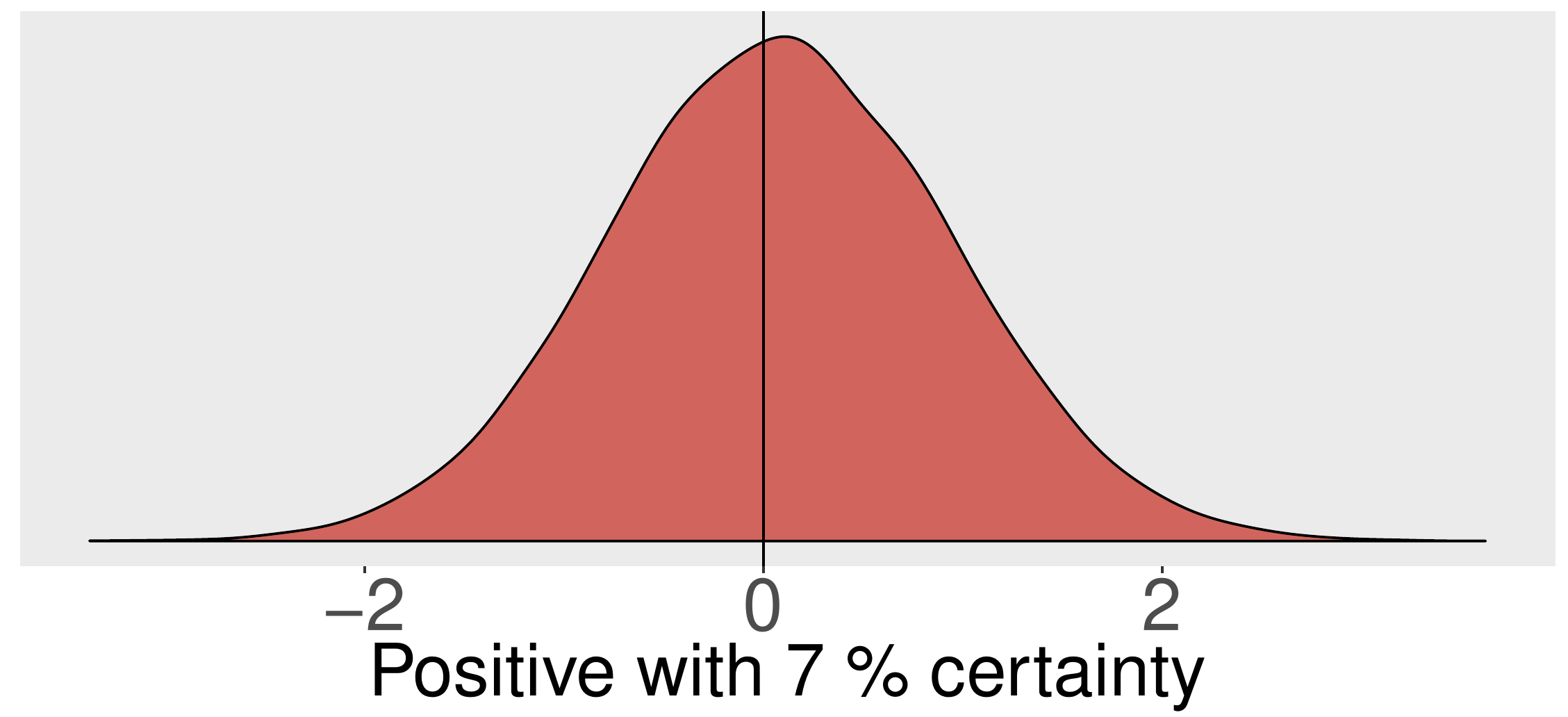}
    \caption*{dep\_PL} 
  \end{minipage}
  \hfill
  \begin{minipage}[b]{0.32\linewidth}
    \centering
    \includegraphics[width=\linewidth]{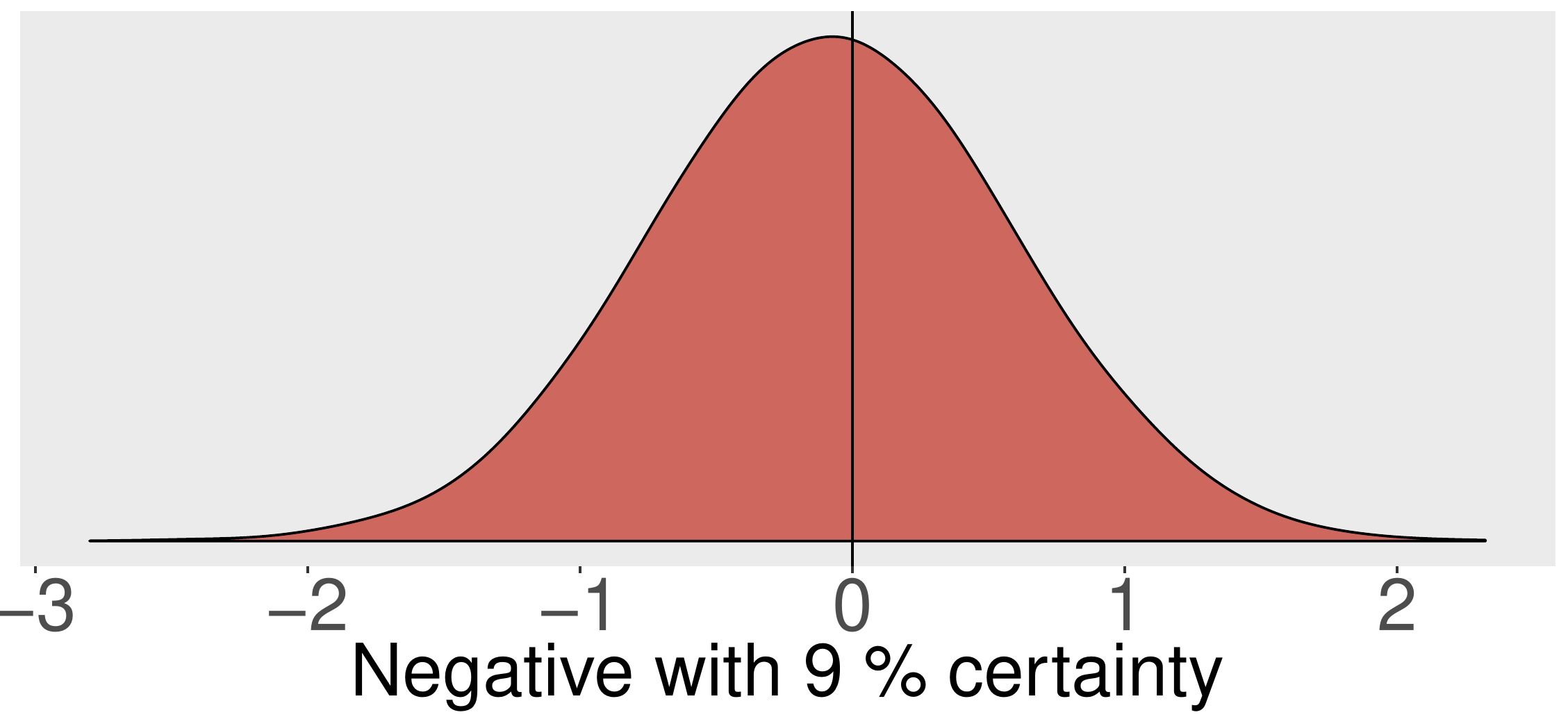}
    \caption*{verbArity0} 
  \end{minipage}
  \hfill
  \begin{minipage}[b]{0.32\linewidth}
    \centering
    \includegraphics[width=\linewidth]{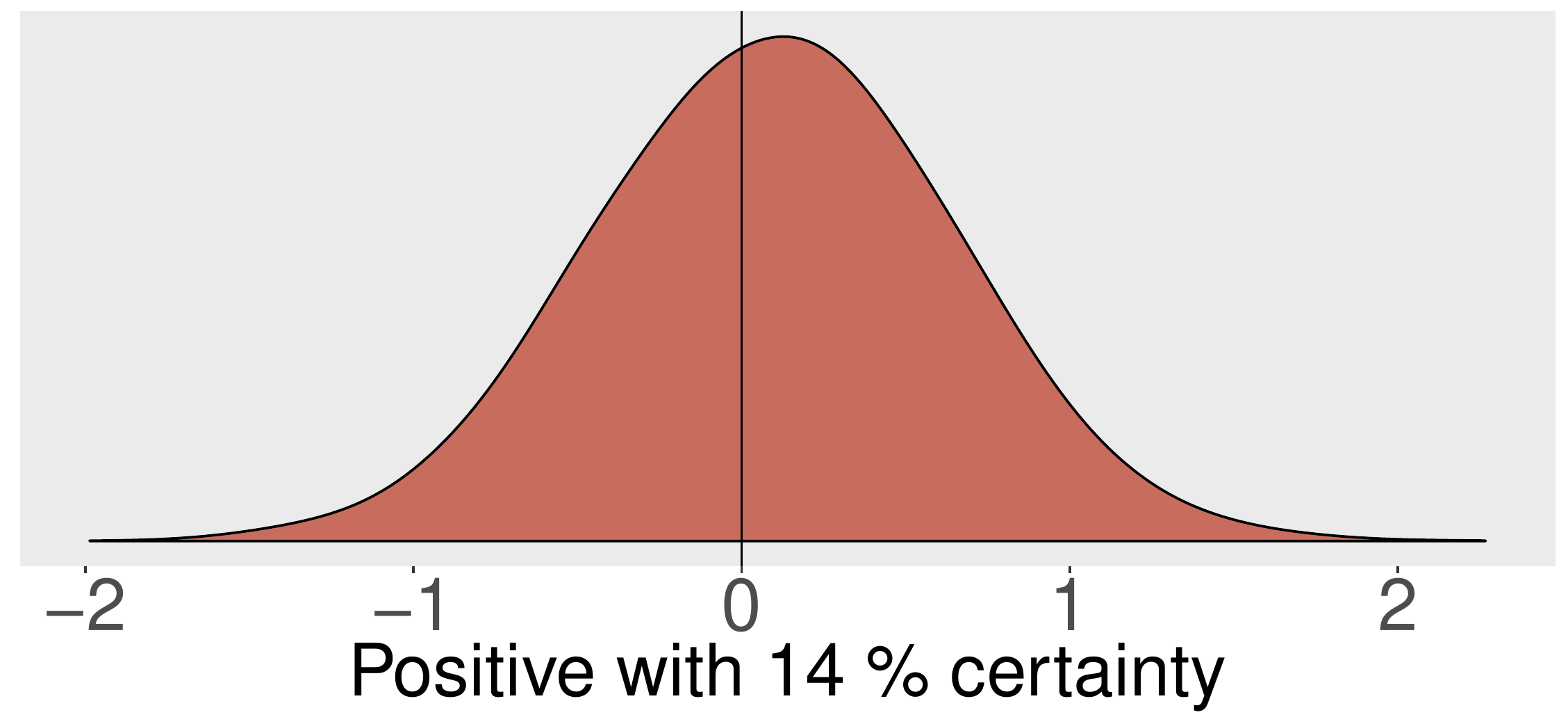}
    \caption*{dep\_FP} 
  \end{minipage}
  \hfill
  \begin{minipage}[b]{0.32\linewidth}
    \centering
    \includegraphics[width=\linewidth]{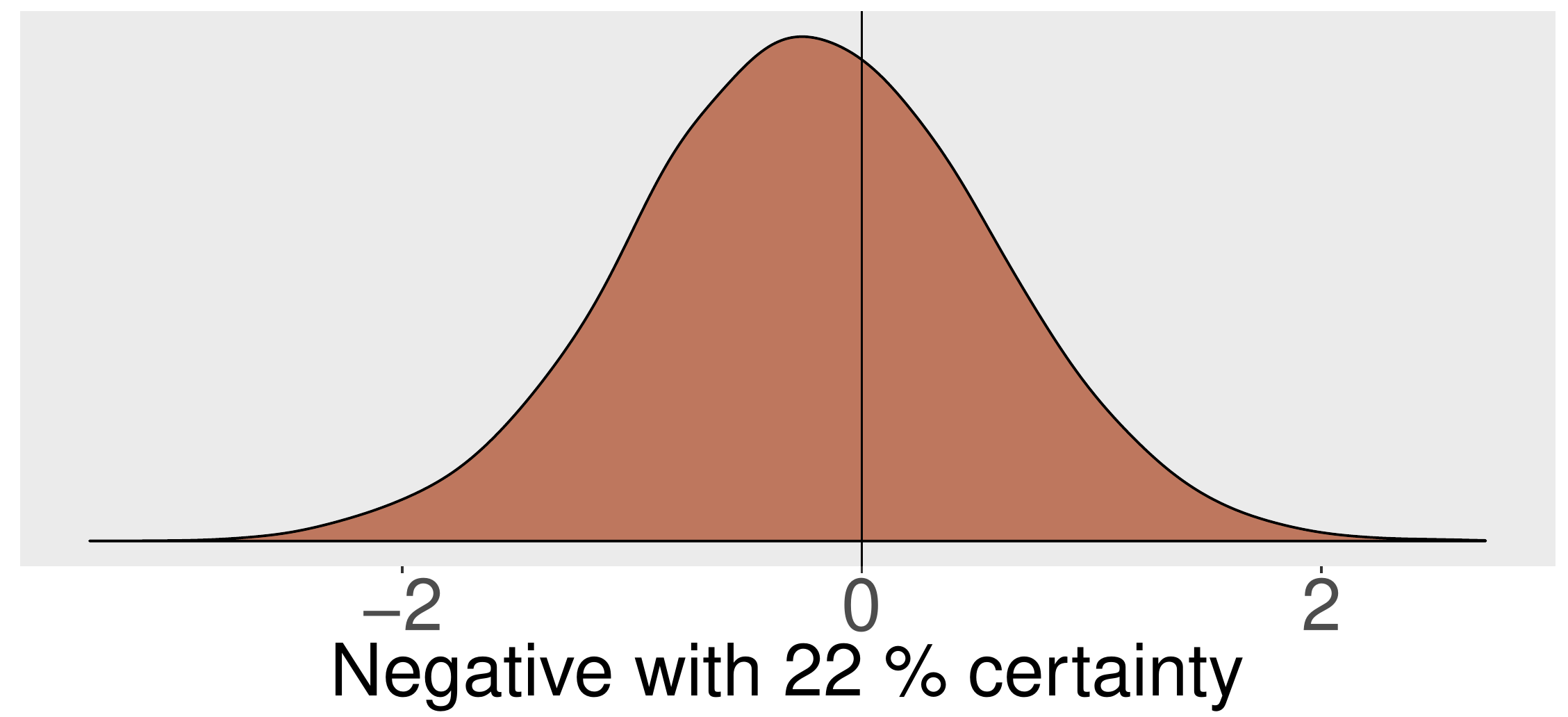}
    \caption*{dep\_DB} 
  \end{minipage}
  \hfill
  \begin{minipage}[b]{0.32\linewidth}
    \centering
    \includegraphics[width=\linewidth]{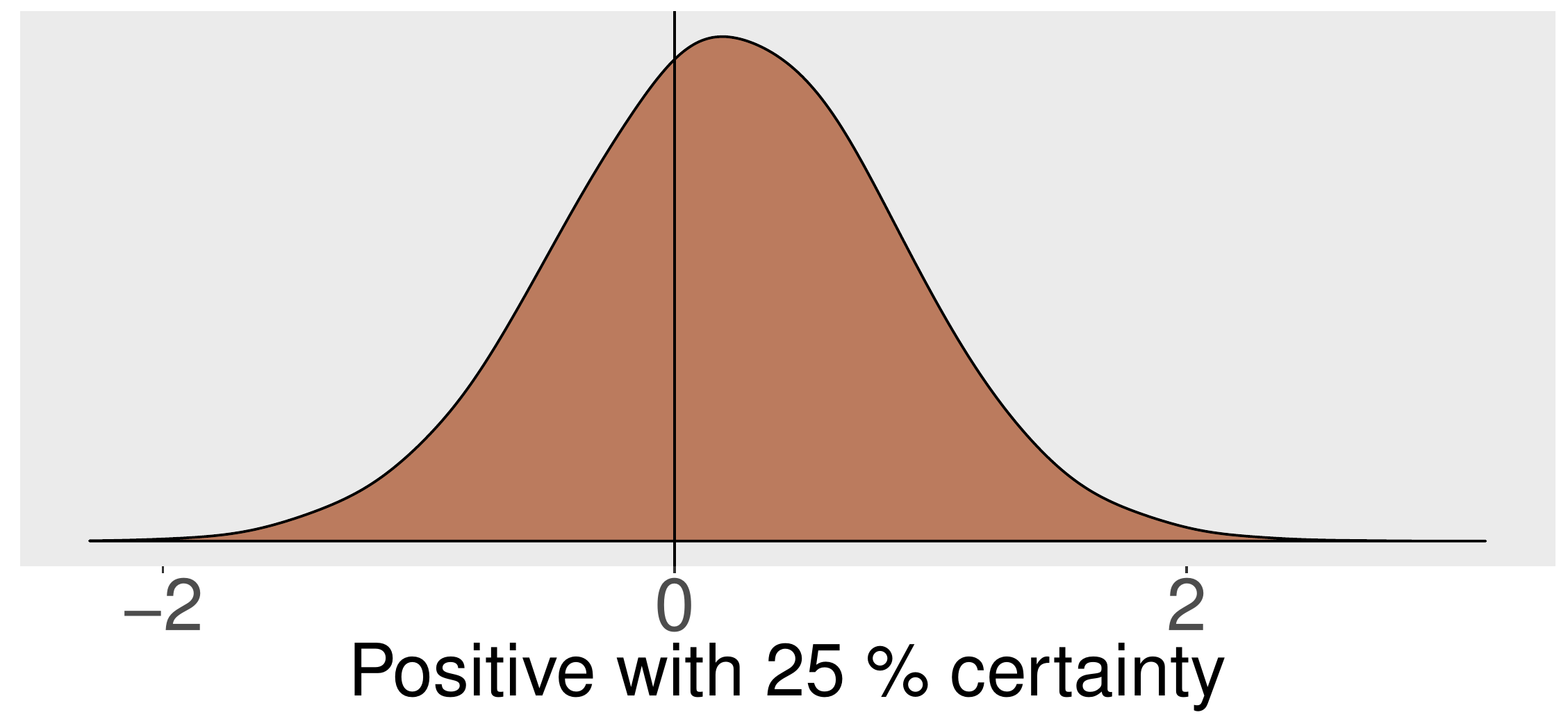}
    \caption*{pos\_KN} 
  \end{minipage}
  \end{center}
\end{figure}

\begin{figure}
\ContinuedFloat
  \begin{center}
  \begin{minipage}[b]{0.32\linewidth}
    \centering
    \includegraphics[width=\linewidth]{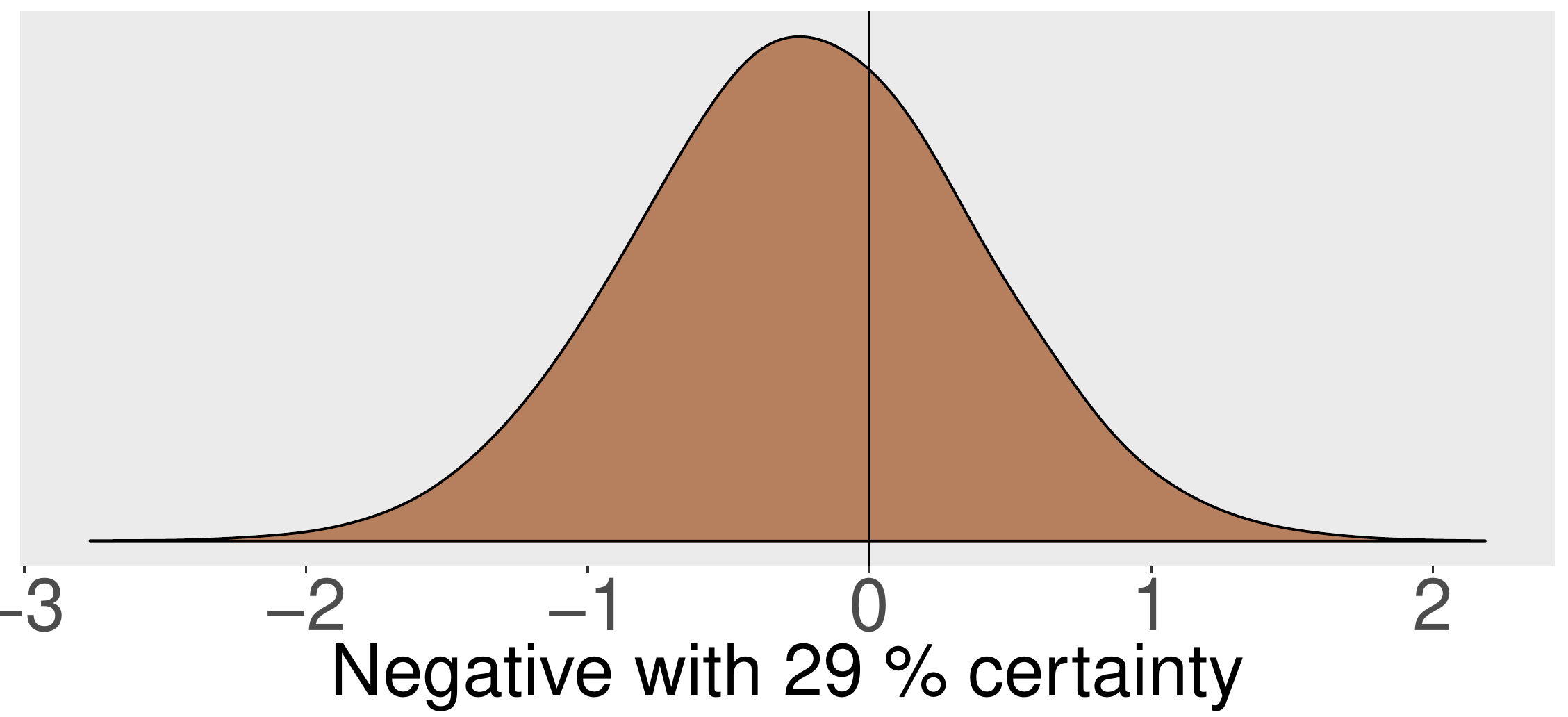}
    \caption*{pos\_HA} 
  \end{minipage}
  \hfill
  \begin{minipage}[b]{0.32\linewidth}
    \centering
    \includegraphics[width=\linewidth]{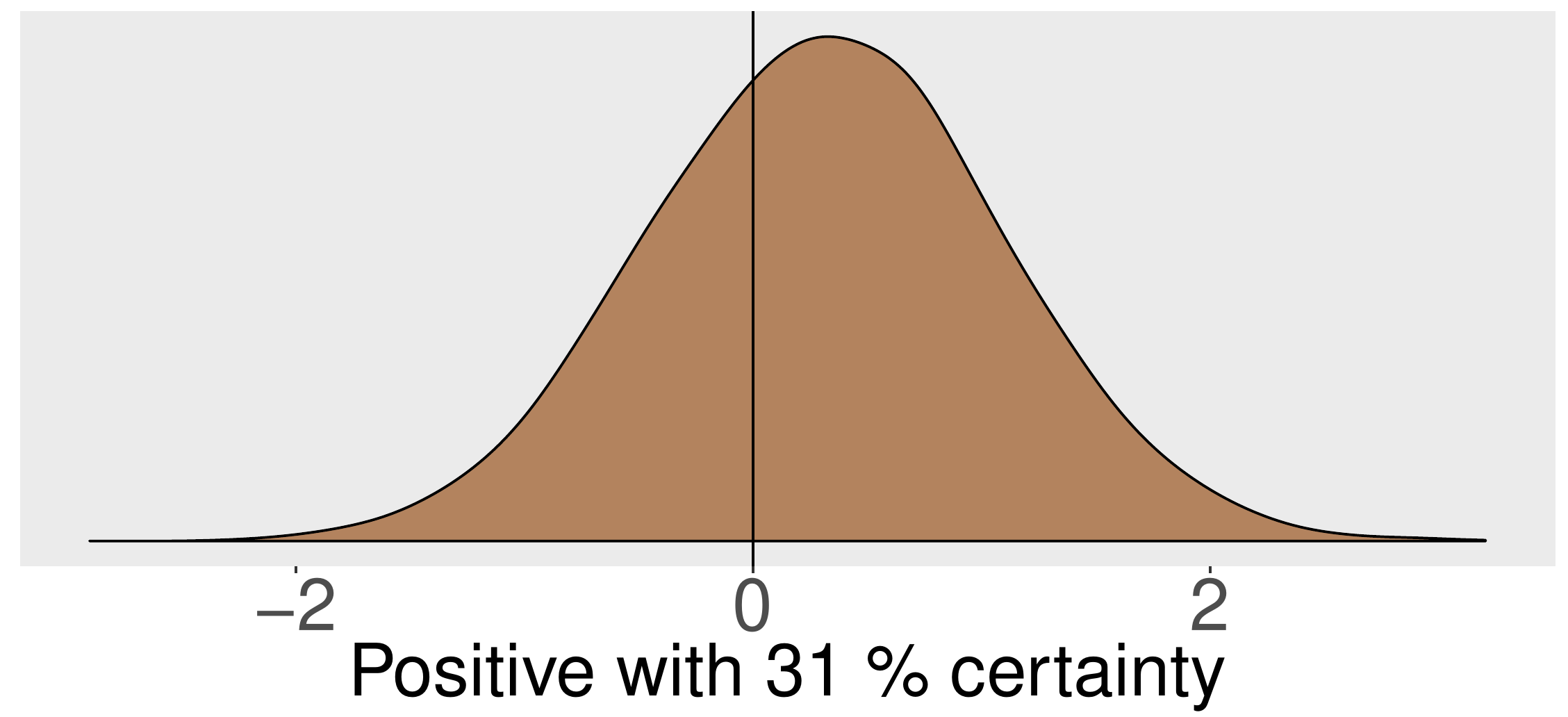}
    \caption*{verbArity2} 
  \end{minipage}
  \hfill
  \begin{minipage}[b]{0.32\linewidth}
    \centering
    \includegraphics[width=\linewidth]{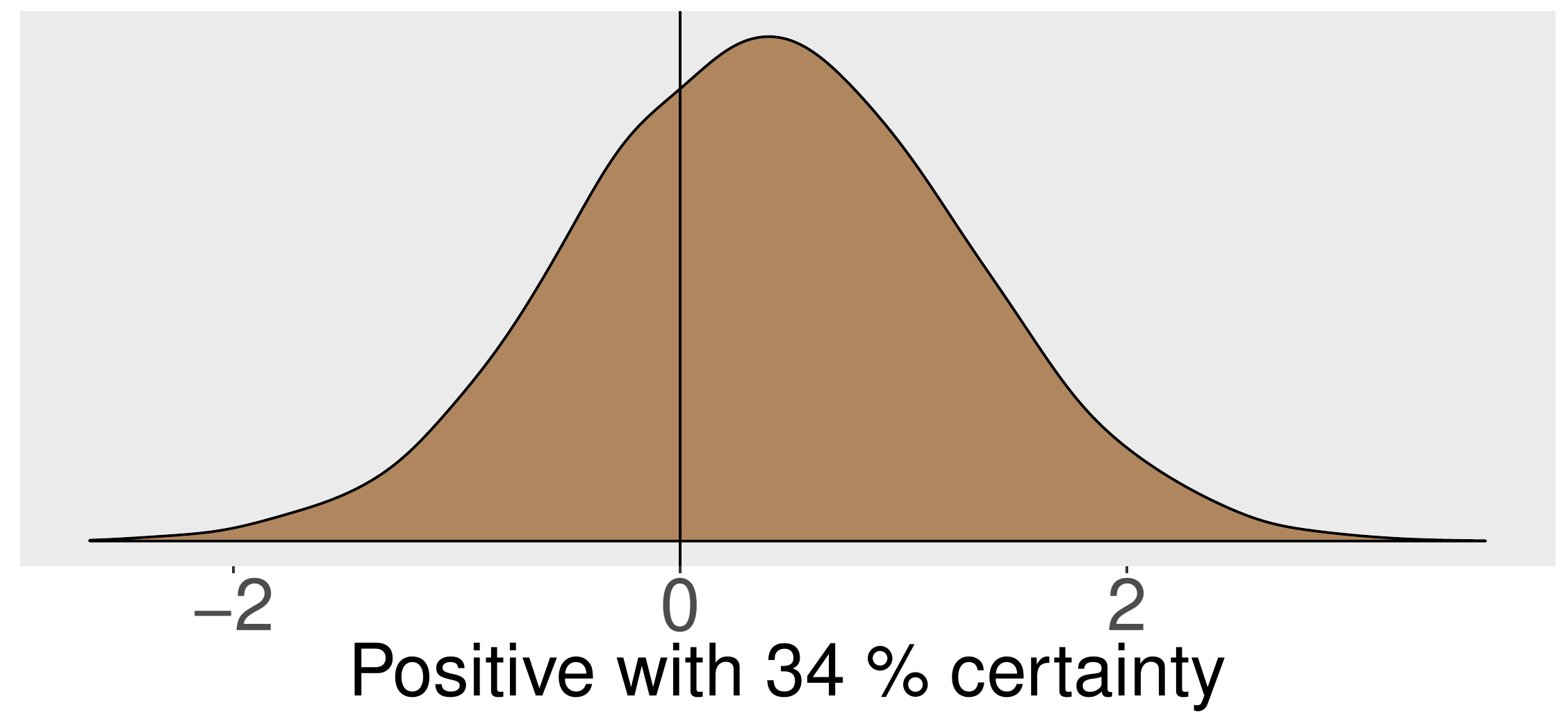}
    \caption*{pos\_MID} 
  \end{minipage}
  \hfill
  \begin{minipage}[b]{0.32\linewidth}
    \centering
    \includegraphics[width=\linewidth]{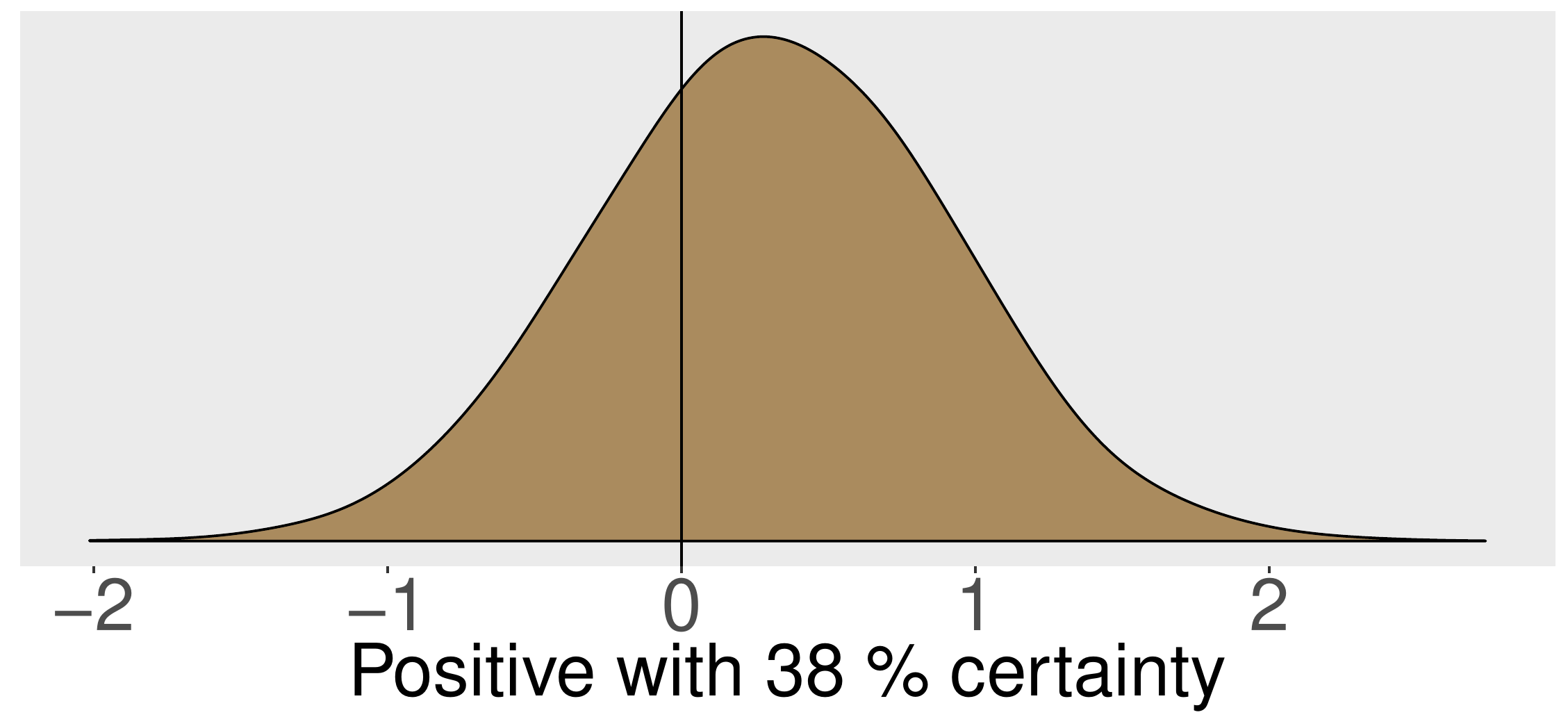}
    \caption*{pos\_PS} 
  \end{minipage}
  \hfill
  \begin{minipage}[b]{0.32\linewidth}
    \centering
    \includegraphics[width=\linewidth]{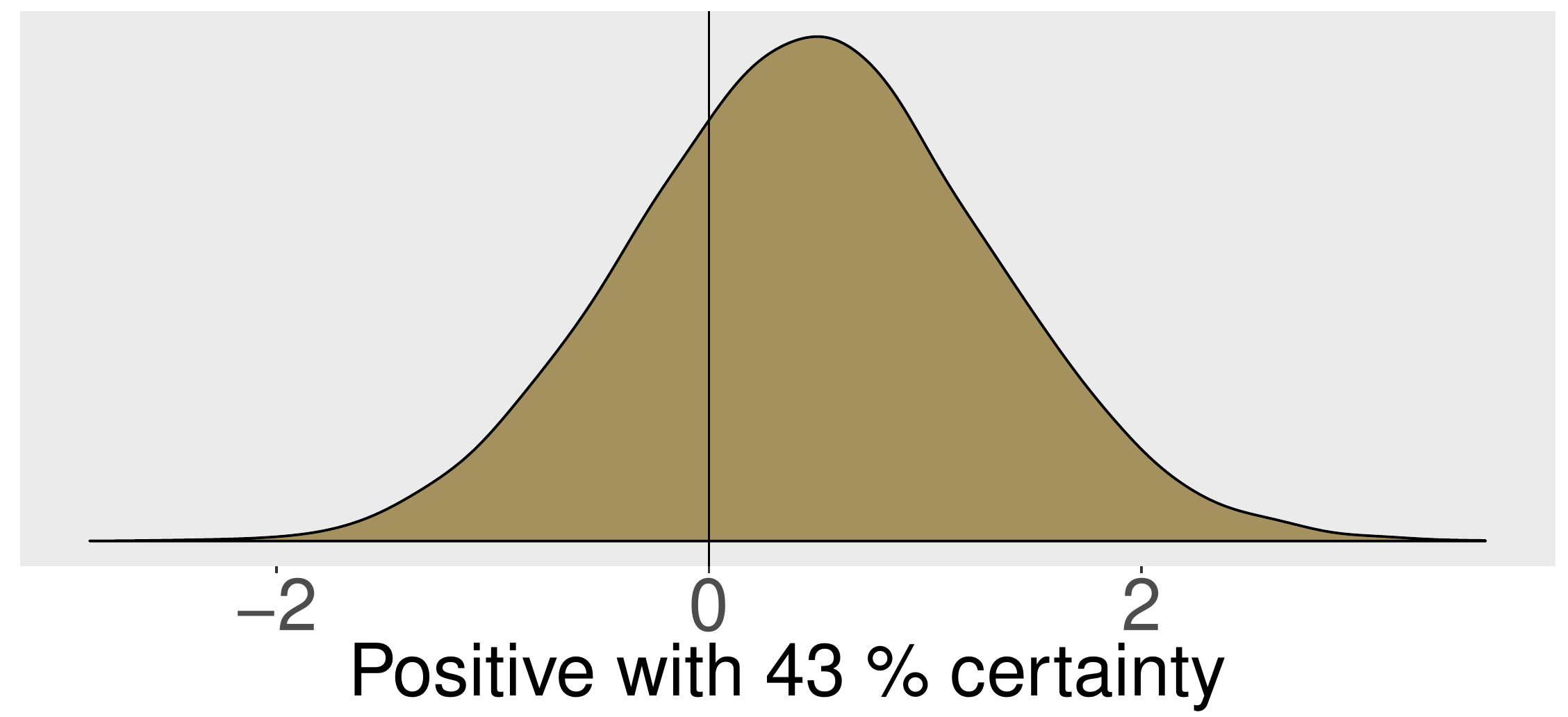}
    \caption*{dep\_FS} 
  \end{minipage}
  \hfill
  \begin{minipage}[b]{0.32\linewidth}
    \centering
    \includegraphics[width=\linewidth]{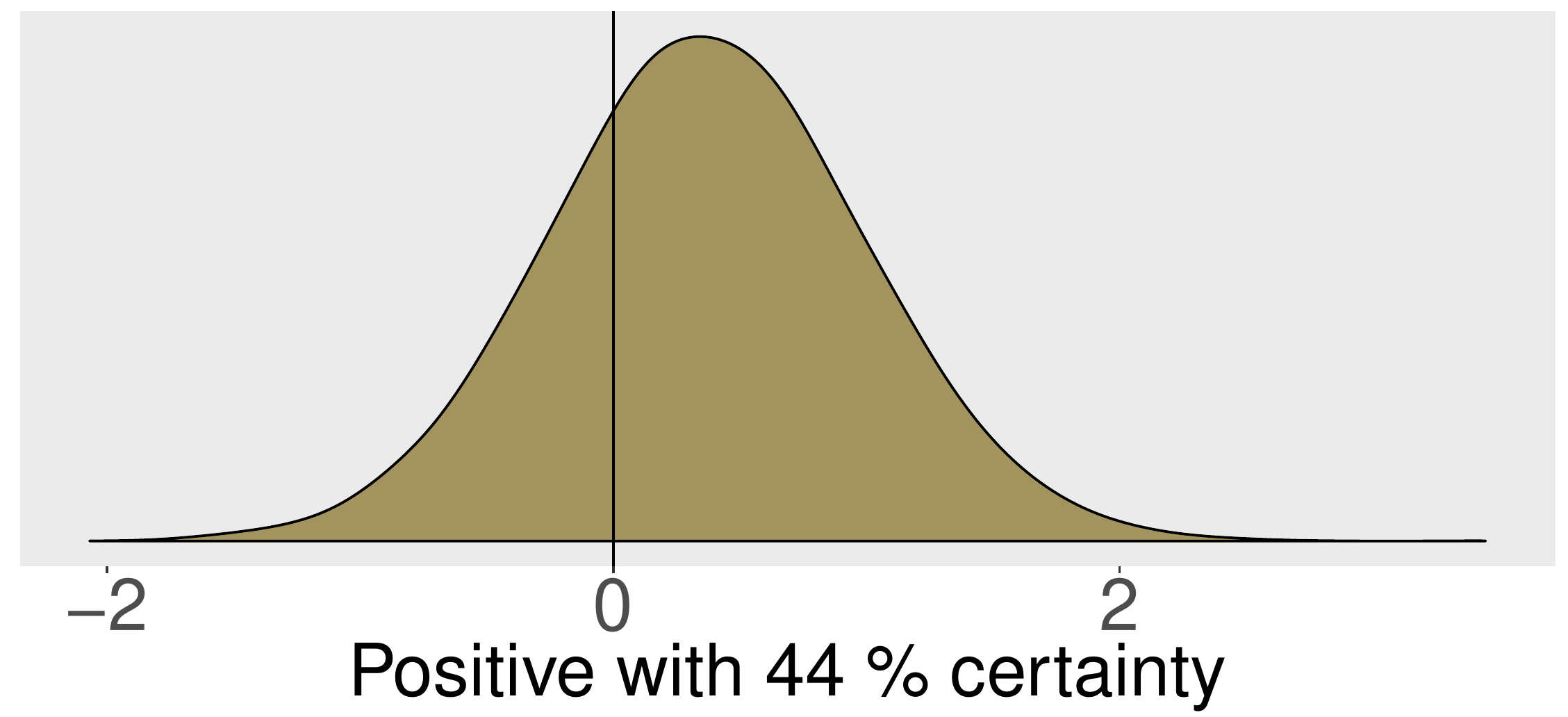}
    \caption*{pos\_RO} 
  \end{minipage}
  \hfill
  \begin{minipage}[b]{0.32\linewidth}
    \centering
    \includegraphics[width=\linewidth]{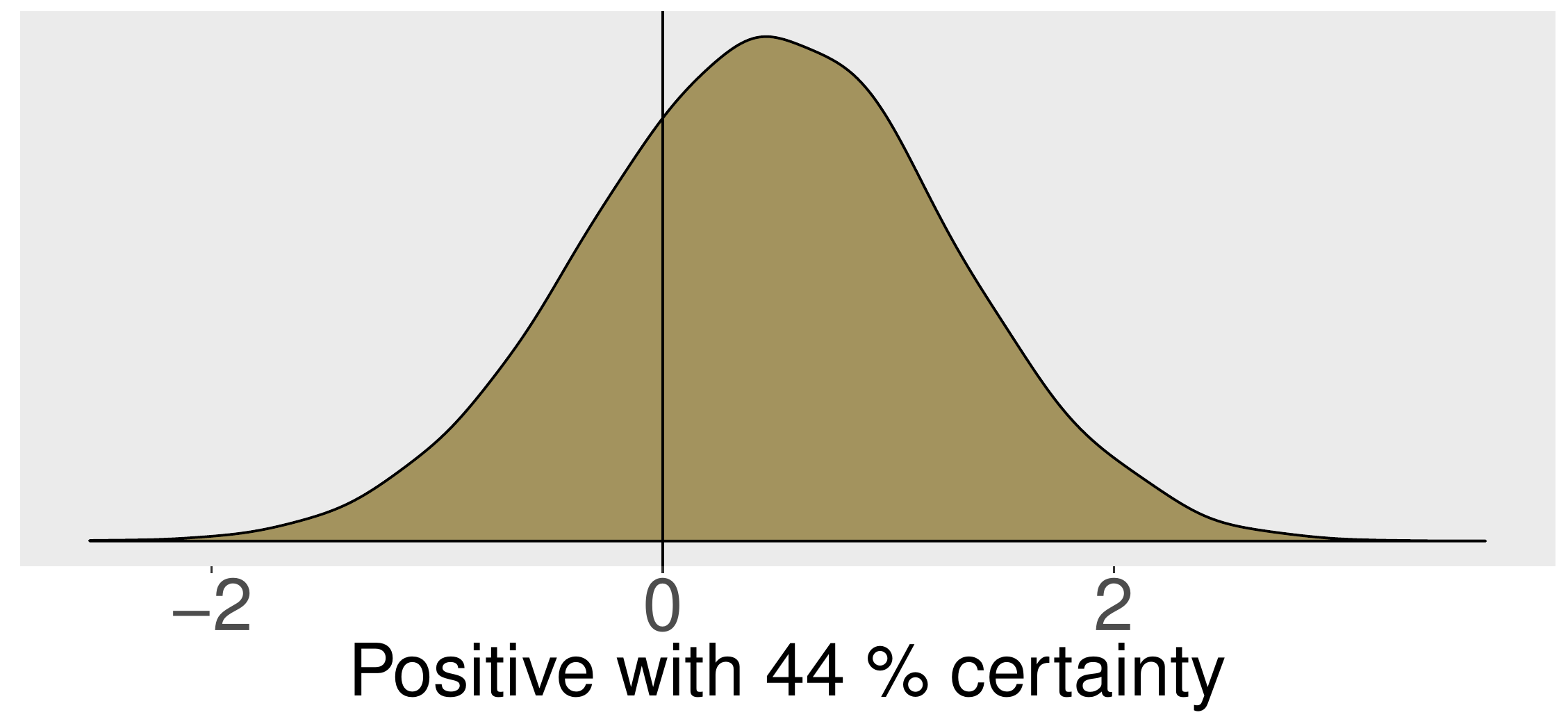}
    \caption*{dep\_AG} 
  \end{minipage}
  \hfill
  \begin{minipage}[b]{0.32\linewidth}
    \centering
    \includegraphics[width=\linewidth]{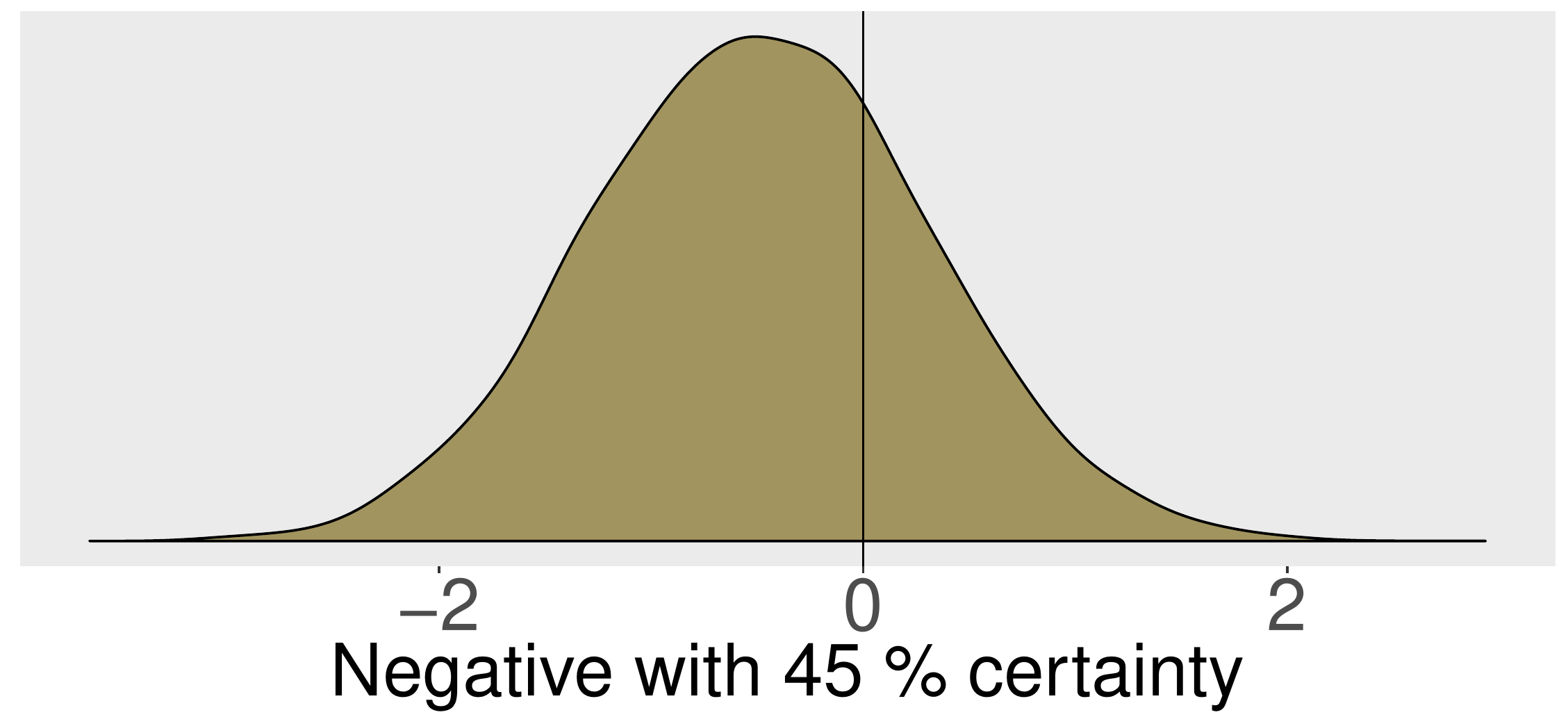}
    \caption*{ratioSweVocD} 
  \end{minipage}
  \hfill
  \begin{minipage}[b]{0.32\linewidth}
    \centering
    \includegraphics[width=\linewidth]{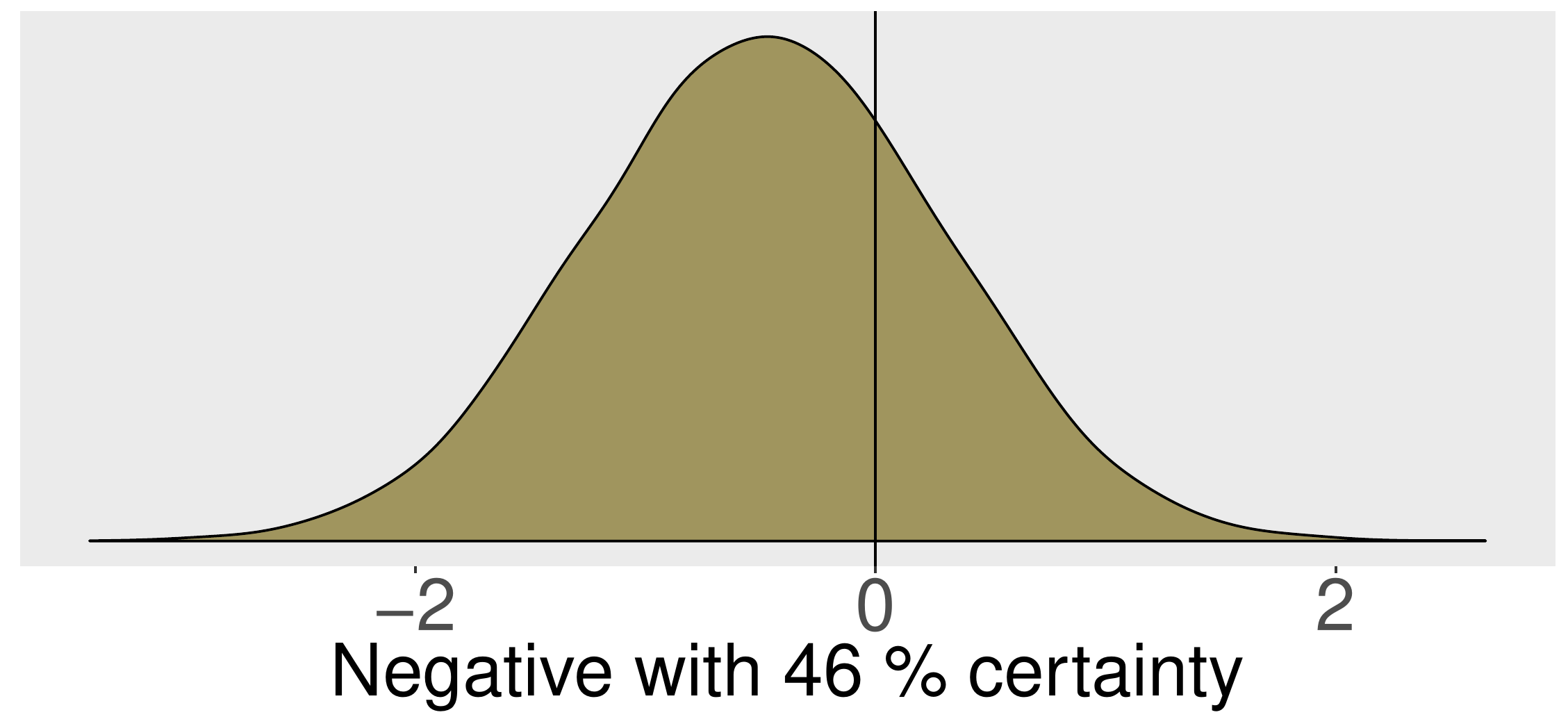}
    \caption*{pos\_PM} 
  \end{minipage}
  \hfill
  \begin{minipage}[b]{0.32\linewidth}
    \centering
    \includegraphics[width=\linewidth]{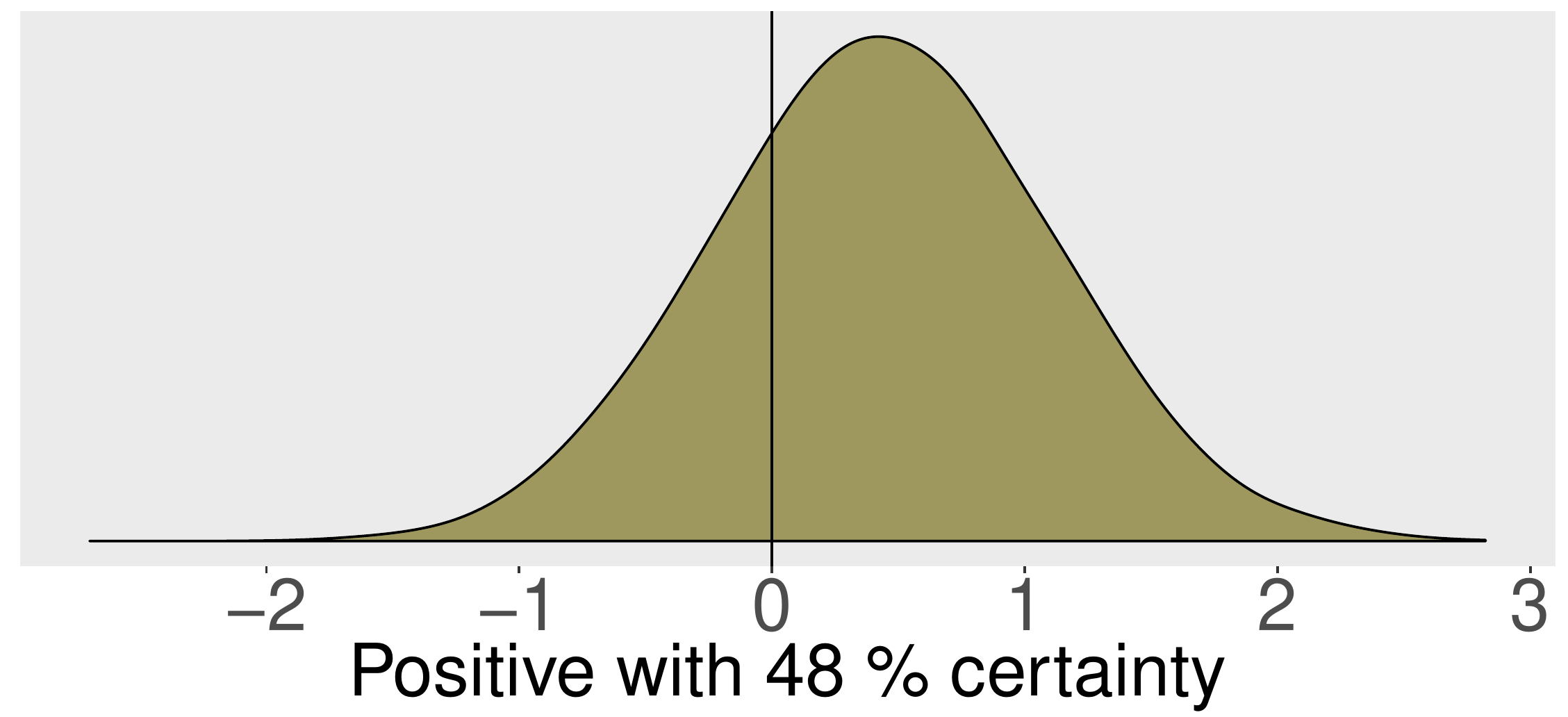}
    \caption*{dep\_OA} 
  \end{minipage}
  \hfill
  \begin{minipage}[b]{0.32\linewidth}
    \centering
    \includegraphics[width=\linewidth]{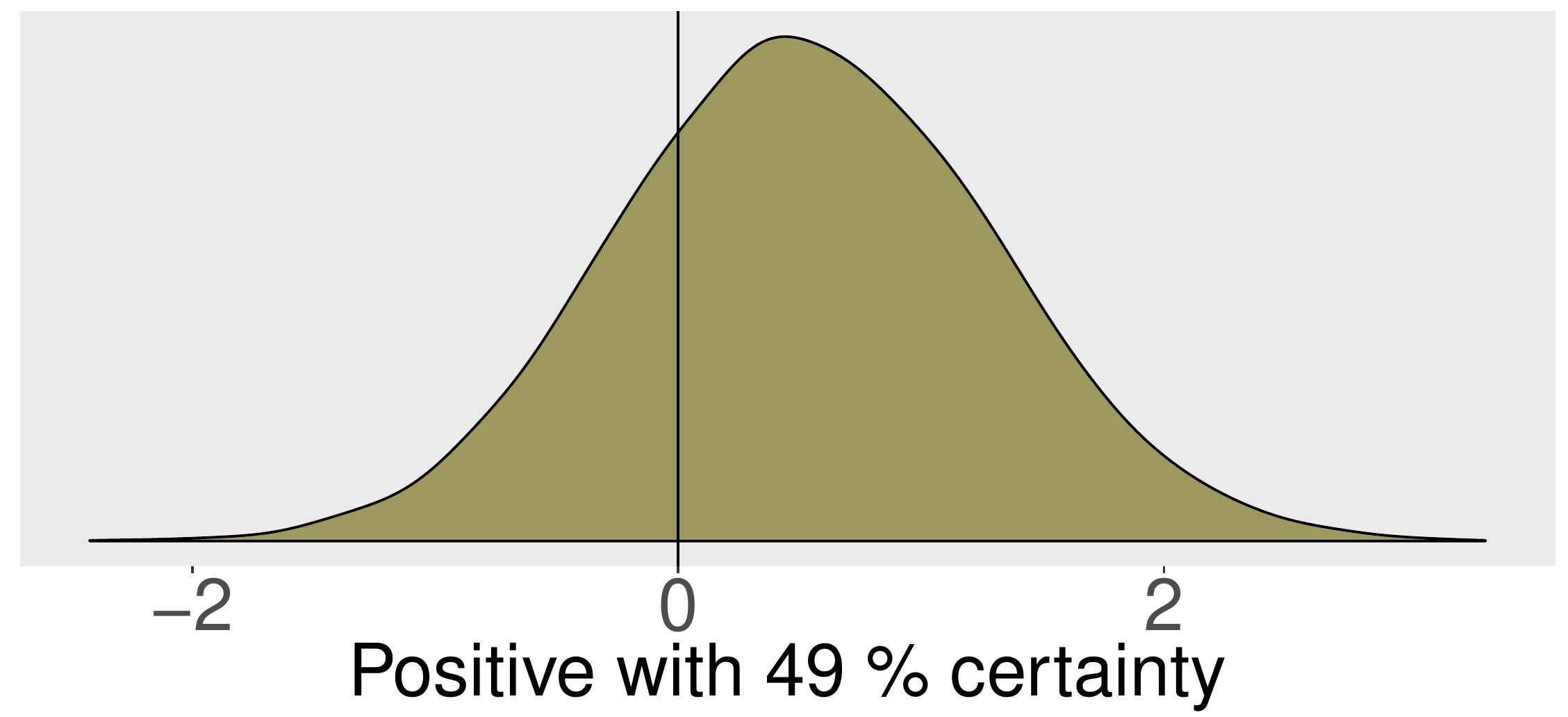}
    \caption*{pos\_HD} 
  \end{minipage}
  \hfill
  \begin{minipage}[b]{0.32\linewidth}
    \centering
    \includegraphics[width=\linewidth]{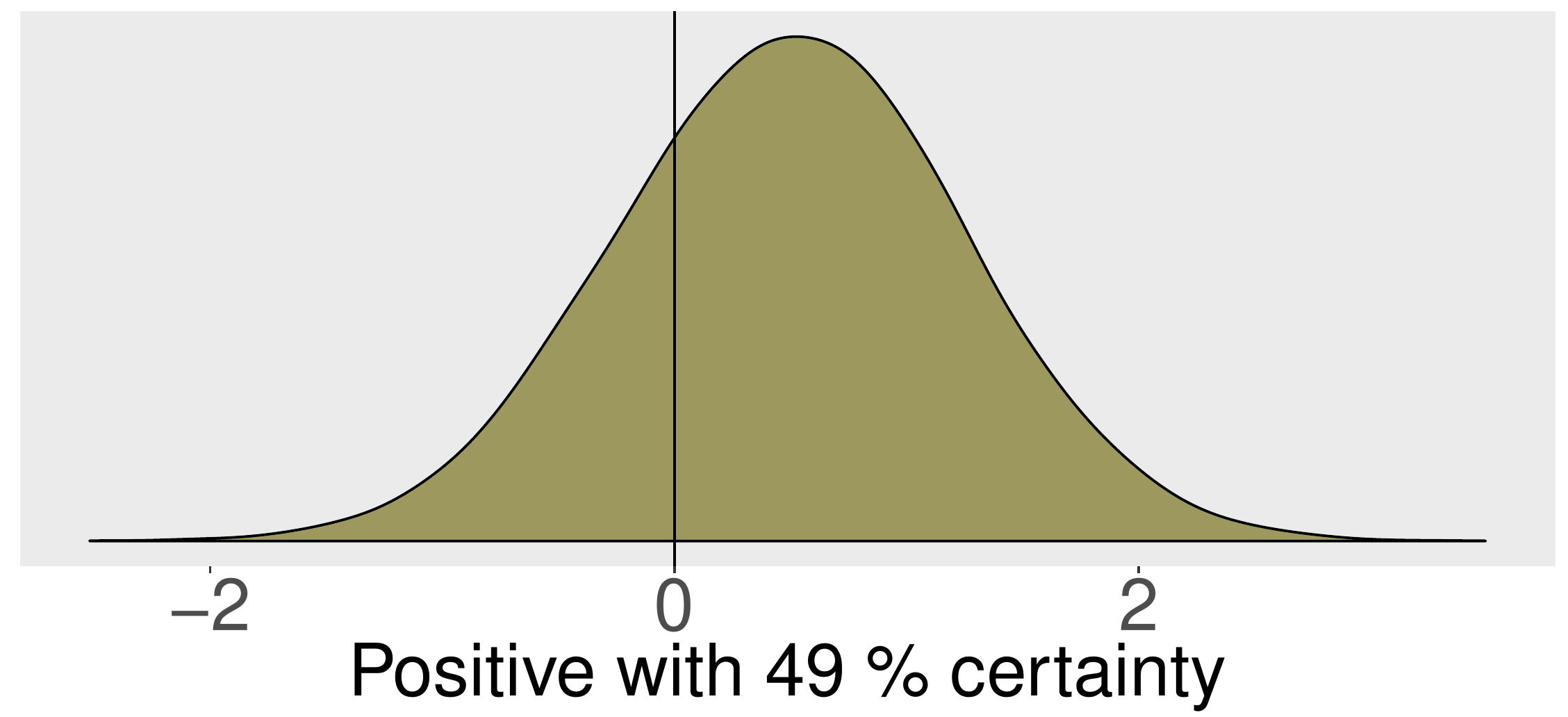}
    \caption*{ratioSweVocH} 
  \end{minipage}
  \hfill
  \begin{minipage}[b]{0.32\linewidth}
    \centering
    \includegraphics[width=\linewidth]{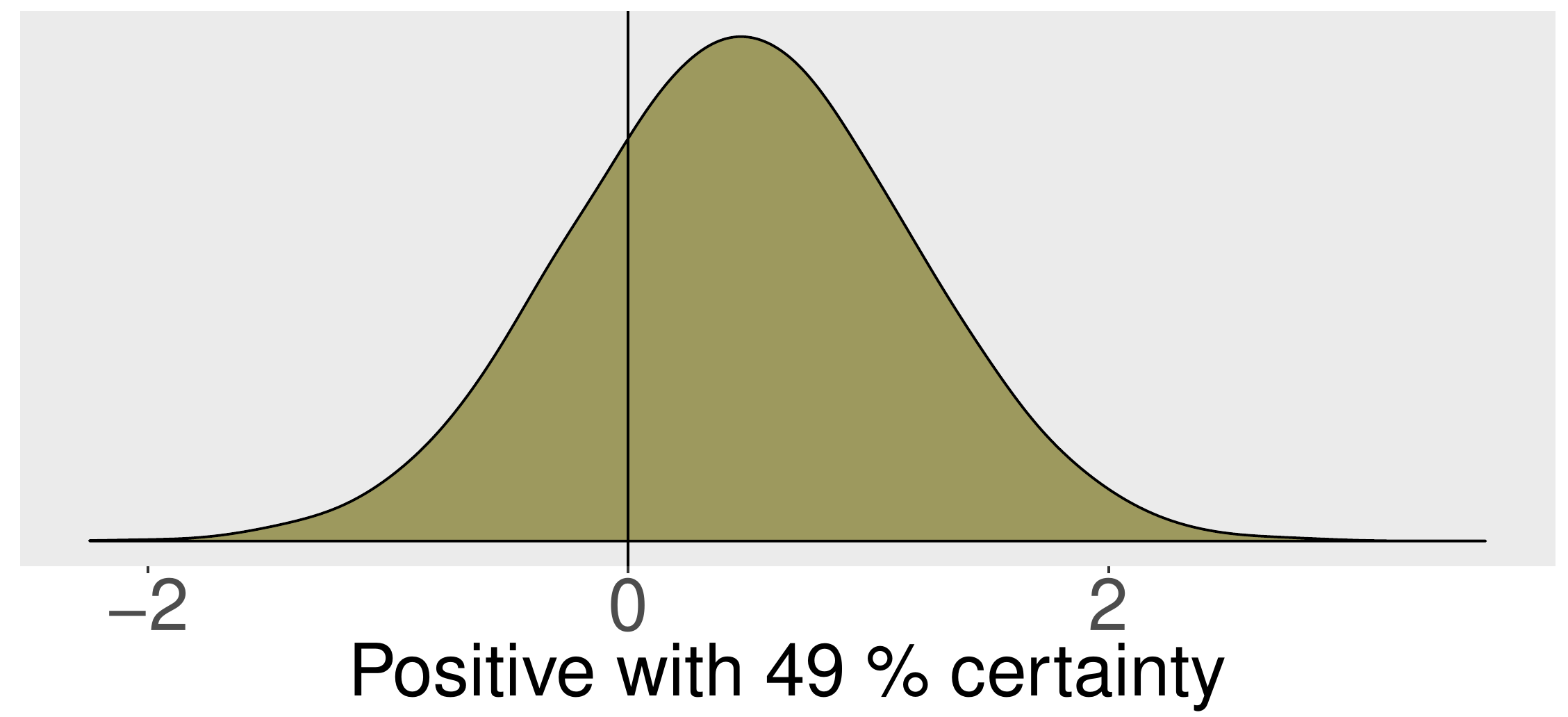}
    \caption*{dep\_.A} 
  \end{minipage}
  \hfill
  \begin{minipage}[b]{0.32\linewidth}
    \centering
    \includegraphics[width=\linewidth]{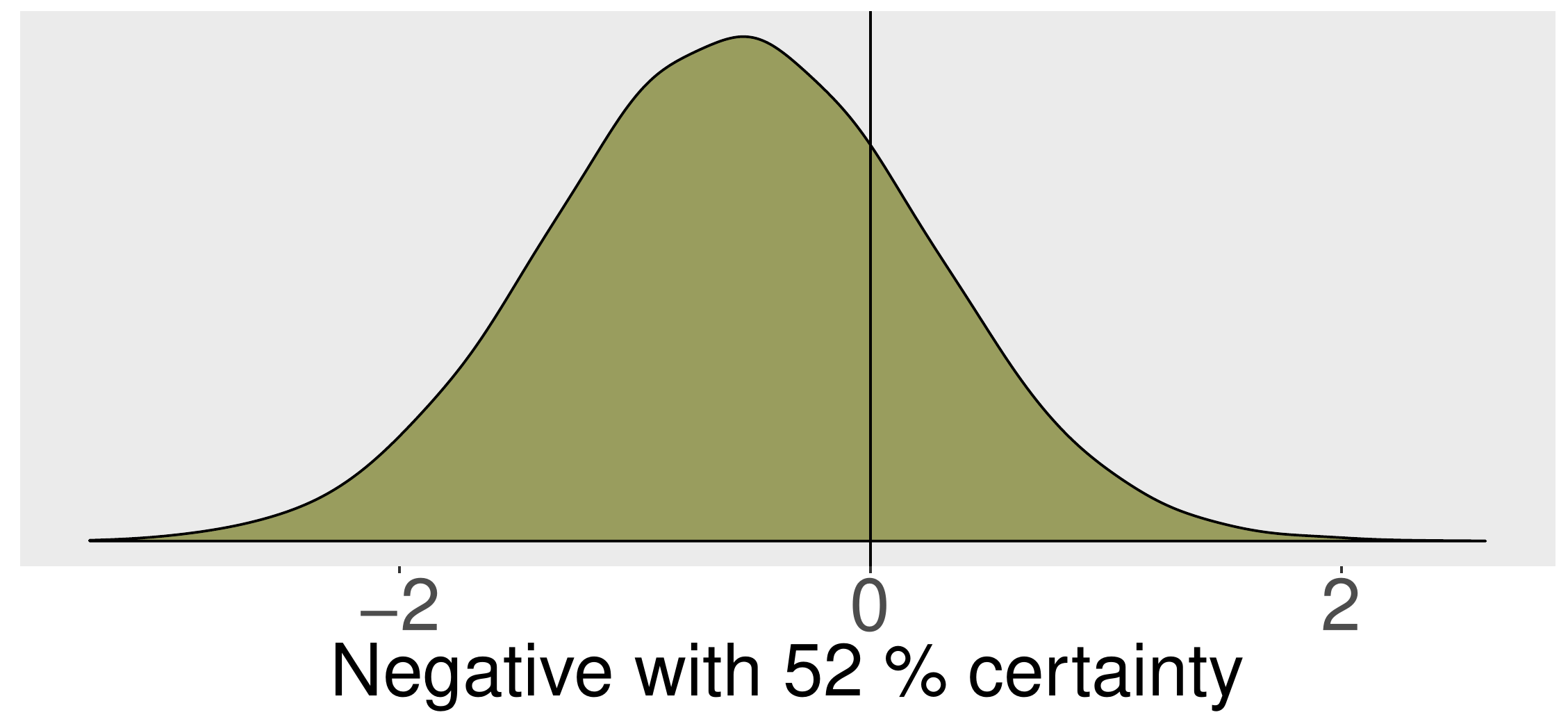}
    \caption*{dep\_..} 
  \end{minipage}
  \hfill
  \begin{minipage}[b]{0.32\linewidth}
    \centering
    \includegraphics[width=\linewidth]{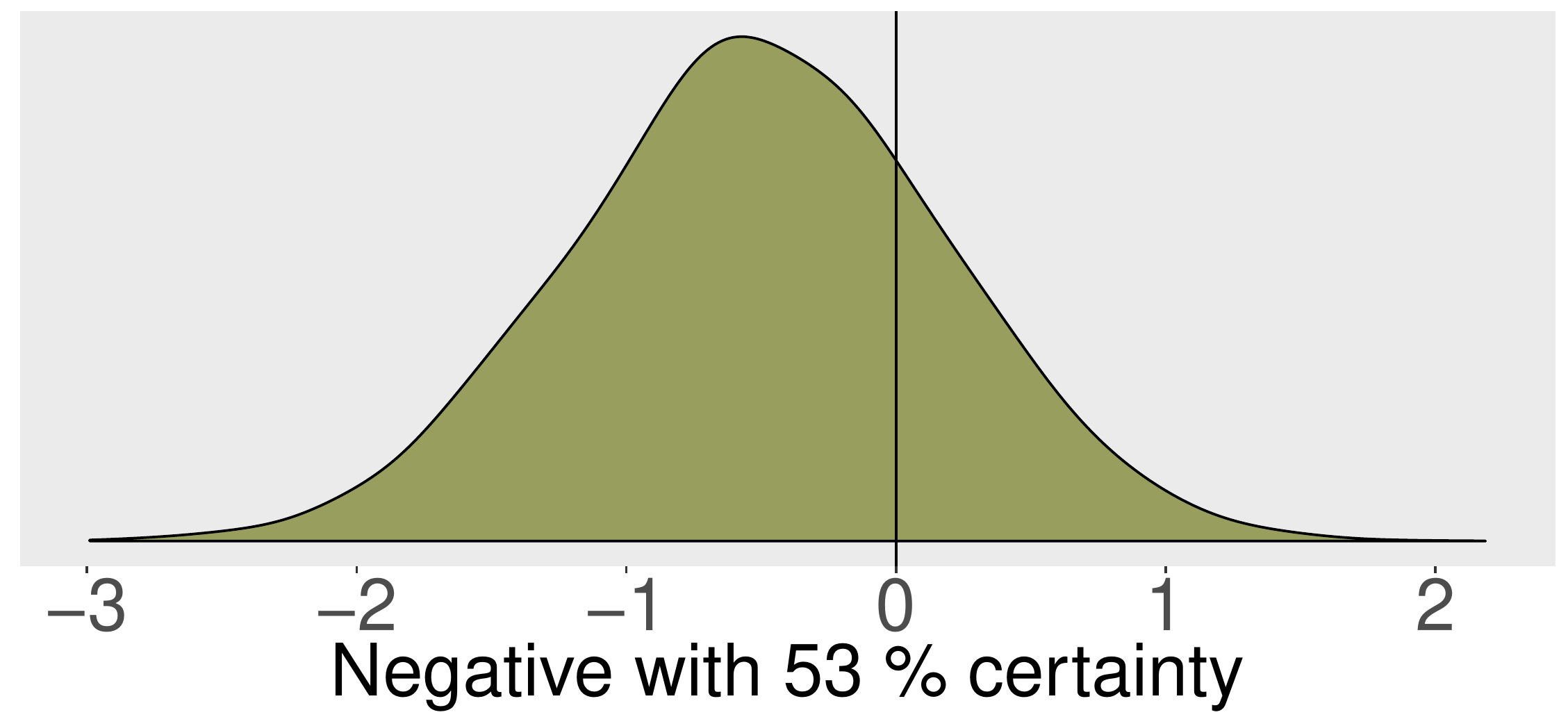}
    \caption*{dep\_CA} 
  \end{minipage}
  \hfill
  \begin{minipage}[b]{0.32\linewidth}
    \centering
    \includegraphics[width=\linewidth]{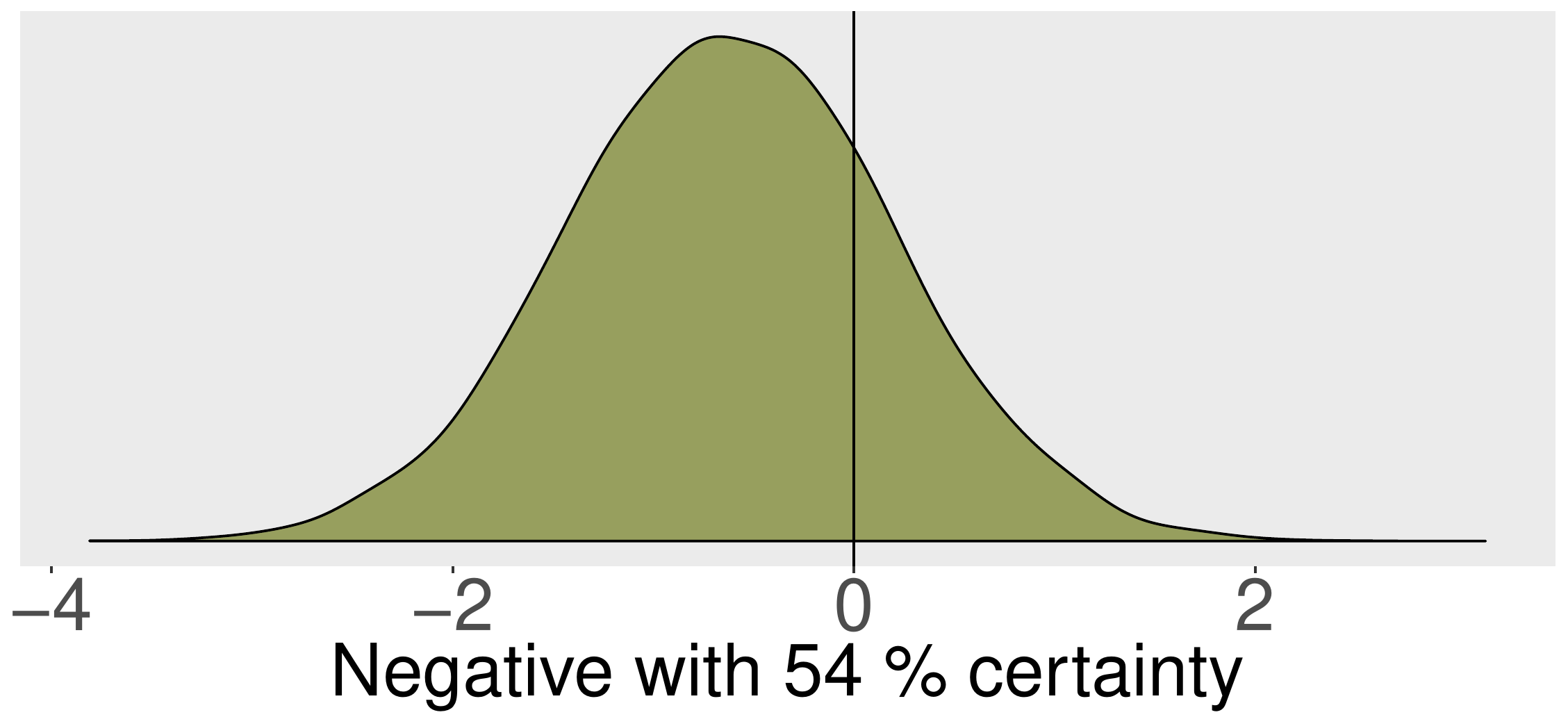}
    \caption*{dep\_SP} 
  \end{minipage}
  \hfill
  \begin{minipage}[b]{0.32\linewidth}
    \centering
    \includegraphics[width=\linewidth]{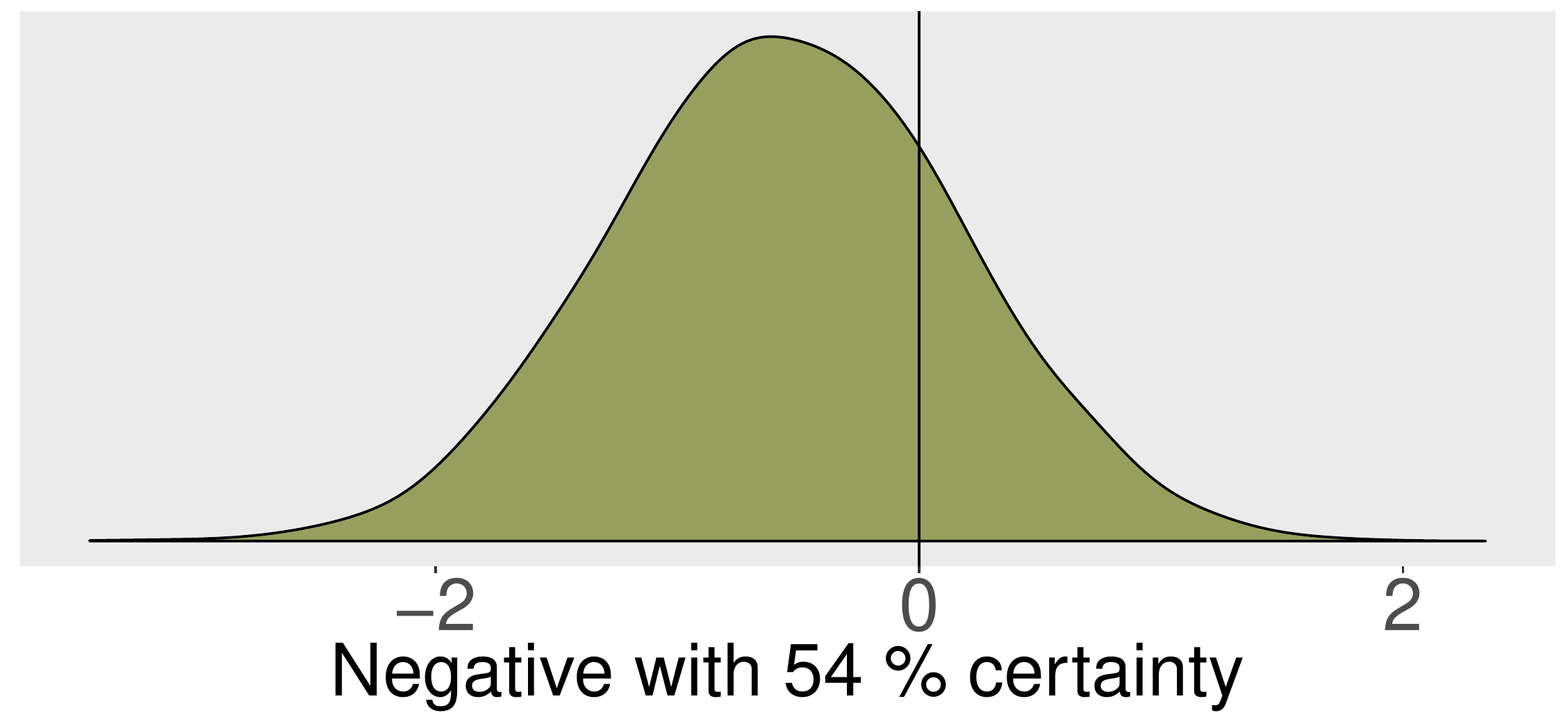}
    \caption*{dep\_I.} 
  \end{minipage}
  \hfill
  \begin{minipage}[b]{0.32\linewidth}
    \centering
    \includegraphics[width=\linewidth]{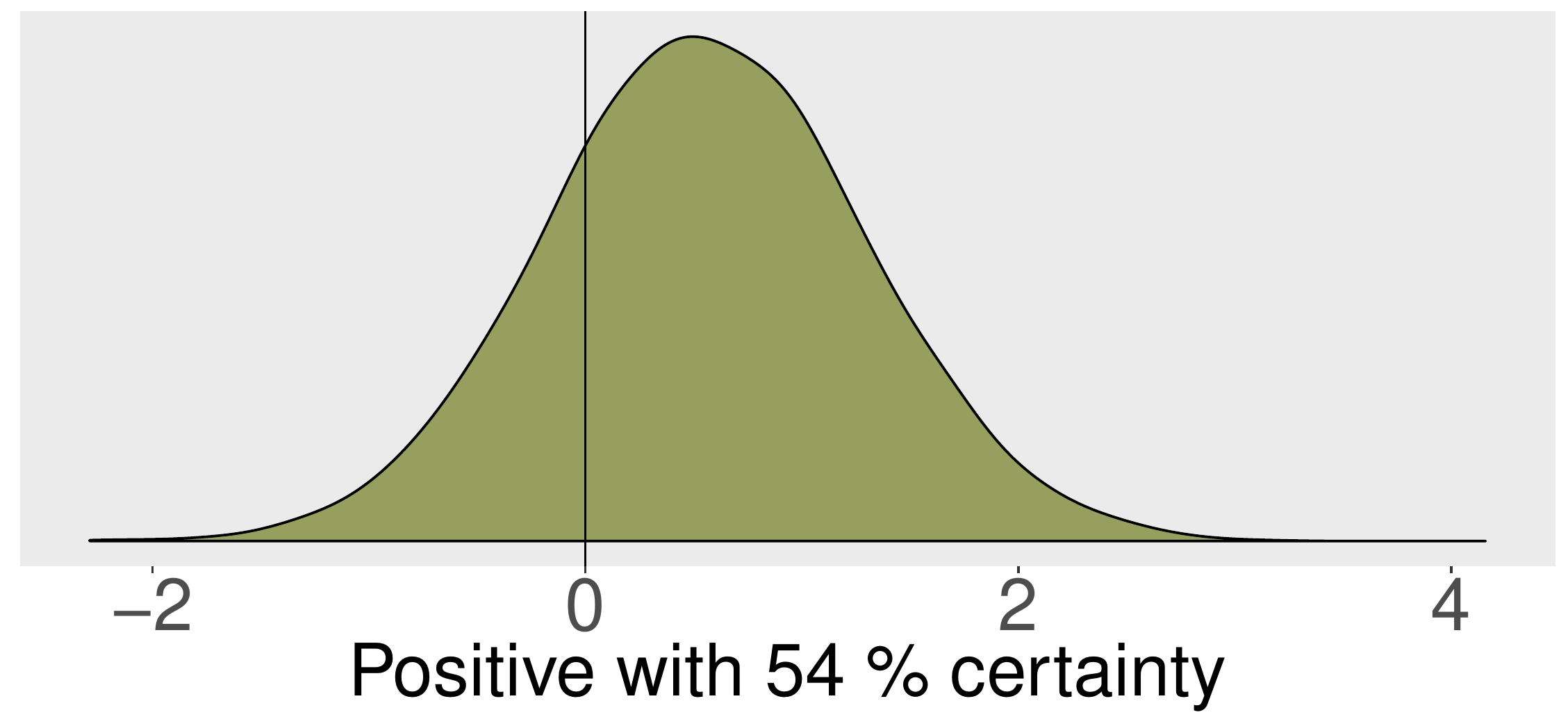}
    \caption*{dep\_.F} 
  \end{minipage}
  \hfill
  \begin{minipage}[b]{0.32\linewidth}
    \centering
    \includegraphics[width=\linewidth]{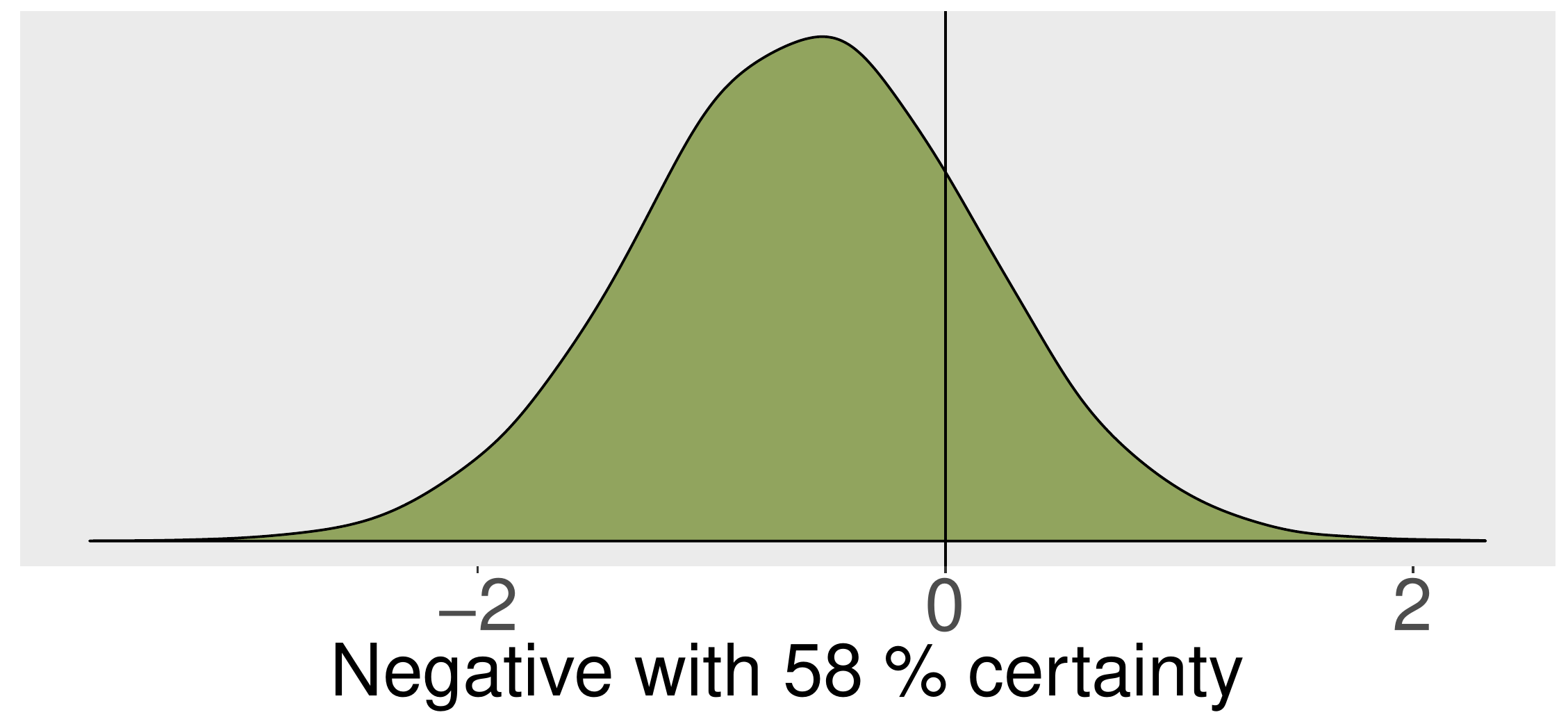}
    \caption*{dep\_RA} 
  \end{minipage}
  \hfill
  \begin{minipage}[b]{0.32\linewidth}
    \centering
    \includegraphics[width=\linewidth]{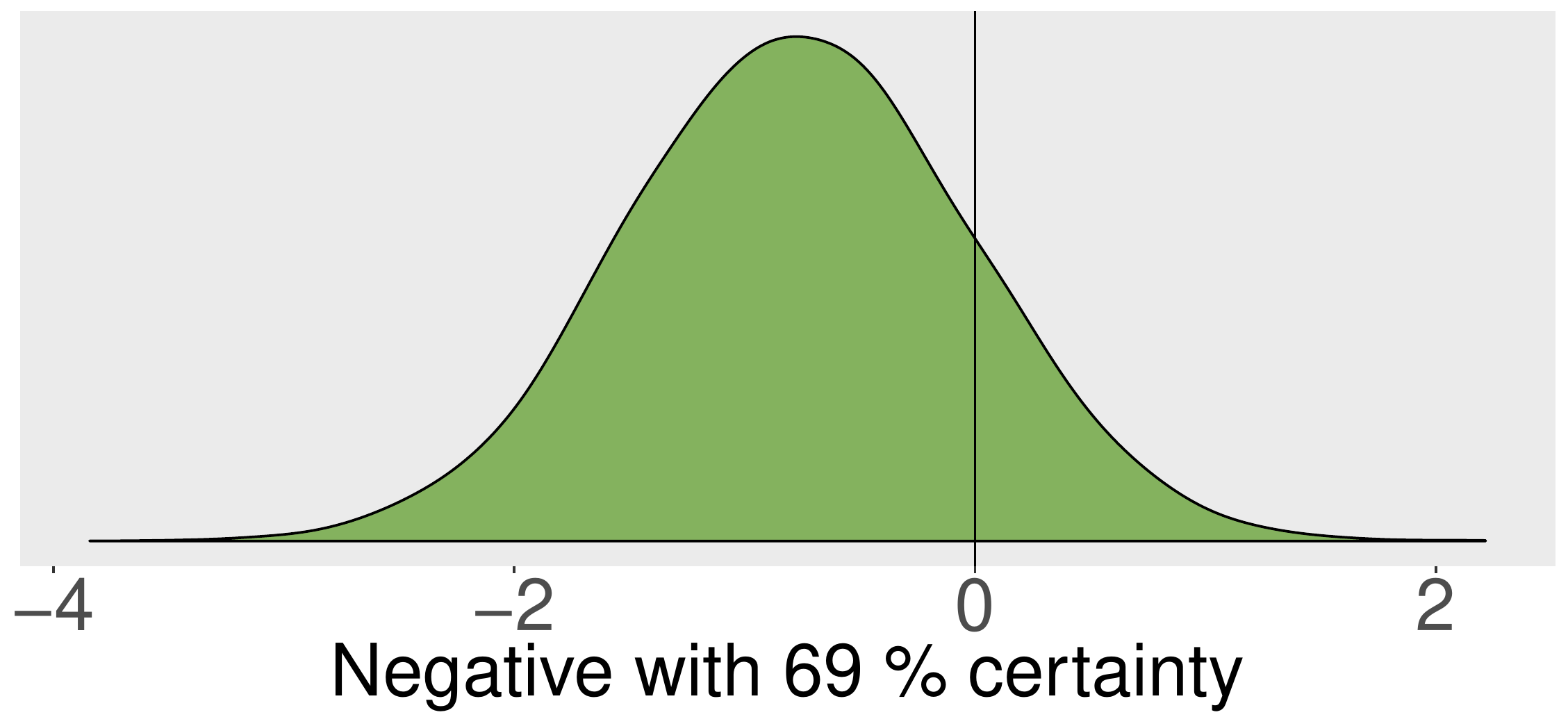}
    \caption*{dep\_VS} 
  \end{minipage}
  \hfill
  \begin{minipage}[b]{0.32\linewidth}
    \centering
    \includegraphics[width=\linewidth]{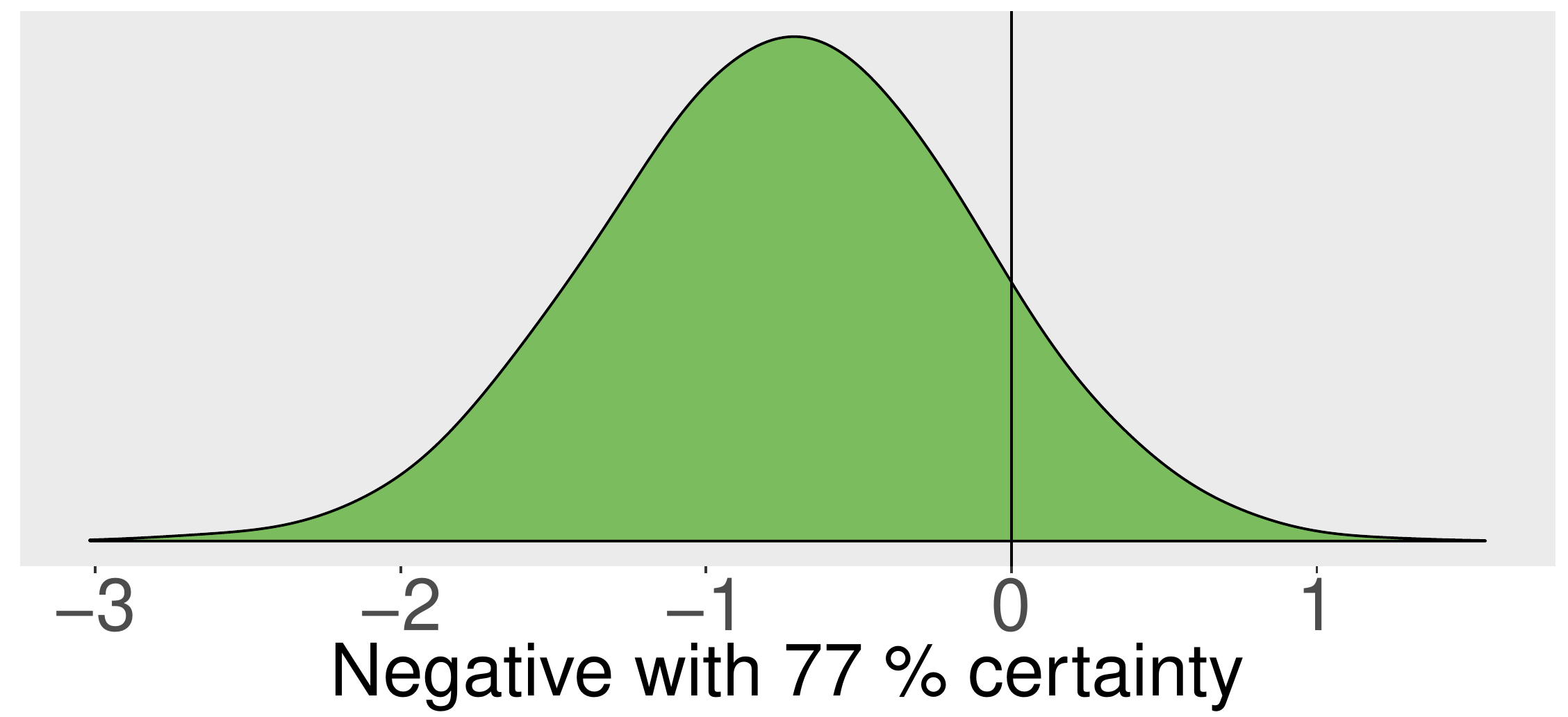}
    \caption*{dep\_XX} 
  \end{minipage}
  \hfill
  \begin{minipage}[b]{0.32\linewidth}
    \centering
    \includegraphics[width=\linewidth]{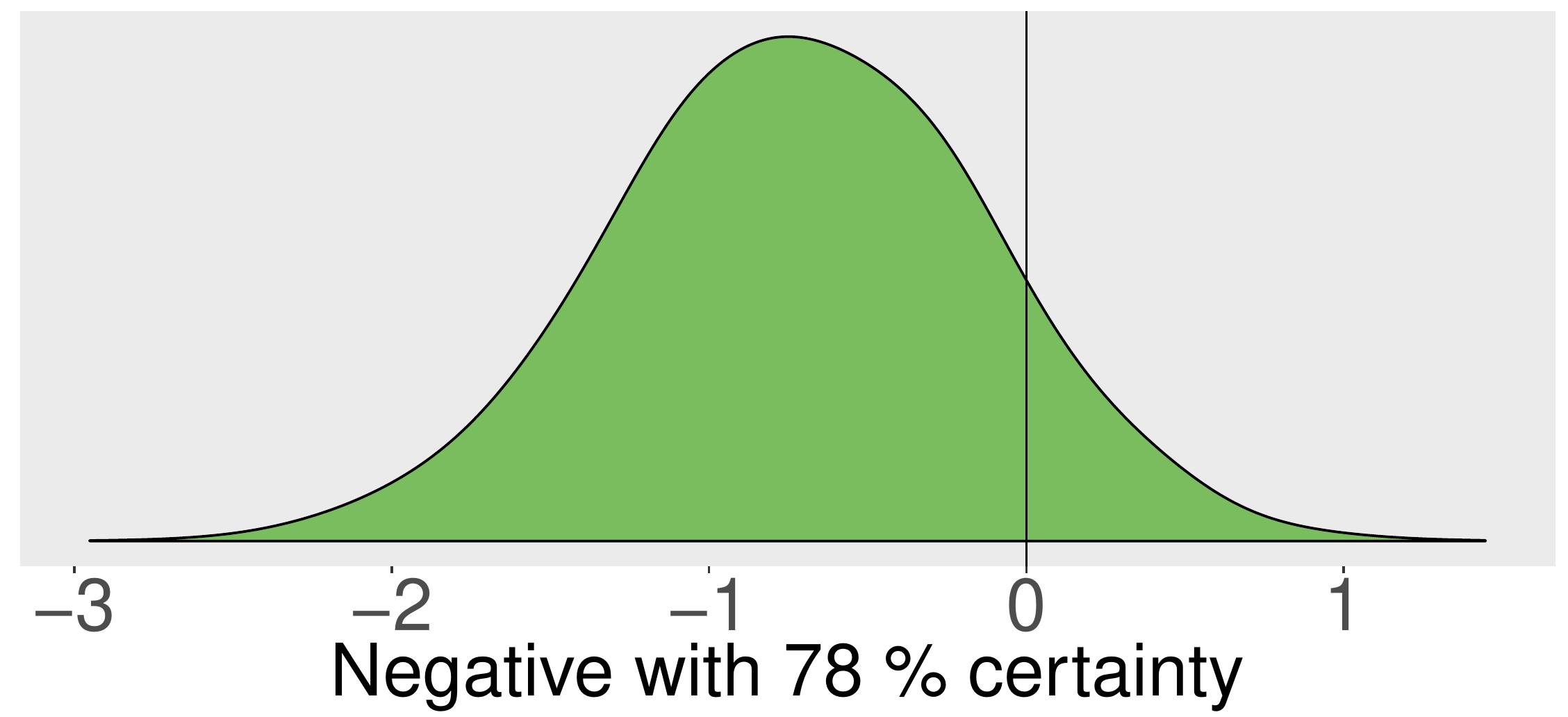}
    \caption*{dep\_IO} 
  \end{minipage}
  \hfill
  \begin{minipage}[b]{0.32\linewidth}
    \centering
    \includegraphics[width=\linewidth]{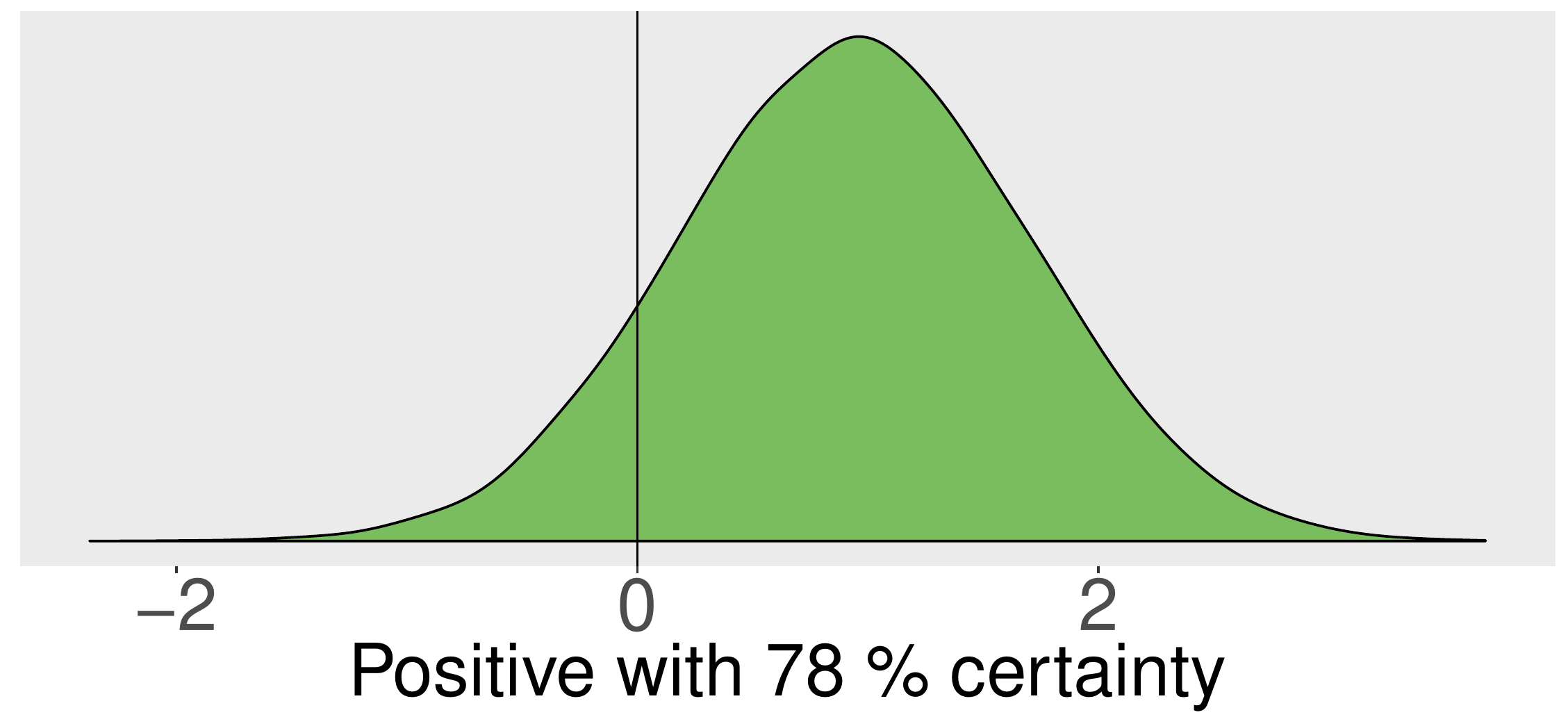}
    \caption*{dep\_HD} 
  \end{minipage}
  \hfill
  \begin{minipage}[b]{0.32\linewidth}
    \centering
    \includegraphics[width=\linewidth]{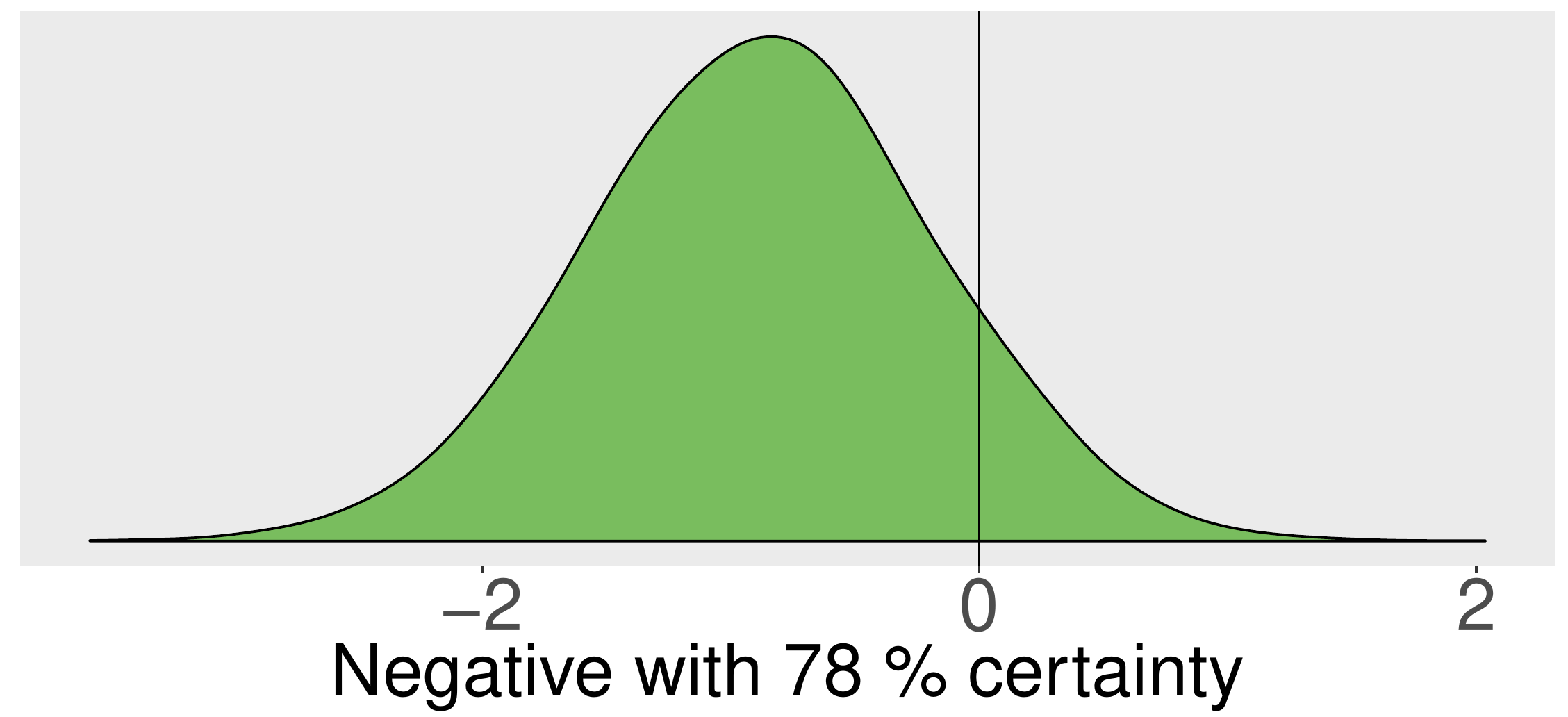}
    \caption*{verbArity1} 
  \end{minipage}
  \end{center}
\end{figure}

\begin{figure}[t]
\ContinuedFloat
  \begin{center}
  \begin{minipage}[b]{0.32\linewidth}
    \centering
    \includegraphics[width=\linewidth]{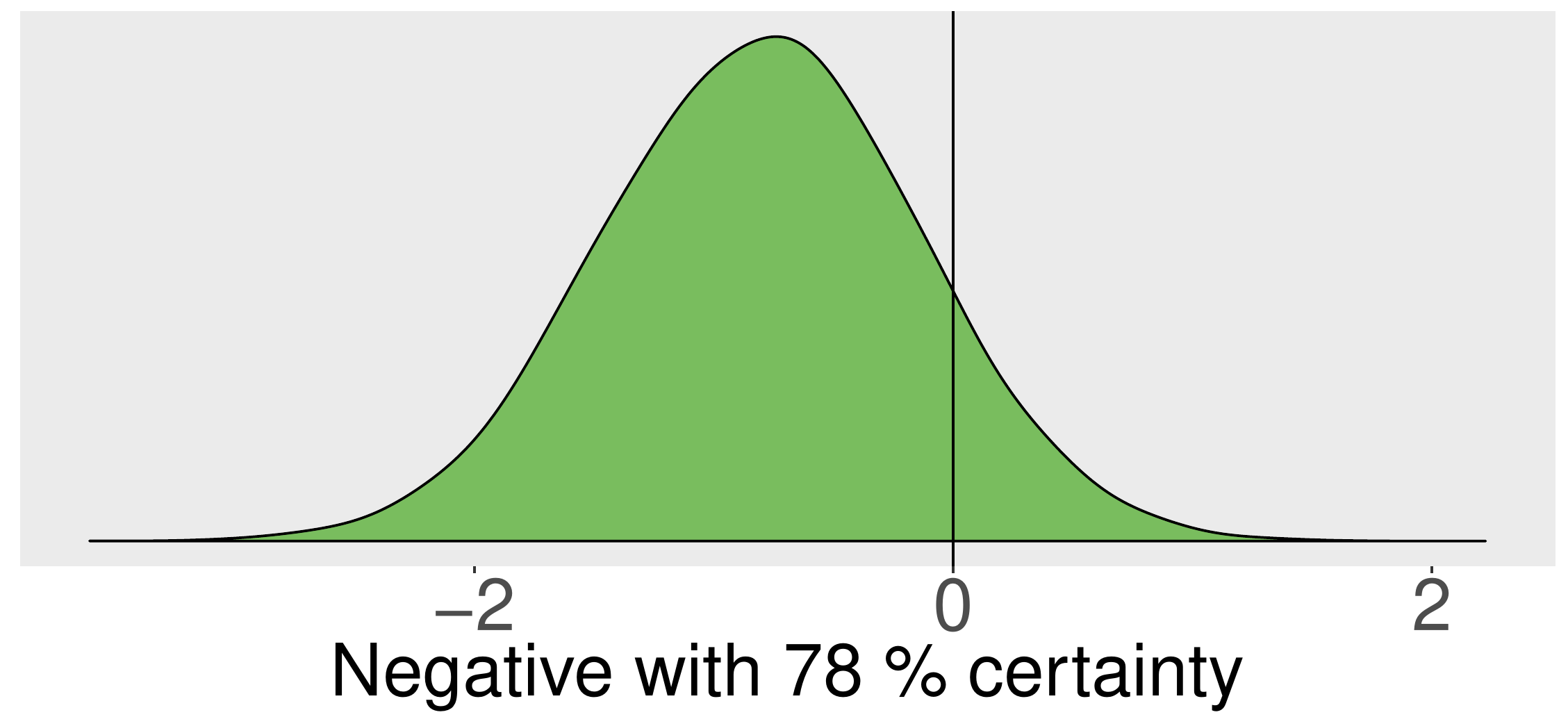}
    \caption*{lexicalDensity} 
  \end{minipage}
  \hfill
  \begin{minipage}[b]{0.32\linewidth}
    \centering
    \includegraphics[width=\linewidth]{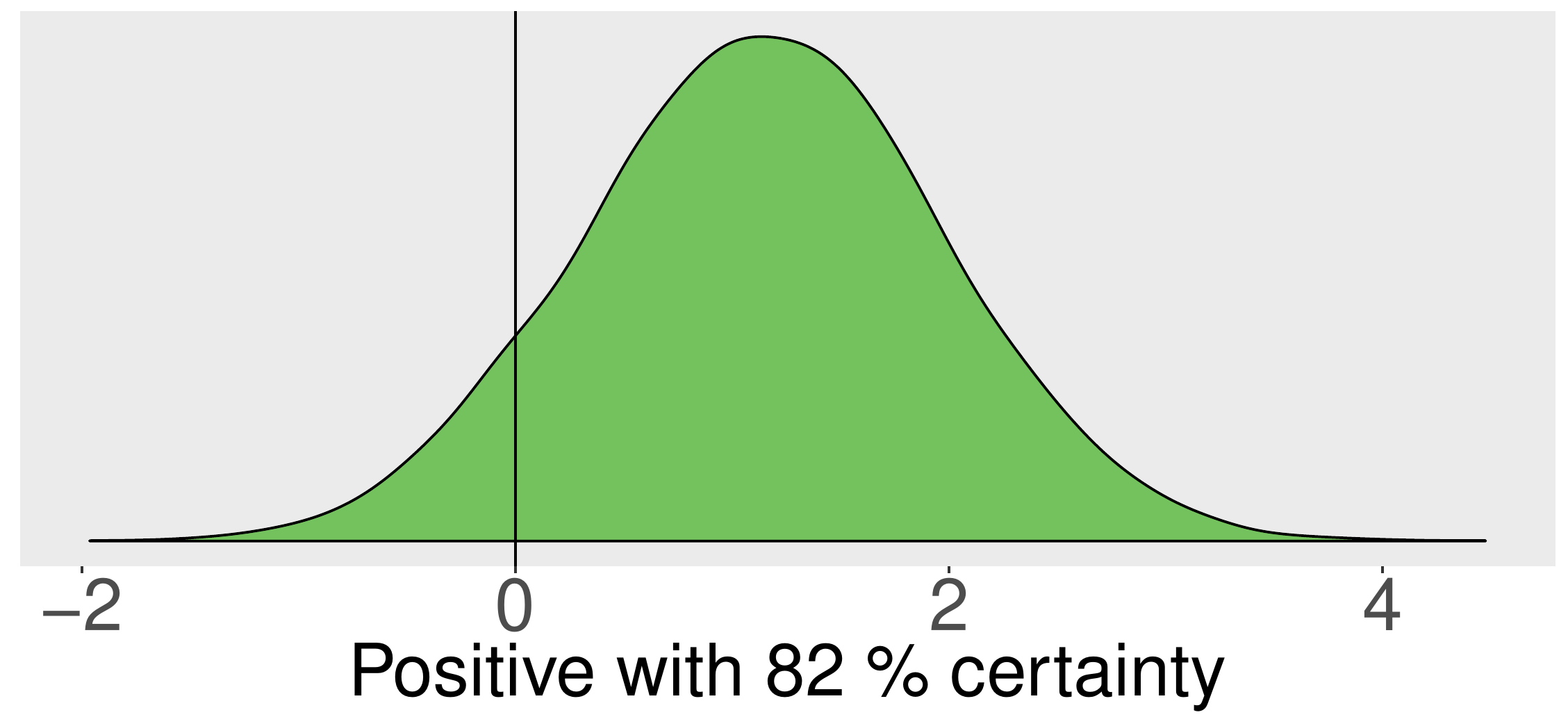}
    \caption*{dep\_IQ} 
  \end{minipage}
  \hfill
  \begin{minipage}[b]{0.32\linewidth}
    \centering
    \includegraphics[width=\linewidth]{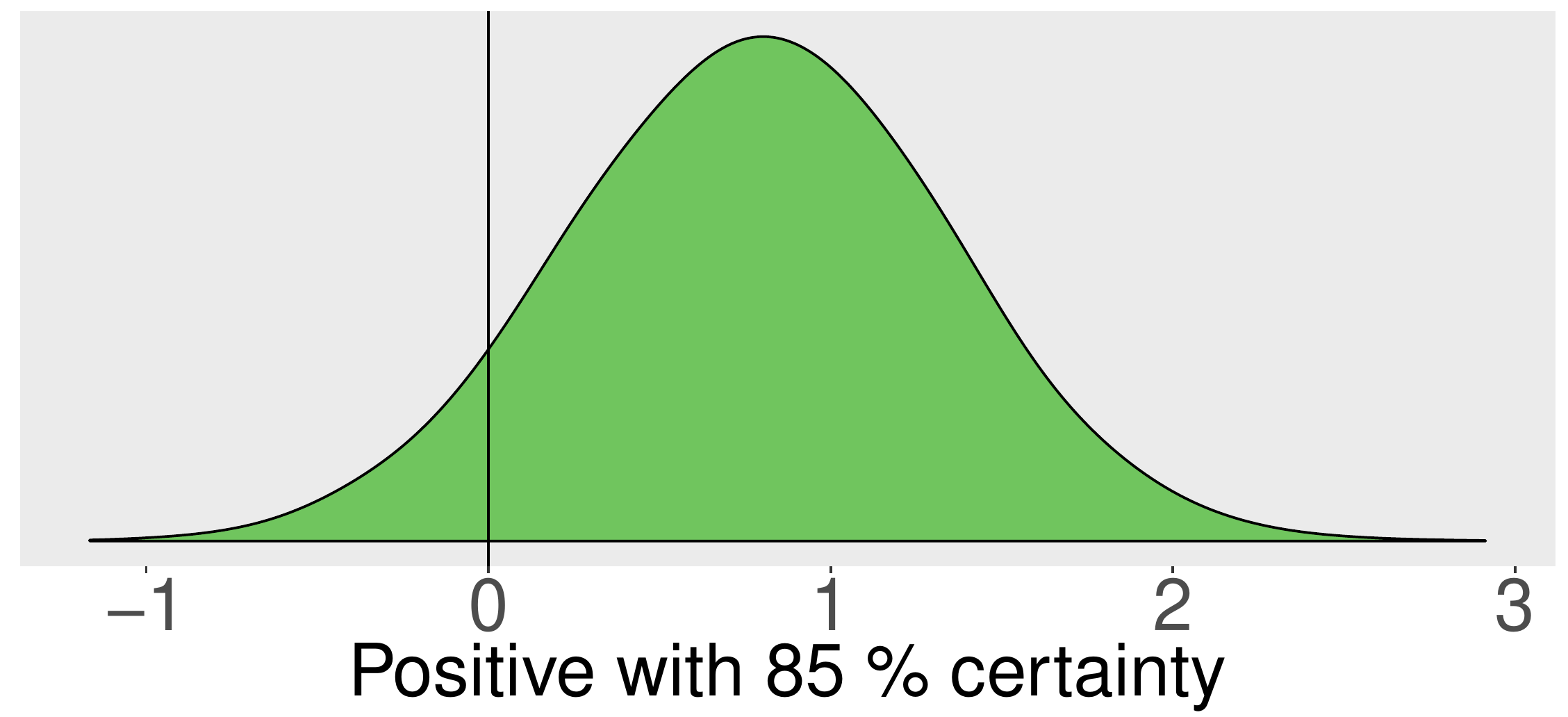}
    \caption*{dep\_XF} 
  \end{minipage}
  \hfill
  \begin{minipage}[b]{0.32\linewidth}
    \centering
    \includegraphics[width=\linewidth]{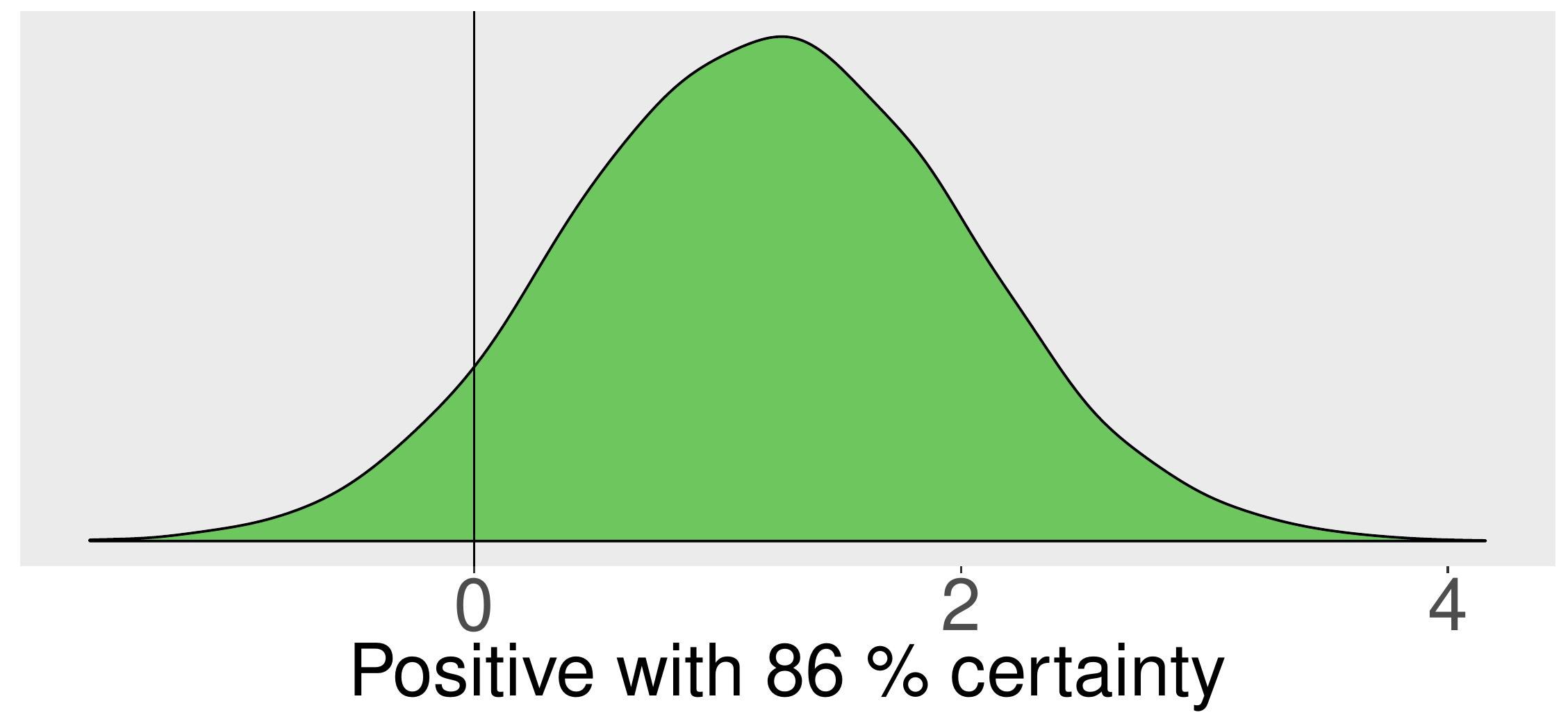}
    \caption*{dep\_AA} 
  \end{minipage}
  \hfill
  \begin{minipage}[b]{0.32\linewidth}
    \centering
    \includegraphics[width=\linewidth]{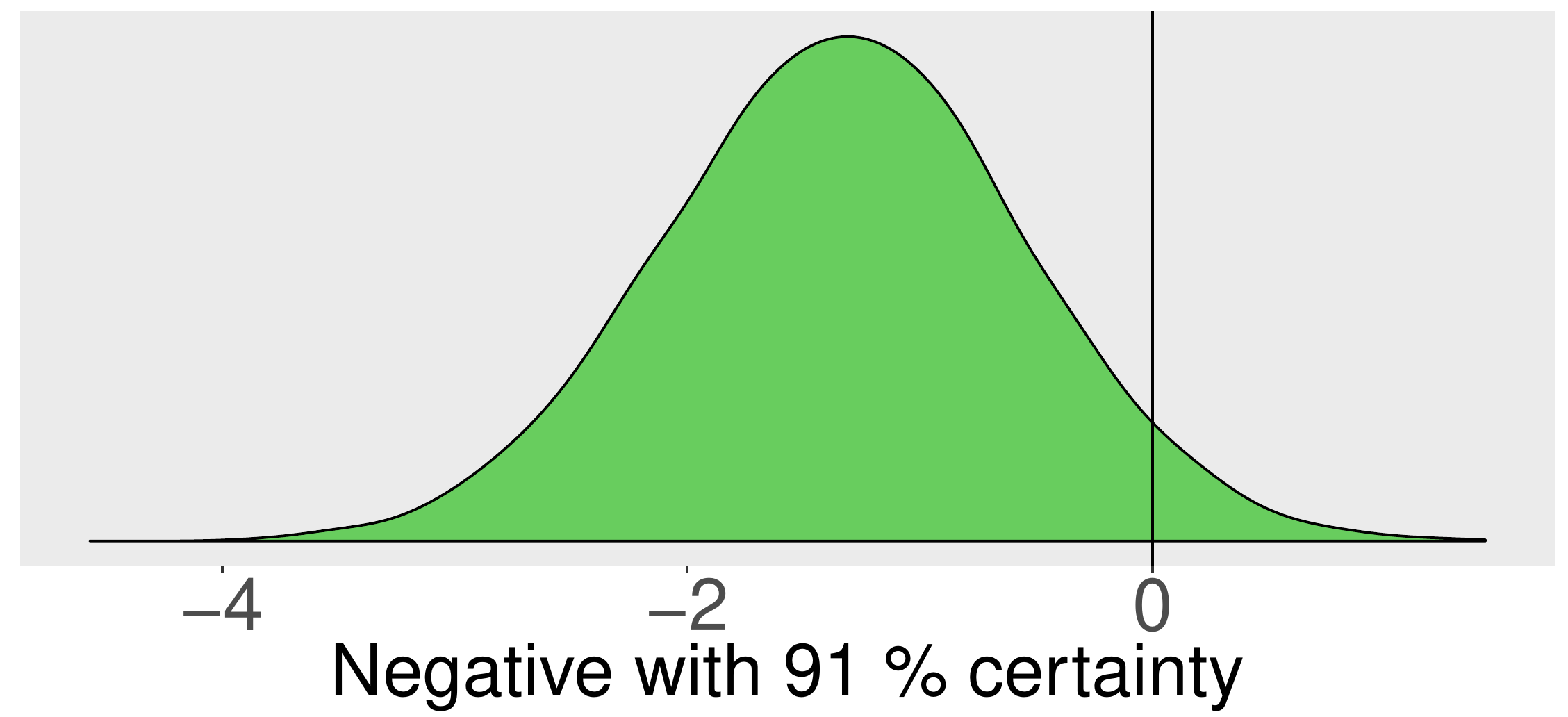}
    \caption*{dep\_TA} 
  \end{minipage}
  \hfill
  \begin{minipage}[b]{0.32\linewidth}
    \centering
    \includegraphics[width=\linewidth]{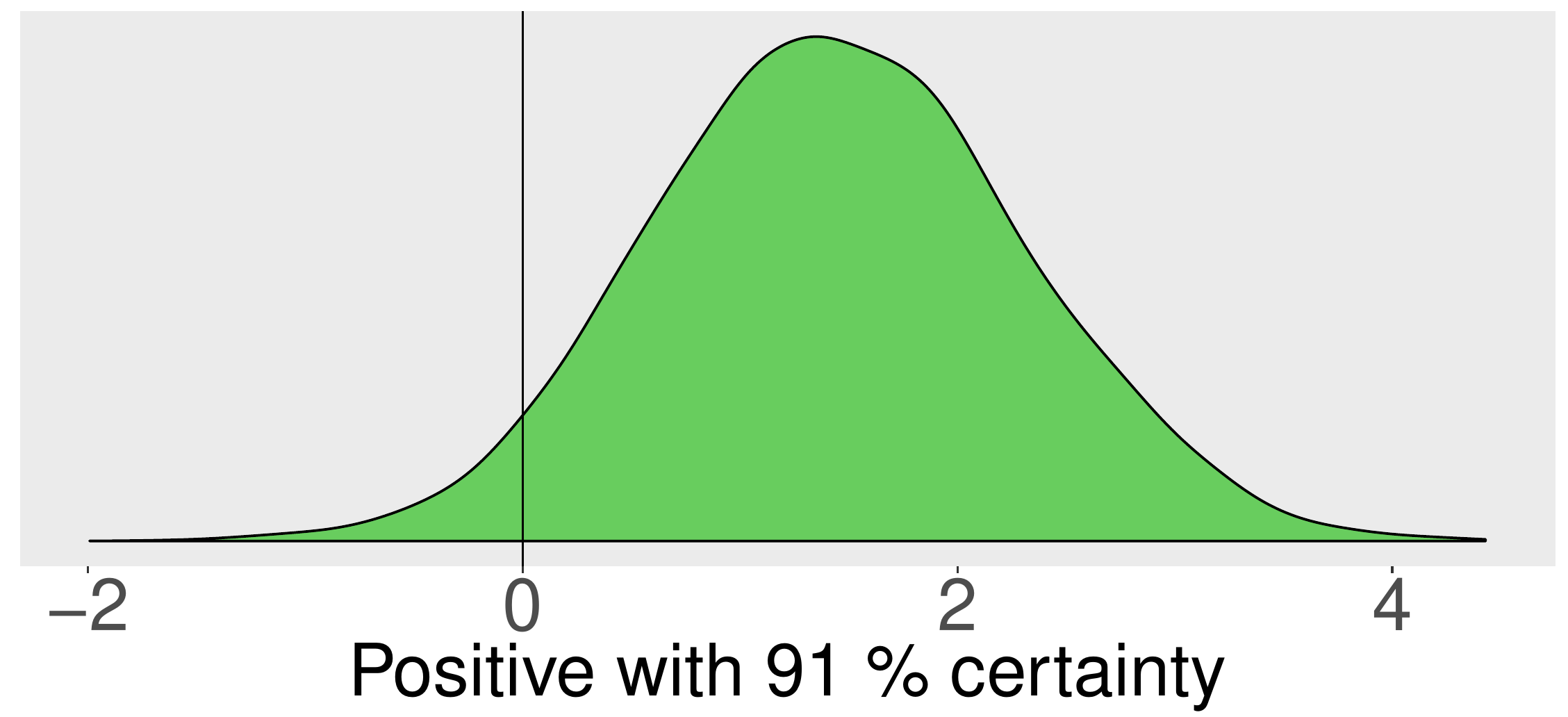}
    \caption*{verbArity6} 
  \end{minipage}
  \hfill
  \begin{minipage}[b]{0.32\linewidth}
    \centering
    \includegraphics[width=\linewidth]{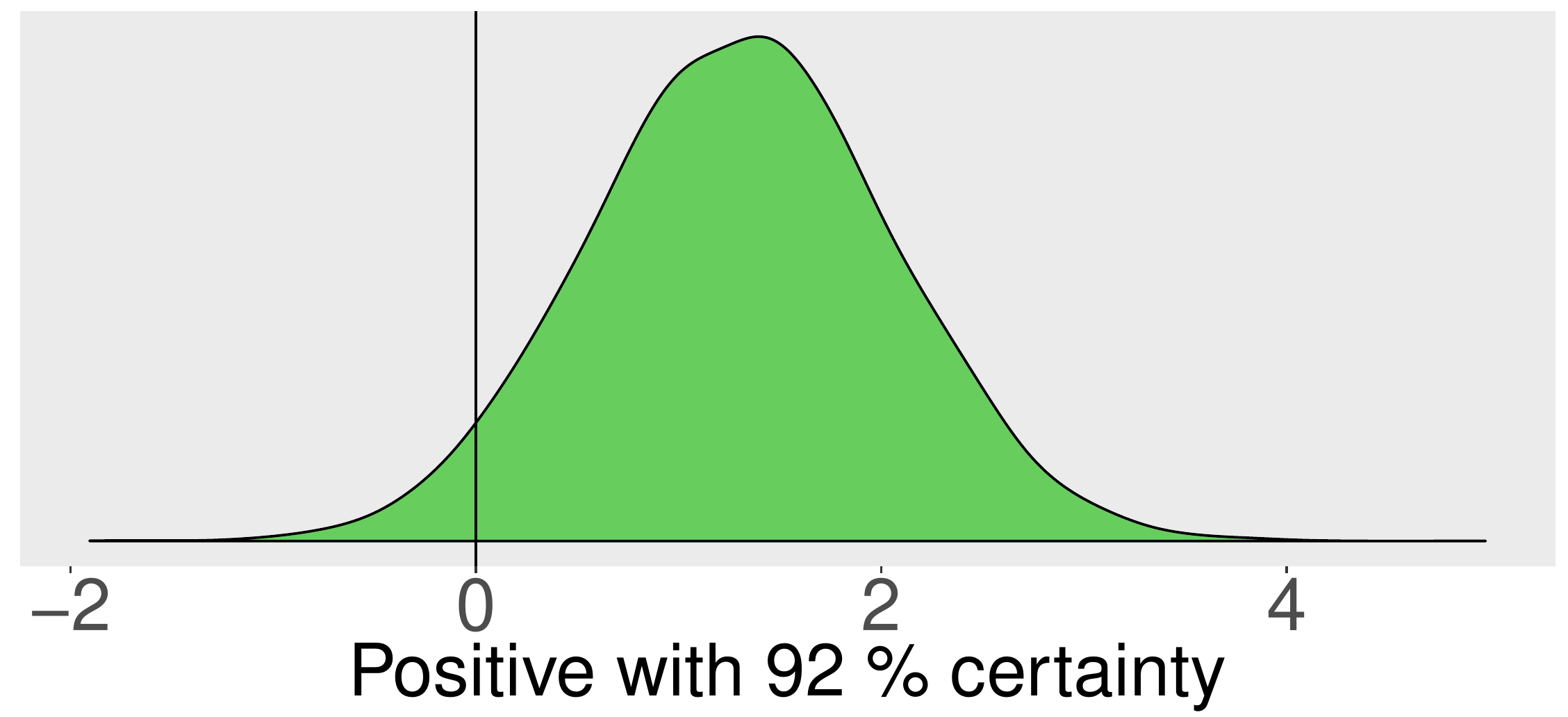}
    \caption*{dep\_IF} 
  \end{minipage}
  \hfill
  \begin{minipage}[b]{0.32\linewidth}
    \centering
    \includegraphics[width=\linewidth]{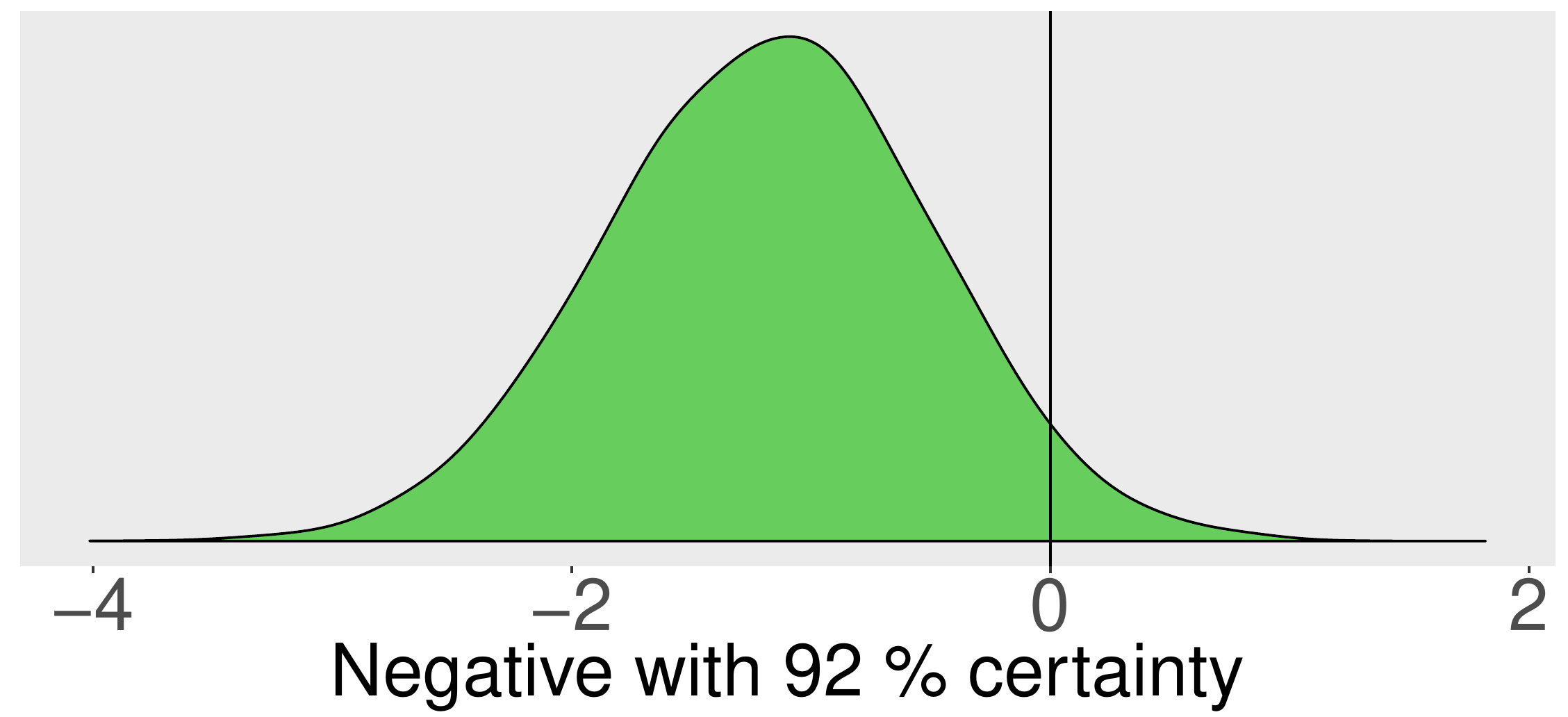}
    \caption*{dep\_IU} 
  \end{minipage}
  \hfill
  \begin{minipage}[b]{0.32\linewidth}
    \centering
    \includegraphics[width=\linewidth]{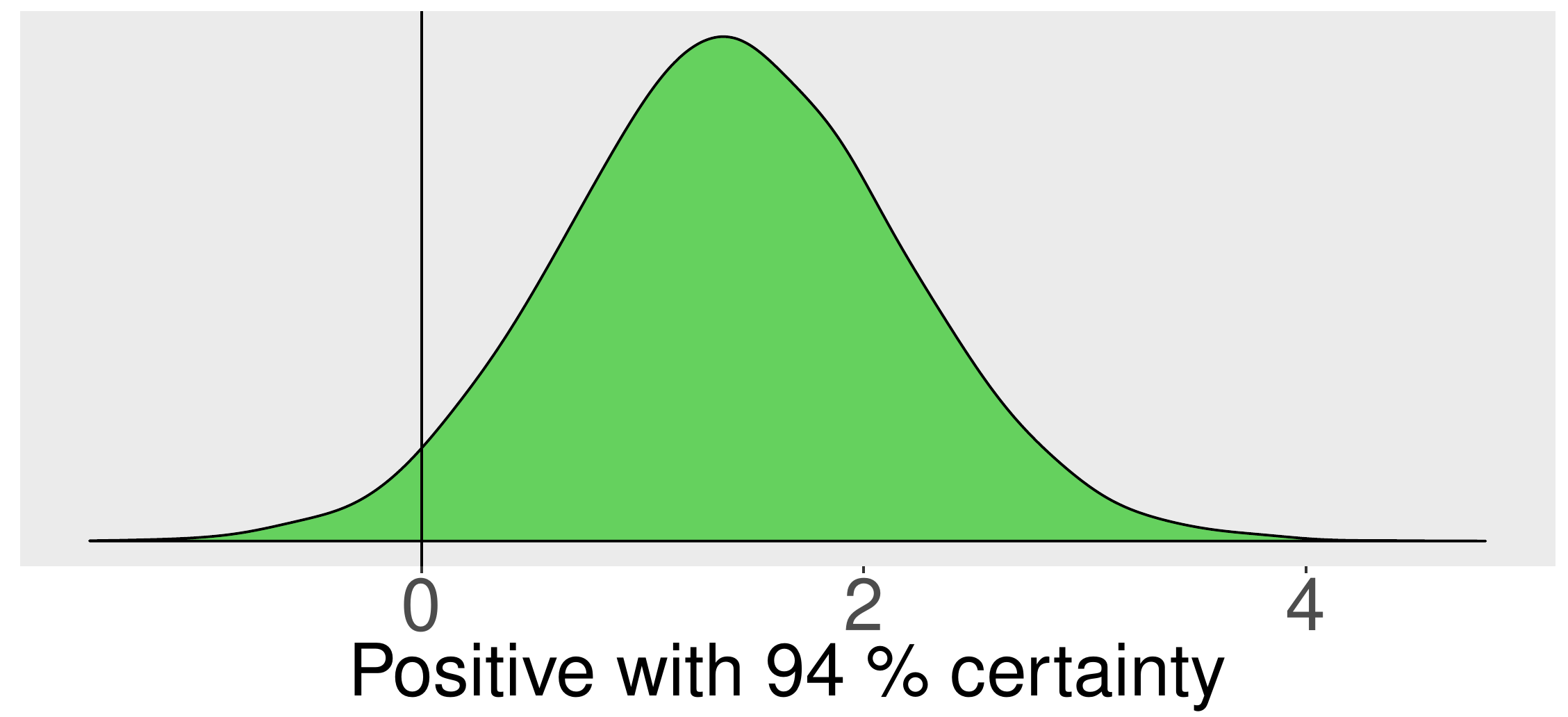}
    \caption*{dep\_OP} 
  \end{minipage}
  \hfill
  \begin{minipage}[b]{0.32\linewidth}
    \centering
    \includegraphics[width=\linewidth]{images/beta_posterior_96_ratioRightDeps.pdf}
    \caption*{ratioRightDeps} 
  \end{minipage}
  \hfill
  \begin{minipage}[b]{0.32\linewidth}
    \centering
    \includegraphics[width=\linewidth]{images/beta_posterior_97_ratioSweVocTotal.pdf}
    \caption*{ratioSweVocTotal} 
  \end{minipage}
  \hfill
  \begin{minipage}[b]{0.32\linewidth}
    \centering
    \includegraphics[width=\linewidth]{images/beta_posterior_98_pos_HP.pdf}
    \caption*{pos\_HP} 
  \end{minipage}
    \end{center}
  \caption{Marginal posteriors for all coefficients in $\boldsymbol{\beta}$.}
  \label{fig:beta_posterior_uncertain_all}
\end{figure}

\end{document}